\documentclass{emulateapj}

\usepackage{url}
\usepackage{epstopdf}
\usepackage{multirow}
\usepackage{amsmath}
\usepackage{natbib}
\usepackage{color}
\usepackage{longtable}

\newcommand\name[1]{{\small\sc #1}}

\slugcomment{}

\shorttitle{The long-term centimeter variability of AGNs}
\shortauthors{Park \& Trippe}

\begin{document}

\title{The long-term centimeter variability of active galactic nuclei:\\ A new relation between variability timescale and accretion rate\altaffilmark{1}}

\author{Jongho Park \& Sascha Trippe\altaffilmark{2}}
\affil{Department of Physics and Astronomy, Seoul National University, Gwanak-gu, Seoul 08826, South Korea}
\email{jhpark@astro.snu.ac.kr (Jongho Park), trippe@astro.snu.ac.kr (Sascha Trippe)}
\received{...}
\accepted{...}

\altaffiltext{1}{Based on observations obtained by the University of Michigan Radio Astronomy Observatory (UMRAO)}
\altaffiltext{2}{Corresponding author}

\begin{abstract}
We study the long-term ($\approx$30 years) radio variability of 43 radio bright AGNs by exploiting the data base of the University of Michigan Radio Astronomy Observatory (UMRAO) monitoring program. We model the periodograms (temporal power spectra) of the observed lightcurves as simple power-law noise (red noise, spectral power $P(f)\propto f^{-\beta}$) using Monte Carlo simulations, taking into account windowing effects (red-noise leak, aliasing). The power spectra of 39 (out of 43) sources are in good agreement with the models, yielding a range in power spectral index ($\beta$) from $\approx$1 to $\approx$3. We fit a Gaussian function to each flare in a given lightcurve to obtain the flare duration. We discover a correlation between $\beta$ and the median duration of the flares. We use the derivative of a lightcurve to obtain a characteristic variability timescale which does not depend on the assumed functional form of the flares, incomplete fitting, and so on. We find that, once the effects of relativistic Doppler boosting are corrected for, the variability timescales of our sources are proportional to the accretion rate to the power of $0.25\pm 0.03$ over five orders of magnitude in accretion rate, regardless of source type. We further find that modelling the periodograms of four of our sources requires the assumption of broken powerlaw spectra. From simulating lightcurves as superpositions of exponential flares we conclude that strong overlap of flares leads to featureless simple power-law periodograms of AGNs at radio wavelengths in most cases.
\end{abstract}

\keywords{Galaxies: active --- radiation mechanisms: non-thermal --- methods: statistical}

\section{Introduction \label{sect_ntro}}

Active Galactic Nuclei (AGNs) are characterized by strong temporal flux variability, which can provide valuable information on the complex physical processes of accretion and plasma outflows of AGNs (see, e.g., \citealt{Ulrich1997} for a review). A number of studies have found that various types of AGNs, from Seyfert galaxies (e.g., \citealt{Lawrence1987}) to quasars (e.g., \citealt{Kelly2009}) and radio bright AGNs (e.g., \citealt{Hovatta2007}), show ubiquitous aperiodic variability across various wavebands. The temporal power spectra or periodograms (see \citealt{Priestley1981} for an exhaustive review of time series analysis) -- i.e., the square moduli of the Fourier transforms -- of lightcurves have been employed to quantify the statistical properties of AGN variability (e.g., \citealt{Abramowicz1991, Fan1999, Benlloch2001, Aller2003, Do2009, Rani2009, Rani2010, Trippe2011, Gupta2012}). In many cases, their power spectra globally follow power laws $P(f)\propto f^{-\beta}$ with $\beta > 0$, corresponding to \emph{red noise}\footnote{Formally, the term ``red noise'' is reserved for the specific case $\beta = 2$; in astronomy however, ``red noise'' is conventionally used for any $\beta > 0$ (cf. \citealt{Vaughan2005, Vaughan2010}). We adopt the astronomical nomenclature in this paper.} \citep{Press1978}. We note that, in time series theory, the term ``noise'' generically refers to random intrinsic variations in brightness (i.e., not to measurement errors or instrumental noise). 

Even though the fact that AGN lightcurves show red noise power spectra was discovered almost 40 years ago \citep{Press1978} and many studies have confirmed since then that this is a generic property of AGNs, it is still unclear why different AGNs show different characteristic variability patterns. Especially interesting have been sources that show a break in their power spectra, resulting in different slopes ($\beta$) below and above a certain frequency (the \emph{break frequency}) (e.g., \citealt{Uttley2002, McHardy2004}). Such behaviour is often seen in optical and/or X-ray variability of Seyfert galaxies, quasars, and even galactic black holes (GBHs; \citealt{Uttley2002, McHardy2004, Kelly2009, Kelly2011}). The break frequencies for both GBHs and AGNs show an anti-correlation with the black hole mass \citep{McHardy2004, UM2005, Kelly2009, Kelly2011} and are also related to the accretion rate \citep{McHardy2006}. The presence of break frequencies in the power spectra indicates that there is a characteristic timescale that governs the variability. Candidate timescales are the light crossing timescale, the orbital timescale, the disk thermal timescale, and the disk viscous timescale (see e.g., \citealt{Kelly2009, Kelly2011} for more details). These timescales are functions of the size of emitting regions such as accretion disks and hot coronae. Thus, it makes sense that they scale with the black hole mass because each length scale is proportional to the Schwarzschild radius, though the actual underlying relations must involve geometry and other physical properties of the emitting system.

Compared to the optical/X-ray variability of AGNs, the understanding of radio variability of AGNs is poor. The variability mechanism of \emph{radio bright}\footnote{We use the term \emph{radio bright AGNs} because not all of our sources might be radio loud, i.e., have a radio-to-optical flux density ratio higher than a certain threshold value. All sources we discuss show strong activity at radio bands, including multiple flares during the time of observation.} AGNs is quite different from that of radio faint, optical and/or X-ray bright AGNs. They emit strong non-thermal emission that is usually thought to originate from relativistic jets \citep{BK1979}. Characteristic variability patterns are seen, especially flares or outbursts in the lightcurves (e.g., \citealt{Valtaoja1999}), which have been associated with shocks in jets (e.g., \citealt{Marscher1985, Hughes1985}, see also \citealt{Fromm2011}). Blazars, a subset of AGNs that comprises most of the radio bright AGNs show violent flux variability across the entire electromagnetic spectrum. This special property of blazars is arguably related to relativistic jets (almost) aligned with the line of sight (e.g., \citealt{Jorstad2005}), implying that relativistic Doppler boosting plays an important role. Accordingly, a number of physical parameters and various mechanisms are involved in generating the complicated variability of radio bright AGNs at radio wavelengths.

Additional difficulties arise from limited sampling of lightcurves of AGNs and statistical analyses that do not consider the red-noise properties intrinsic to AGN lightcurves. As already noted by \cite{Park2014}, Monte Carlo simulations of red-noise lightcurves are essential to reveal the intrinsic statistical properties of AGN lightcurves that are usually masked by uneven, finite sampling. Despite the importance (a) of the effects of limited sampling or windowing (e.g., \citealt{Uttley2002}, see also \citealt{Isobe2015}), which are often described as \emph{red-noise leak} and \emph{aliasing}, (b) of using goodness-of-fit tests correctly \citep{Papadakis1993}, and (c) of deriving the statistical significance of supposed quasi-periodic oscillation (QPO) signals in red noise power spectra \citep{Benlloch2001, Vaughan2005, Vaughan2010}, multiple studies of AGN radio variability focused on apparent characteristic variability timescales while using the incorrect assumption of constant (as function of $f$) significance levels in power spectra, which is only valid when $\beta$ is close to 0 (\citealt{Ciaramella2004, Hovatta2007, Nieppola2009}, but, see \citealt{MaxM2014a, MaxM2014b, Ramakrishnan2015} for recent progress). In line with this, the characteristic timescales derived from using structure functions (SFs; \citealt{Simonetti1985, Hughes1992}) have been interpreted as physical variability timescales of AGNs \citep{Ciaramella2004, Hovatta2007, Nieppola2009}. However, as argued by \cite{Emmanoulopoulos2010}, this approach is probably misleading; in their study, peaks or breaks appeared in the SFs for all their simulated lightcurves even though there was no intrinsic characteristic timescale. Artificial signals appeared at timescales close to the length of the timeseries. The signals observed by \citet{Ciaramella2004, Hovatta2007, Nieppola2009} typically correspond to timescales smaller than $1/10$ of the length of the timeseries, leaving the possibility that they are indeed physical. Given the substantial uncertainties in previous timeseries studies, alternative ways to extract variability timescales from lightcurves or power spectra need to be explored.

\begin{deluxetable*}{cccccccccccccc}[t!]

\tablecaption{Properties of the sample \label{Information}}
\tablehead{
\colhead{Object} & \colhead{Type} &
\colhead{$z$} & \colhead{$T$ [yr]} &
\colhead{$\beta_{4.8}$} & \colhead{$\beta_{8.0}$} & \colhead{$\beta_{14.5}$} &
\colhead{$\tau_{\rm med}^{4.8}$} & \colhead{$\tau_{\rm med}^{8.0}$} & \colhead{$\tau_{\rm med}^{14.5}$} &
\colhead{$\sigma_{\mathrm{der}}^{4.8}$} & \colhead{$\sigma_{\mathrm{der}}^{8.0}$} & \colhead{$\sigma_{\mathrm{der}}^{14.5}$} &
\colhead{$\delta_z$}\\
& (1) & & (2) & & (3) & & & (4) & & & (5) & & (6)
}
\startdata
0133+476 & FSRQ  & 0.859 & 25.45 & $1.8^{+0.21}_{-0.21}$ & $2.0^{+0.21}_{-0.21}$ & $2.2^{+0.05}_{-0.21}$ & 0.87 & 0.71 & 0.98 & 3.49 & 3.29 & 2.76 & 11.1\\
0235+164 & FSRQ & 0.940 & 32.28 & $1.5^{+0.05}_{-0.11}$ & $1.4^{+0.11}_{-0.05}$ & $1.4^{+0.05}_{-0.05}$ & 0.60 & 0.55 & 0.43 & 3.40 & 3.72 & 4.40 & 12.4\\
0316+413 & GAL  & 0.018 & 32.44 & $3.3^{+0.60}_{-0.50}$ & $2.6^{+0.40}_{-0.05}$ & $2.6^{+0.11}_{-0.05}$ & 6.54 & 5.45 & 3.82 & 0.33 & 0.71 & 1.14 & 0.3\\
0333+321 & FSRQ  & 1.258 & 25.41 & ... & $2.1^{+0.21}_{-0.05}$ & ... & ... & 0.97 & ... & ... & 2.98 & ... & 9.8\\
0336$-$019 & FSRQ  & 0.852 & 22.29 & ... & $1.4^{+0.11}_{-0.11}$ & ... & ... & 0.60 & ... & ... & 4.70 & ... & 9.4\\
0355+508 & FSRQ  & 1.520 & 25.56 & ... & $2.3^{+0.30}_{-0.21}$ & $2.0^{+0.30}_{-0.05}$ & ... & 1.81 & 2.17 & ... & 1.27 & 1.11 & 5.2\\
0415+379 & GAL  & 0.049 & 25.42 & ... & $1.6^{+0.21}_{-0.11}$ & ... & ... & 0.87 & ... & ... & 2.15 & ... & 2.7\\
0420$-$014 & FSRQ  & 0.916 & 32.40 & $1.7^{+0.21}_{-0.11}$ & $1.6^{+0.11}_{-0.05}$ & $1.7^{+0.11}_{-0.05}$ &1.31 & 0.92 & 0.93 & 2.28 & 2.73 & 2.58 & 10.4\\
0422+004 & BLO  & 0.310 & 25.59 & ... & $1.4^{+0.11}_{-0.11}$ & ... & ... & 0.63 & ... & ... & 3.64 & ... & ...\\
0430+052 & GAL  & 0.033 & 25.33 & $1.8^{+0.11}_{-0.05}$ & $1.7^{+0.21}_{-0.11}$ & $1.7^{+0.11}_{-0.11}$ & 0.75 & 0.59 & 0.55 & 2.89 & 4.34 & 4.42 & 4.2\\
0528+134 & FSRQ  & 2.060 & 24.18 & ... & $2.2^{+0.30}_{-0.30}$ & ... & ... & 0.99 & ... & ... & 1.80 & ... & 12.2\\
0607$-$157 & FSRQ  & 0.323 & 25.28 & $1.6^{+0.11}_{-0.11}$ & $1.8^{+0.11}_{-0.11}$ & $1.9^{+0.11}_{-0.11}$ & 0.89 & 0.88 & 0.93 & 2.62 & 2.64 & 3.03 & ...\\
0716+714 & BLO  & 0.300 & 24.29 & ... & ... & $1.4^{+0.05}_{-0.11}$ & ... & ... & 0.38 & ... & ... & 5.54 & 8.4\\
0735+178 & BLO  & 0.424 & 25.47 & ... & $2.2^{+0.21}_{-0.30}$ & $2.5^{+0.30}_{-0.40}$ & ... & 1.20 & 1.23 & ... & 1.61 & 1.70 & 2.7\\
0851+202 & BLO  & 0.306 & 25.47 & $1.6^{+0.05}_{-0.05}$ & $1.5^{+0.11}_{-0.05}$ & $1.6^{+0.11}_{-0.11}$ & 0.52 & 0.40 & 0.41& 3.77 & 4.40 & 4.69 & 11.6\\
0923+392 & FSRQ  & 0.695 & 25.50 & $2.4^{+0.21}_{-0.11}$ & $2.6^{+0.05}_{-0.40}$ &$2.3^{+0.21}_{-0.05}$ & 2.03 & 1.71 & 2.16 & 1.24 & 1.68 & 1.47 & 2.5\\
1055+018 & FSRQ  & 0.890 & 24.13 & ... & $1.7^{+0.11}_{-0.11}$ & $1.7^{+0.21}_{-0.11}$ & ... & 1.04 & 0.91 & ... & 3.66 & 3.69 & 6.5\\
1101+384 & BLO  & 0.030 & 25.71 & ... & $0.9^{+0.21}_{-0.11}$ & $1.1^{+0.21}_{-0.11}$ & ... & 0.57 & 0.44 & ... & 10.04 & 8.42 & 3$^b$\\
1156+295 & FSRQ  & 0.725 & 24.53 & $1.4^{+0.11}_{-0.11}$ & $1.5^{+0.11}_{-0.05}$ & $1.6^{+0.11}_{-0.11}$ & 0.66 & 0.60 & 0.61 & 3.42 & 5.07 & 3.72 & 16.5\\
1226+023 & FSRQ  & 0.158 & 25.55 & $1.9^{+0.21}_{-0.11}$ & $1.7^{+0.11}_{-0.11}$ & $2.0^{+0.11}_{-0.11}$ & 0.88 & 0.99 & 0.86 & 2.56 & 2.21 & 2.17 & 11.6\\
1253$-$055 & FSRQ  & 0.536 & 32.47 & $2.5^{+0.40}_{-0.21}$ & $2.2^{+0.11}_{-0.21}$ & $1.6^{+0.11}_{-0.05}$ & 1.01 & 1.04 & 0.95 & 1.45 & 1.57 & 1.85 & 15.1\\
1308+326 & FSRQ  & 0.998 & 25.42 & $2.3^{+0.21}_{-0.21}$ & $1.8^{+0.11}_{-0.11}$ & $1.9^{+0.21}_{-0.21}$ & 1.31 & 0.89 & 0.81 & 1.55 & 2.80 & 2.54 & 7.7\\
1335$-$127 & FSRQ  & 0.539 & 25.48 & $1.6^{+0.11}_{-0.21}$ & $1.8^{+0.05}_{-0.21}$ & $1.6^{+0.11}_{-0.11}$ & 0.72 & 0.74 & 0.61 & 2.81 & 3.46 & 3.06 & ...\\
1413+135 & BLO  & 0.247 & 24.78 & ... & $1.0^{+0.05}_{-0.05}$ & $1.3^{+0.05}_{-0.11}$ & ... & 0.67 & 0.53 & ... & 6.29 & 3.96 & 9.8\\
1418+546 & BLO  & 0.153 & 25.68 & ... & $1.7^{+0.11}_{-0.21}$ & $1.6^{+0.11}_{-0.11}$ & ... & 0.72 & 0.63 & ... & 3.29 & 2.91 & 4.4\\
1510$-$089 & FSRQ  & 0.360 & 25.47 & $1.5^{+0.11}_{-0.11}$ & $1.3^{+0.05}_{-0.11}$ & $1.4^{+0.11}_{-0.05}$ & 0.43 & 0.31 & 0.34 & 4.58 & 6.18 & 6.57 & 20.2\\
1633+382 & FSRQ  & 1.813 & 22.92 & ... & $1.8^{+0.30}_{-0.11}$ & $2.1^{+0.60}_{-0.21}$ & ... & 0.96 & 0.71 & ... & 2.51 & 1.78 & 7.6\\
1641+399 & FSRQ  & 0.593 & 32.43 & $2.3^{+0.30}_{-0.11}$ & $2.5^{+0.11}_{-0.21}$ & $2.1^{+0.21}_{-0.21}$ & 1.72 & 1.55 & 1.63 & 1.72 & 1.89 & 2.16 & 9.4\\
1652+398 & BLO  & 0.034 & 25.74 & ... & $1.5^{+0.21}_{-0.21}$ & ... & ... & 0.42 & ... & ... & 5.83 & ... & 10$^c$\\
1730$-$130 & FSRQ  & 0.902 & 32.34 & ... & $1.7^{+0.11}_{-0.05}$ & $1.6^{+0.05}_{-0.21}$ & ... & 0.98 & 1.04 & ... & 1.93 & 1.64 & 5.6\\
1749+096 & BLO  & 0.322 & 25.45 & $1.3^{+0.11}_{-0.05}$ & $1.6^{+0.05}_{-0.11}$ & $1.4^{+0.05}_{-0.11}$ & 0.49 & 0.47 & 0.37 & 5.20 & 5.11 & 5.13 & 9.1\\
1803+784 & BLO  & 0.680 & 24.54 & ... & ... & $1.4^{+0.21}_{-0.11}$ & ... & ... & 0.63 & ... & ... & 3.68 & 11.4\\
1807+698 & BLO & 0.051 & 25.72 & ... & $1.6^{+0.40}_{-0.21}$ & ... & ... & 0.95 & ... & ... & 5.33 & ... & 1.0\\
1921$-$293 & FSRQ  & 0.353 & 25.57 & ... & $1.5^{+0.11}_{-0.11}$ & $1.7^{+0.21}_{-0.11}$ & ... & 0.90 & 0.73 & ... & 2.41 & 2.53 & ...\\
1928+738 & FSRQ  & 0.302 & 21.98 & $1.7^{+0.30}_{-0.21}$ & ... & $1.3^{+0.21}_{-0.05}$ & 1.00 & ... & 0.88 & 2.74 & ... & 5.24 & 1.5\\
2005+403 & FSRQ  & 1.736 & 25.30 & ... & ... & $1.7^{+0.30}_{-0.11}$ & ... & ... & 1.23 & ... & ... & 2.95 & 4.9\\
2007+777 & BLO  & 0.342 & 22.94 & ... & ... & $1.7^{+0.21}_{-0.11}$ & ... & ... & 0.59 & ... & ... & 2.63 & 5.9\\
2134+004 & FSRQ  & 1.945 & 25.72 & ... & $2.0^{+0.40}_{-0.21}$ & $2.0^{+0.21}_{-0.30}$ & ... & 0.84 & 1.20 & ... & 3.83 & 3.06 & 5.5\\
2145+067 & FSRQ  & 0.990 & 25.39 & ... & $2.2^{+0.40}_{-0.11}$ & $2.0^{+0.40}_{-0.21}$ & ... & 2.37 & 1.22 & ... & 1.27 & 2.35 & 7.8\\
2200+420 & BLO  & 0.069 & 32.44 & $1.6^{+0.11}_{-0.11}$ & $1.6^{+0.11}_{-0.05}$ & $1.6^{+0.05}_{-0.05}$ & 0.47 & 0.42 & 0.43 & 3.96 & 4.29 & 4.48 & 6.6\\
2223$-$052 & FSRQ  & 1.404 & 31.99 & $1.9^{+0.11}_{-0.21}$ & $1.7^{+0.11}_{-0.05}$ & $1.9^{+0.11}_{-0.21}$ & 1.42 & 1.60 & 0.90 & 1.96 & 1.92 & 2.36 & 6.7\\
2230+114 & FSRQ  & 1.037 & 25.50 & ... & $1.6^{+0.21}_{-0.05}$ & $2.0^{+0.21}_{-0.21}$ & ... & 0.89 & 0.77 & ... & 4.44 & 3.40 & 8.4\\
2251+158 & FSRQ  & 0.859 & 25.51 & $1.8^{+0.05}_{-0.05}$ & $2.0^{+0.11}_{-0.21}$ & $1.7^{+0.11}_{-0.11}$ & 1.01 & 0.79 & 0.55 & 2.16 & 2.40 & 2.82 & 14.6
\enddata

\tablecomments{Redshifts are taken from the NED. (1) Source types: FSRQ: flat spectrum radio quasars; BLO: BL Lac objects; GAL: radio galaxies; taken from the MOJAVE web site\footnote{\url{http://www.physics.purdue.edu/astro/MOJAVE/allsources.html}}. (2) Total monitoring time $T$ (in years), averaged over data from different frequencies (see Section~\ref{sect2} for details). (3) Power spectral indices of the best-fit simple power law models (Section~\ref{ssect_lc}). (4) Median duration of flares, obtained from fitting peaks in lightcurves with Gaussian functions (Section~\ref{ssect_fit}). (5) Widths of the distributions of the derivatives of lightcurves (Section~\ref{ssect_der}). (6) Doppler factors from the literature, after correction of cosmological redshifts (Section~\ref{ssect_bb}). $^b$Doppler factor from \cite{Lico2012}, see also \cite{Tavecchio1998}. $^c$Doppler factor from \cite{Tavecchio1998}, see also \cite{Katarzynski2001}.}

\end{deluxetable*}

The Ornstein-Uhlenbeck (OU) process, also referred to as continuous time first-order autoregressive process, or a mixture of several OU processes have been suggested to model the observed lightcurves and power spectra of quasars at optical \citep{Kelly2009}, of Seyferts and a GBH at X-rays \citep{Kelly2011}, and of blazars at $\gamma$-rays \citep{Sobolewska2014}. The OU process describes a time series as a superposition of exponentially decaying outbursts occuring at random times and with random amplitudes. A mixture of OU processes is a linear superposition of OU processes, which has been introduced for a better description of AGN lightcurves. These models were motivated by the ``perturbation'' class of astrophysical models (e.g., \citealt{Lyubarskii1997}) which suggests that the propagation of random accretion rate perturbations through the accretion flow is responsible for the observed variability of AGNs and GBHs. One advantage of these models is that they fit models to lightcurves instead of power spectra, which significantly reduces windowing effects. Another advantage is that they use maximum-likelihood or Bayesian techniques to utilize all the information contained in the data. The tight correlation between timescales and black hole masses seen in \cite{Kelly2009, Kelly2011} suggests that those models accurately extract the relevant timescales from lightcurves. 

However, the emission mechanisms of radio bright and radio faint AGNs are different; the former is dominated by synchrotron radiation from relativistic jets, while the latter is dominated by radiation from geometrically thin accretion disks and/or hot coronae. If the same model is applied to systems with different radiation mechanisms are involved, one needs to explain why and how they can share the same statistical properties (this is partially discussed in \citealt{Sobolewska2014}). We will see that almost all (39 out 43) of our sources do not show indications for a break frequency in their power spectra, which might indicate that they have very long characteristic timescales -- if at all. Thus, we make use of Monte Carlo simulations of red noise lightcurves instead of the OU process in this study, following up on our success in unveiling the intrinsic statistical properties of four radio bright AGNs \citep{Park2014}. 

The format of the paper is as follows. In Section~\ref{sect2}, we describe our data and sample. In Section~\ref{sect3}, we explain how we obtain the statistical properties of our sources and relate them with other physical parameters such as the accretion rate in Section~\ref{sect4}. In Section~\ref{sect5}, we summarize our results and conclude. Throughout the paper, we adopt a cosmology with $H_0 = 70\rm\ km\ s^{-1}\ Mpc^{-1}, \Omega_M = 0.3$, and $\Omega_{\Lambda} = 0.7$. All luminosities used in our paper are corrected to our adopted cosmological parameters.

\section{Sample and Data}
\label{sect2}

We exploited the AGN monitoring database of the 26-meter University of Michigan Radio Astronomy Observatory (UMRAO; see \citealt{Aller1985} for technical details) for our study. For a statistical analysis we selected all lightcurves for which the number of data points exceeds 150 after binning and flagging (cf. Section~\ref{sect3}). The number of data points before binning and flagging was 448 on average. This criterion ensured that at least one data point is available every $\approx$2 months on average. For many sources, only one or two of the three UMRAO bands (4.8, 8, and 14.5 GHz) satisfied this criterion. Our selection left us with a sample of 43 sources (20 sources were available at 4.8 GHz, 38 at 8 GHz, 36 at 14.5 GHz). The minimum source flux was around 0.6 Jy for 1101+384 (Mrk 421), the maximum flux around 35 Jy for 1226+023 (3C~273). Our source list comprised 27 flat spectrum radio quasars (FSRQs), 13 BL Lac objects (BLOs), and 3 radio galaxies (GALs). The lightcurves of eight sources, 0235+164, 0316+413, 0420$-$014, 1253$-$055, 1641+399, 1730$-$130, 2200+420, 2223$-$052, span $\approx$32 years in time from 1980 to around 2012; those of the other sources span $\approx$25 years from 1980 to around 2005. Table~\ref{Information} shows an overview over the basic properties of our sources (partially taken from the NASA/IPAC Extragalactic Database, NED\footnote{\url{http://ned.ipac.caltech.edu/}}).

\begin{figure*}[!t]
\begin{center}
\includegraphics[trim=3mm 0mm 0mm 0mm, clip, width = 85mm]{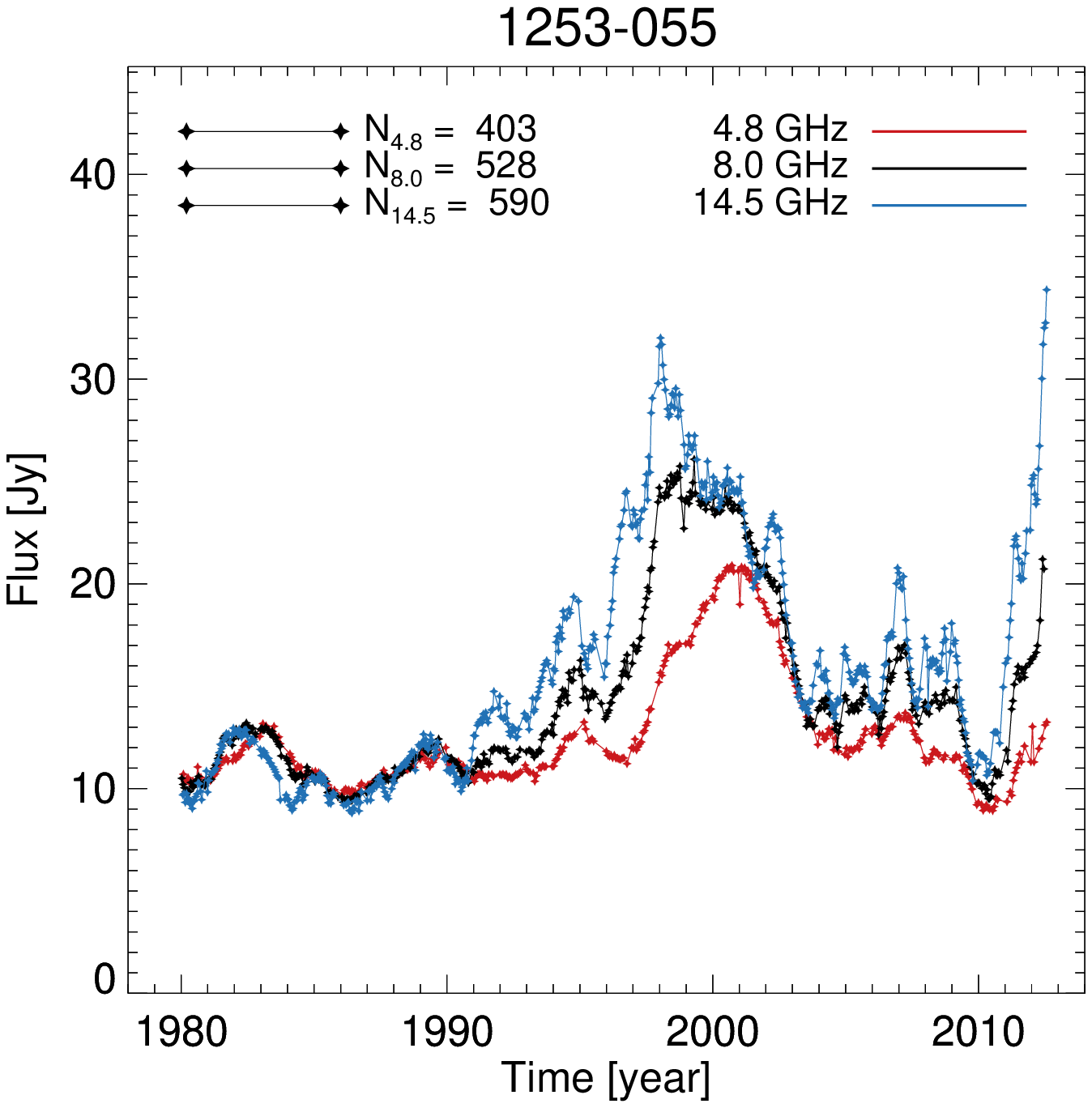}
\includegraphics[trim=0mm 0mm 3mm 0mm, clip, width = 65mm]{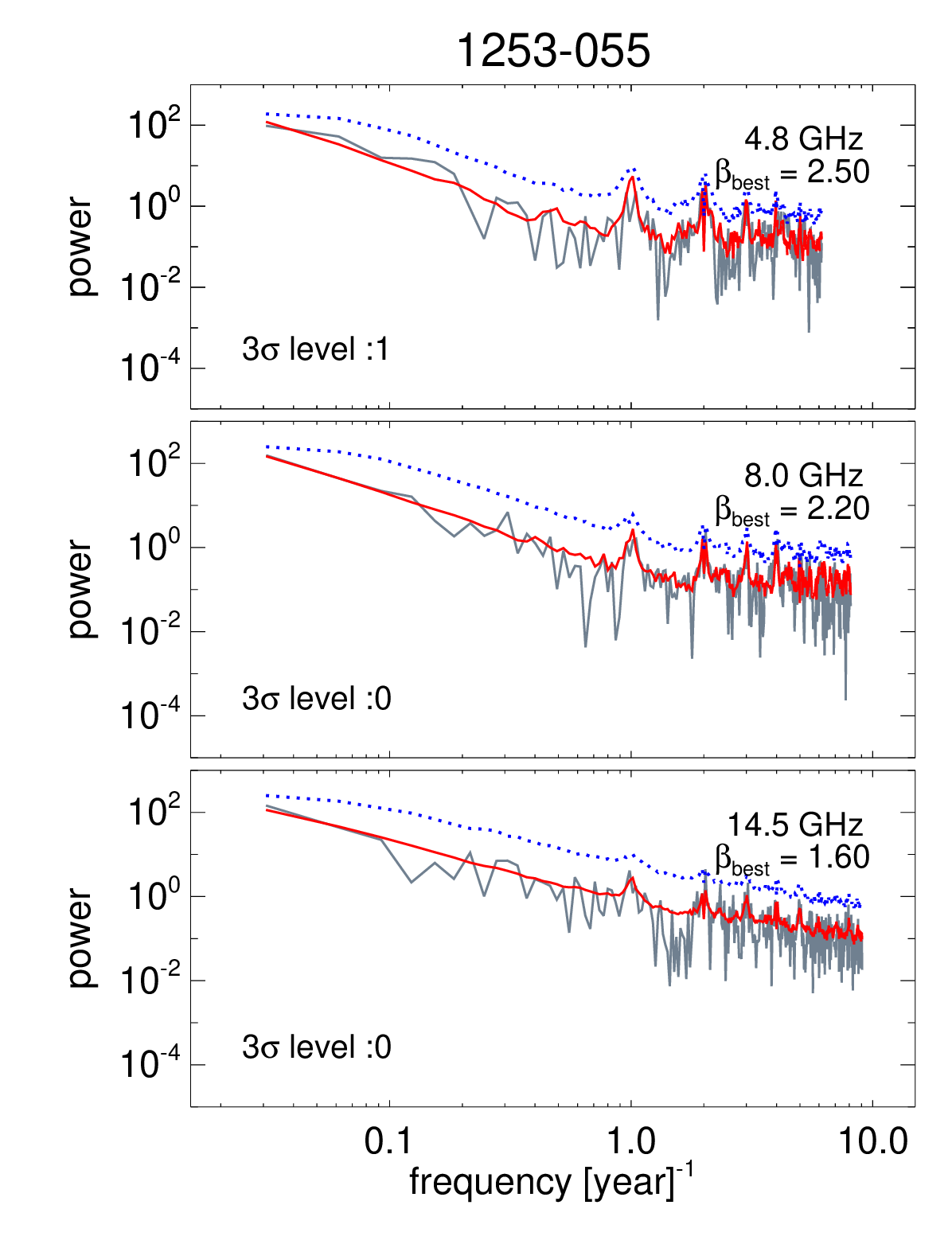}
\caption{Illustration of our flux data and time series analysis. \emph{Left}: Lightcurves of 1253$-$055 (after binning and flagging). Red, black, and blue lines and points are 4.8, 8, and 14.5 GHz data, respectively. The number of data points for each frequency, $N$, is noted. \emph{Right}: Power spectra of the lightcurves in the left panel (black solid lines), the mean power spectra of 5\,000 realizations of our simulation with the best-fit $\beta$ values (red solid lines), and the $3\sigma$ significance levels for possible excess power (blue dotted lines). The best-fit $\beta$ values (``$\beta_{\rm best} = ...$'') and the number of data points that exceed the significance levels (``$3\sigma\ \rm level : ...$'') are noted for each frequency. \label{ex}}
\end{center}
\end{figure*}

\begin{figure*}[!t]
\begin{center}
\includegraphics[trim=4mm 0mm 1mm 0mm, clip, width = 59mm]{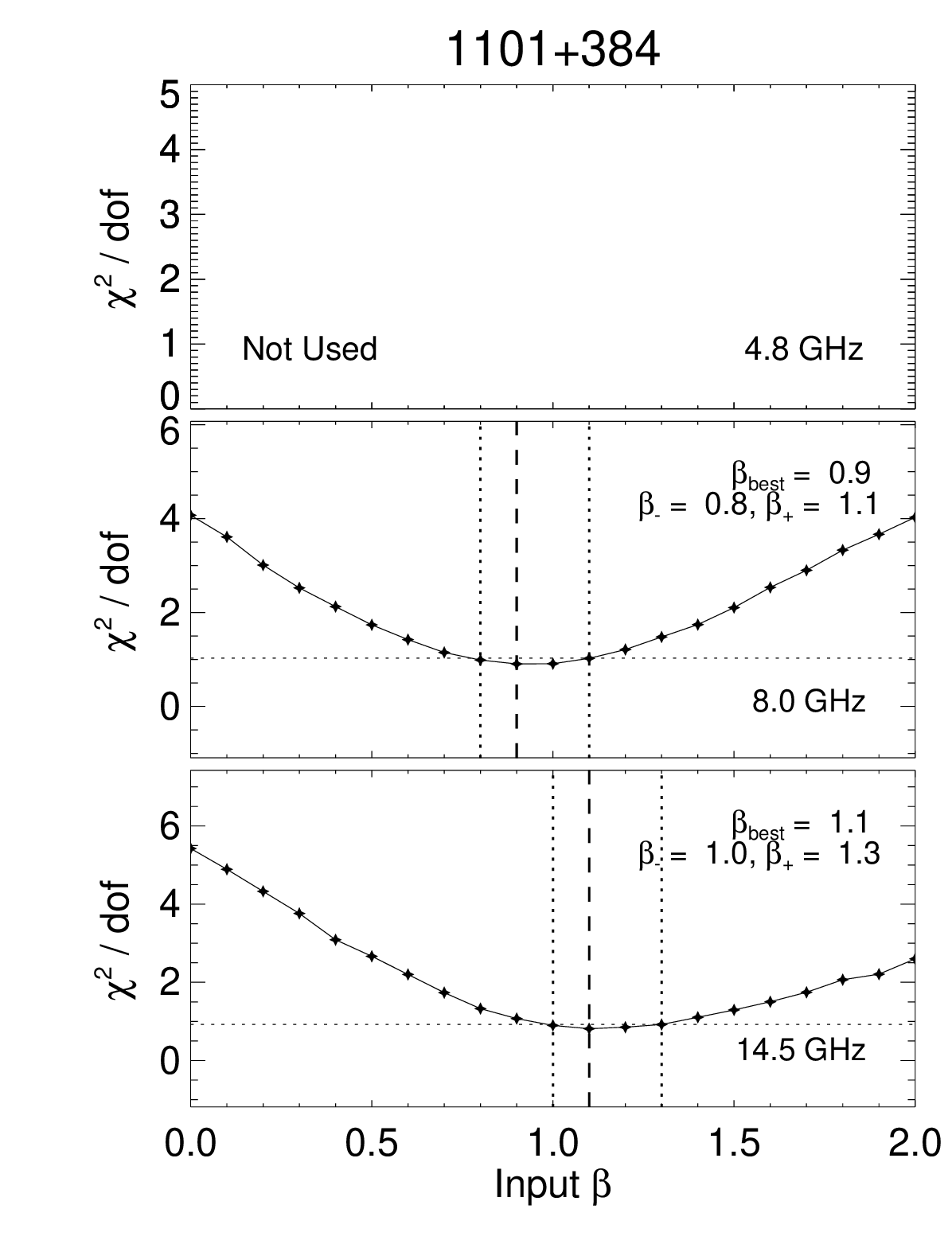}
\includegraphics[trim=4mm 0mm 1mm 0mm, clip, width = 59mm]{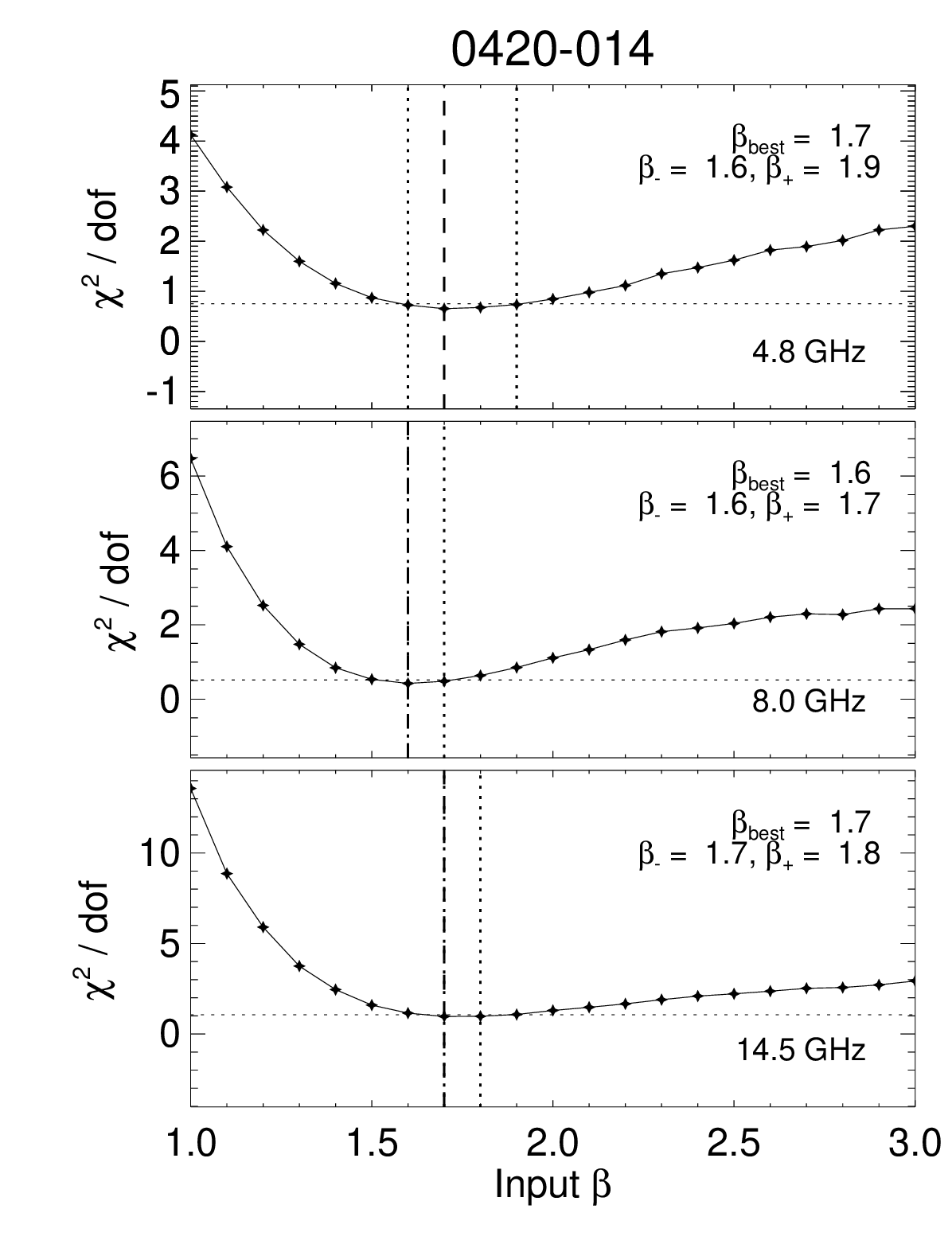}
\includegraphics[trim=4mm 0mm 1mm 0mm, clip, width = 59mm]{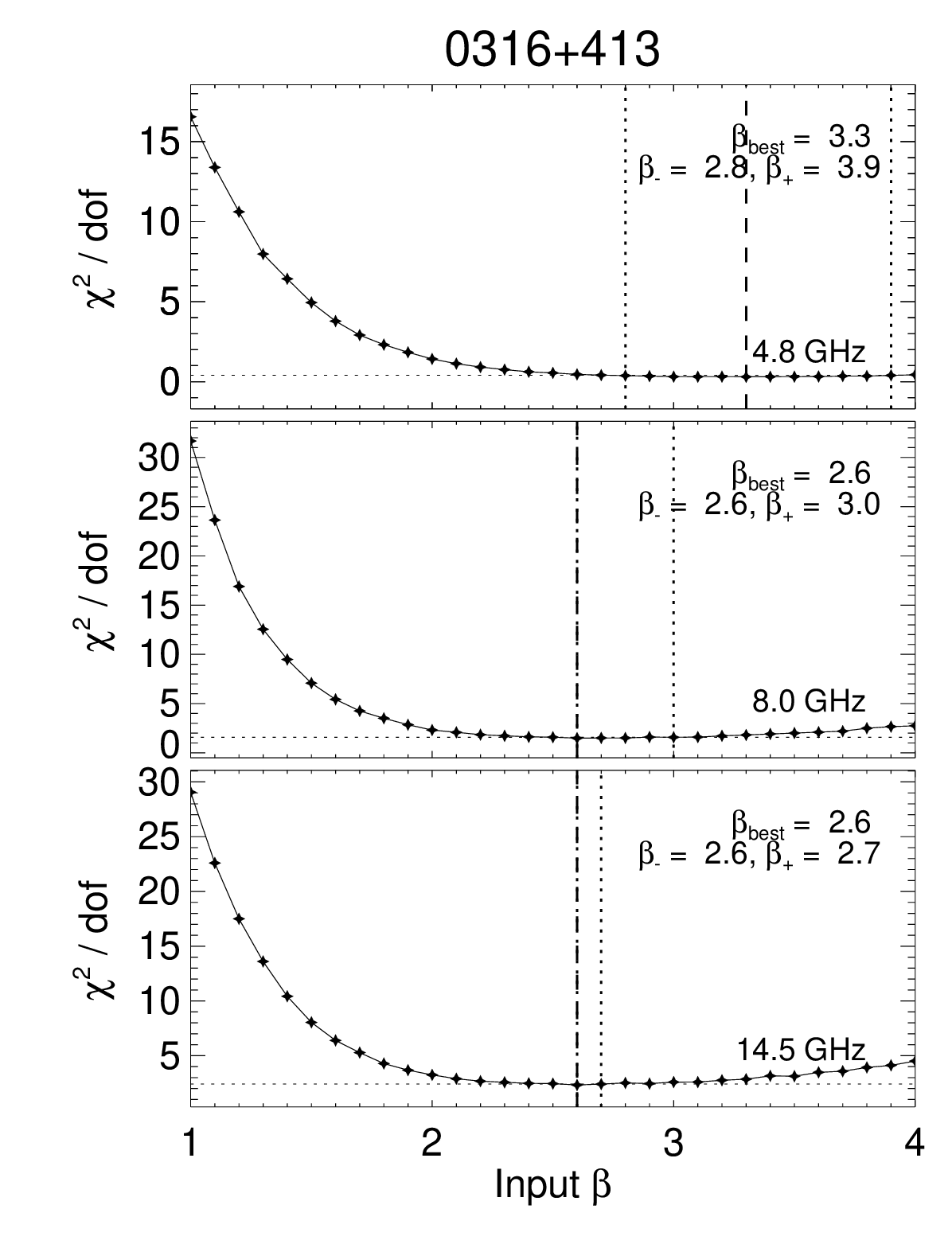}
\caption{$\chi^2 / \rm d.o.f.$ of model power spectra as function of $\beta$ for three sources. `Not Used' means that we excluded the data from our analysis because the number of data points is smaller than 150. Horizontal dotted lines are the $\chi^2_{\rm min} + 1$ lines, i.e., the 68\% significance levels. Vertical dashed lines show the values of $\beta$ that minimize $\chi^2$, $\beta_{\rm best}$. Vertical dotted lines indicate the two $\beta$ values with $\chi^2 = \chi^2_{\rm min} + 1$, $\beta_-$ and $\beta_+$. \label{chi}}
\end{center}
\end{figure*}

\begin{figure*}[!t]
\begin{center}
\includegraphics[trim=10mm 0mm 7mm 0mm, clip, width = 59mm]{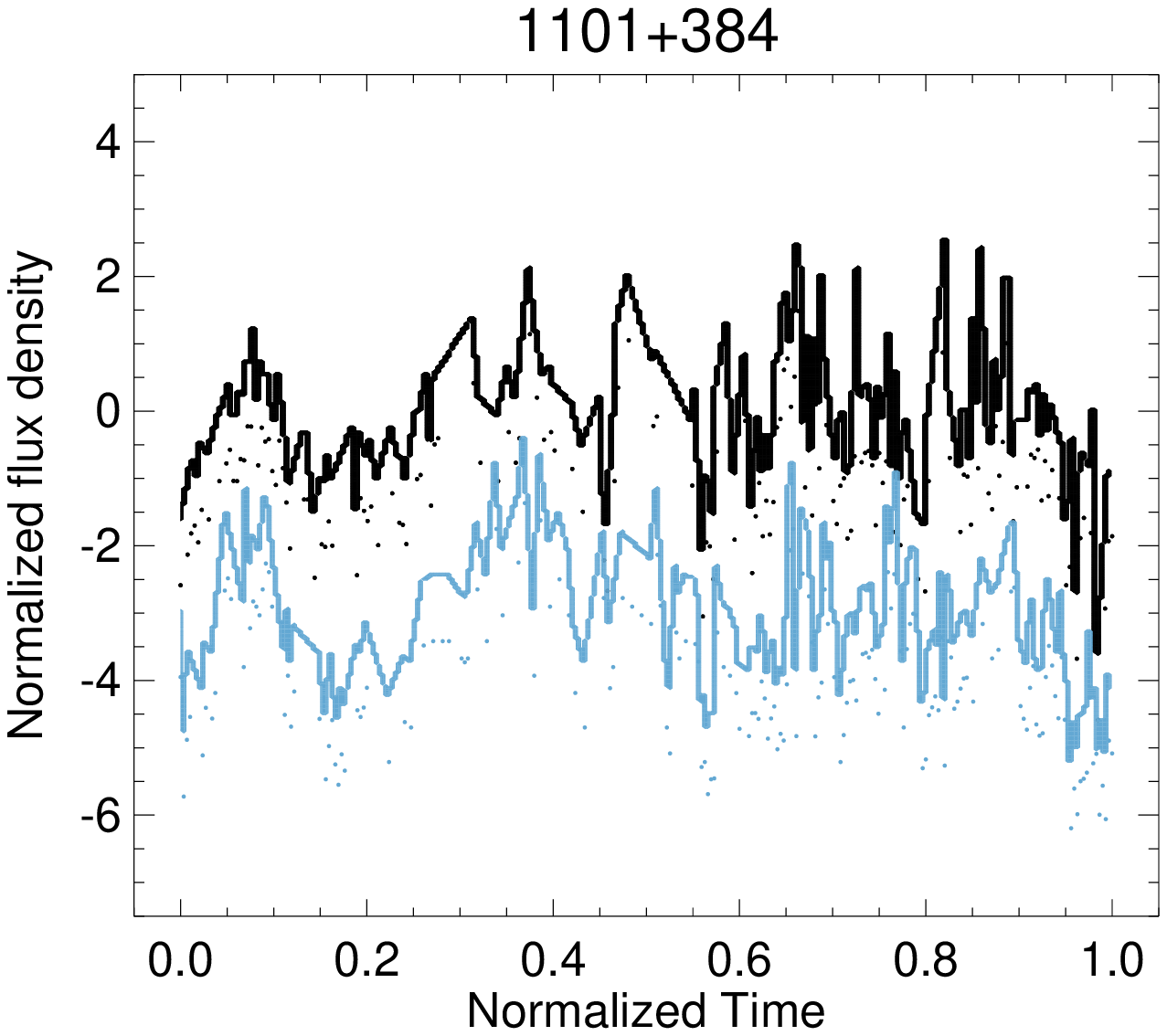}
\includegraphics[trim=10mm 0mm 7mm 0mm, clip, width = 59mm]{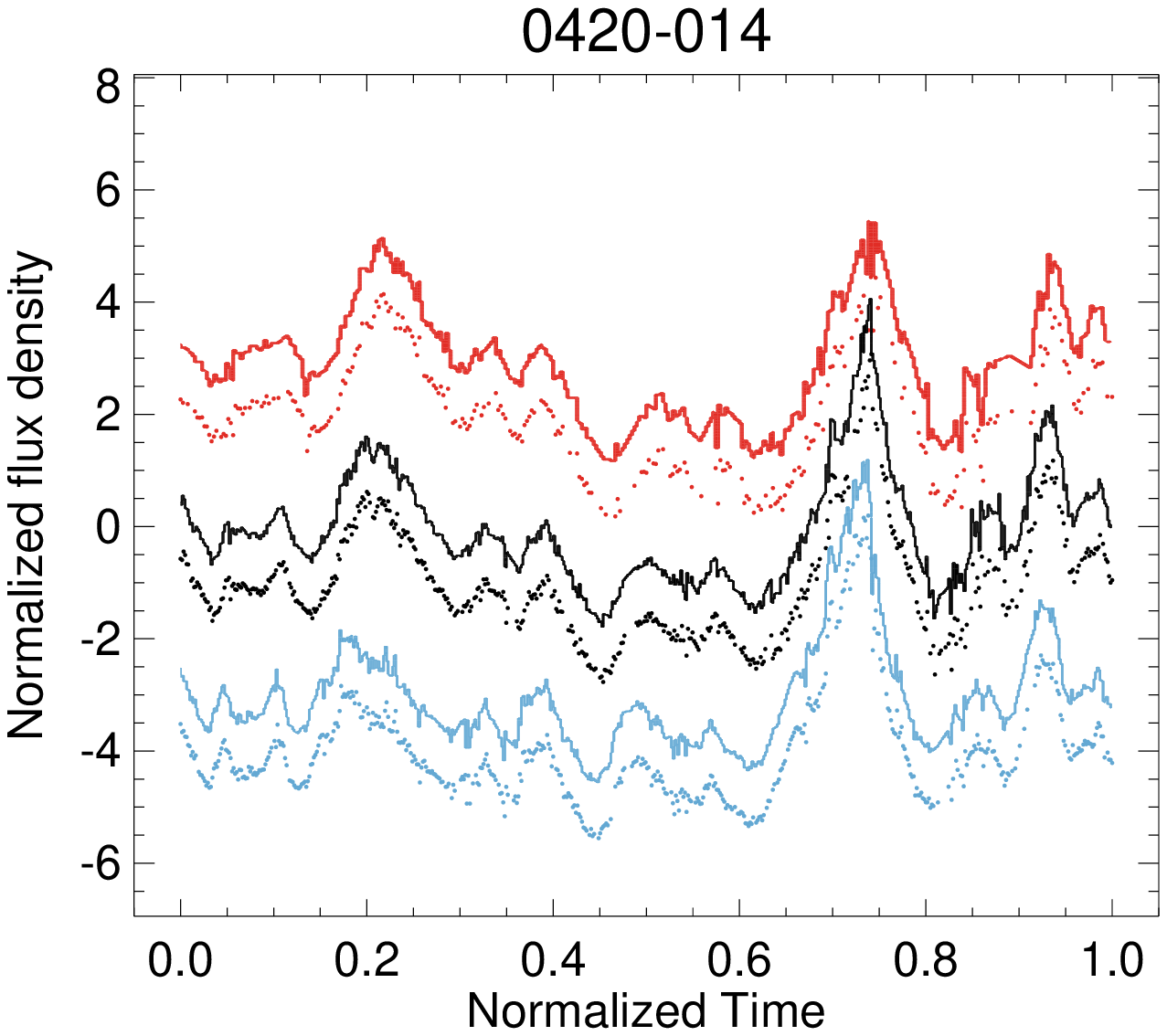}
\includegraphics[trim=10mm 0mm 7mm 0mm, clip, width = 59mm]{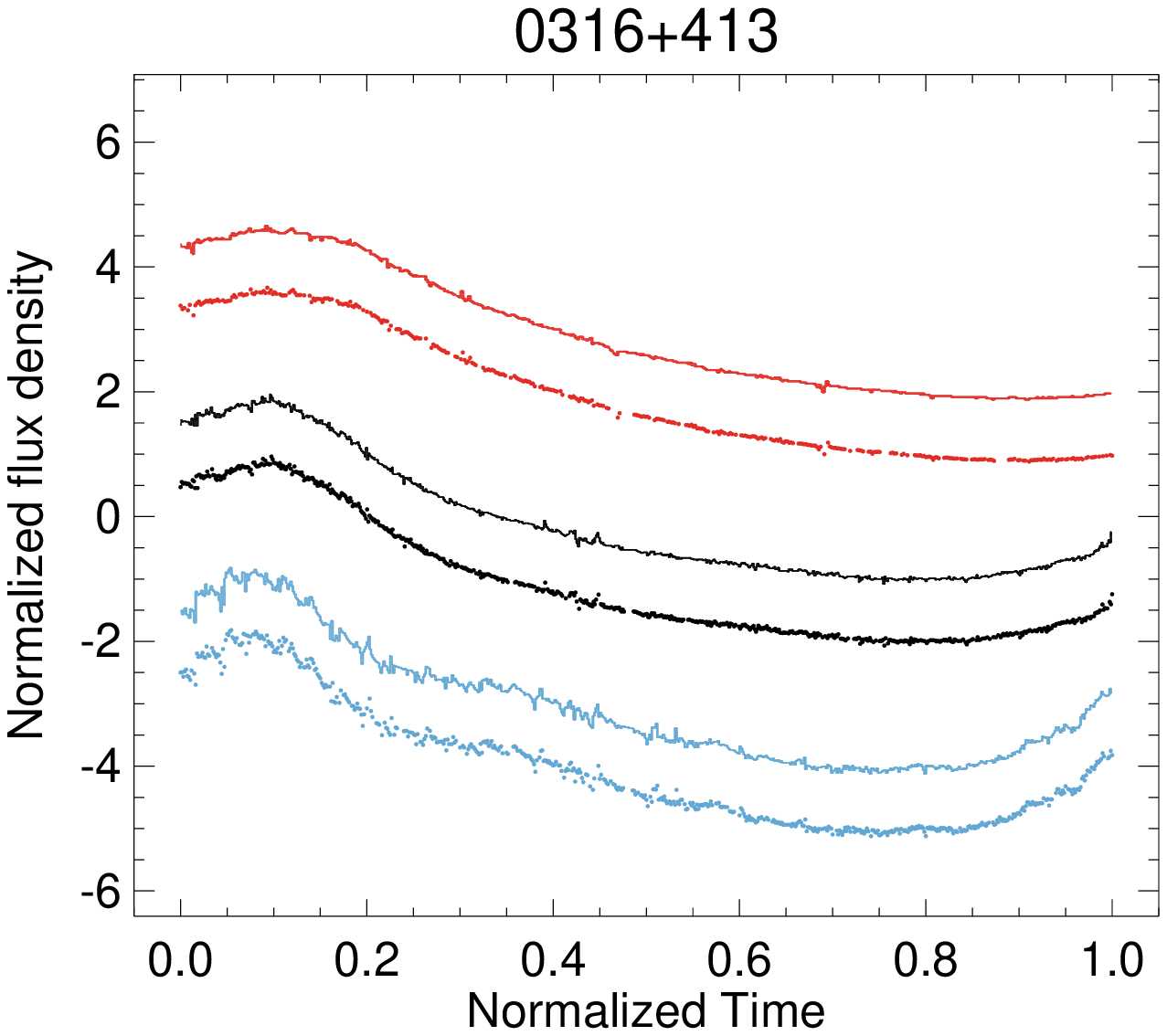}
\caption{Illustration of fractal (box counting) dimension analysis of lightcurves. Each panel shows normalized lightcurves (dotted lines) and corresponding filled grid cells  (solid lines, see Section~\ref{ssect_fractal} for details) in flux density versus time for one of the three sources presented in Figure~\ref{chi}. Red, black, and blue colours indicate the data points at 4.8, 8, 14.5 GHz, respectively. Lightcurves and filled grid cells are offset by 1~Jy for a given frequency, different frequencies are offset by 3~Jy for clarity. \label{fig_fractal}}
\end{center}
\end{figure*}

\section{Analysis}
\label{sect3}
\subsection{Lightcurves and Power Spectra\label{ssect_lc}}

We binned our lightcurves in time to reduce any bias from uneven sampling and flagged outliers that were obvious at visual inspection. We used a bin size of $\Delta t = 2T/N$, where $T$ is the total observing time and $N$ is the number of data points. Binning of lightcurves can change the form of the corresponding power spectra because binning removes power at frequencies higher than the frequency at which binning is performed ($f_{\rm bin}$). However, this effect does not affect our results since the highest sampling frequency in the power spectra, $N_{\rm bin} / 2T$ (where $N_{\rm bin}$ is the number of data points after binning) is always smaller than $f_{\rm bin}$ ($N_{\rm bin}<N$ in our case). Accordingly, our approach does not reduce the power at high sampling frequencies but reduces the maximum sampling frequency. The fraction of flagged data is less than 1\% in most cases, and flagging does not alter the results significantly. We employed the normalized Scargle periodogram for obtaining power spectra from unevenly sampled lightcurves \citep{Scargle1982}. We used the fast algorithm devised by \cite{Press1989} for computing periodograms. We performed Monte Carlo simulations of red-noise lightcurves using the algorithm of \cite{Timmer1995} as we did already in \cite{Park2014}. We summarize the main steps of the simulation process below.

\begin{figure*}[!t]
\begin{center}
\includegraphics[trim=2mm 0mm 3mm 0mm, clip, width = 57mm]{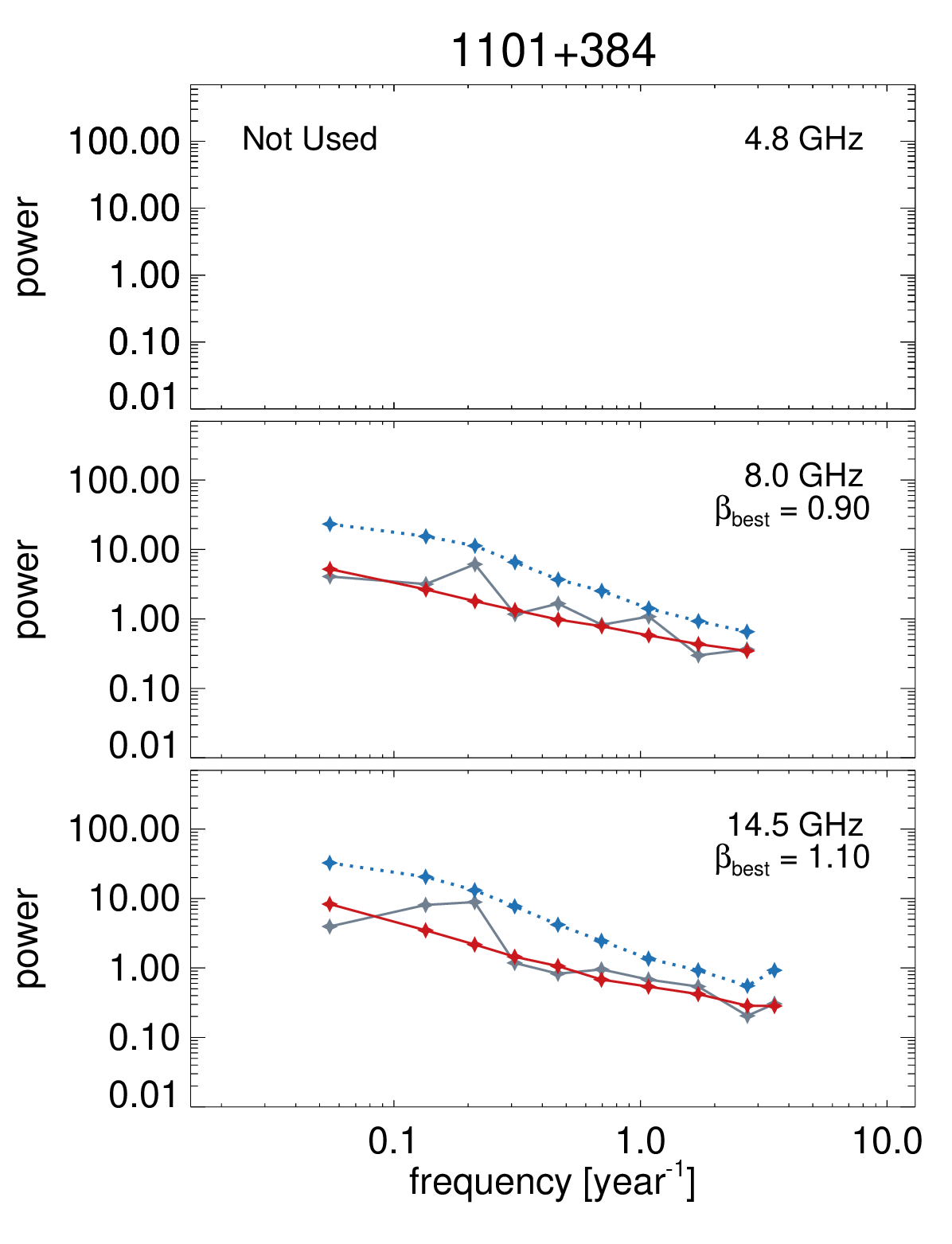}
\includegraphics[trim=2mm 0mm 3mm 0mm, clip, width = 57mm]{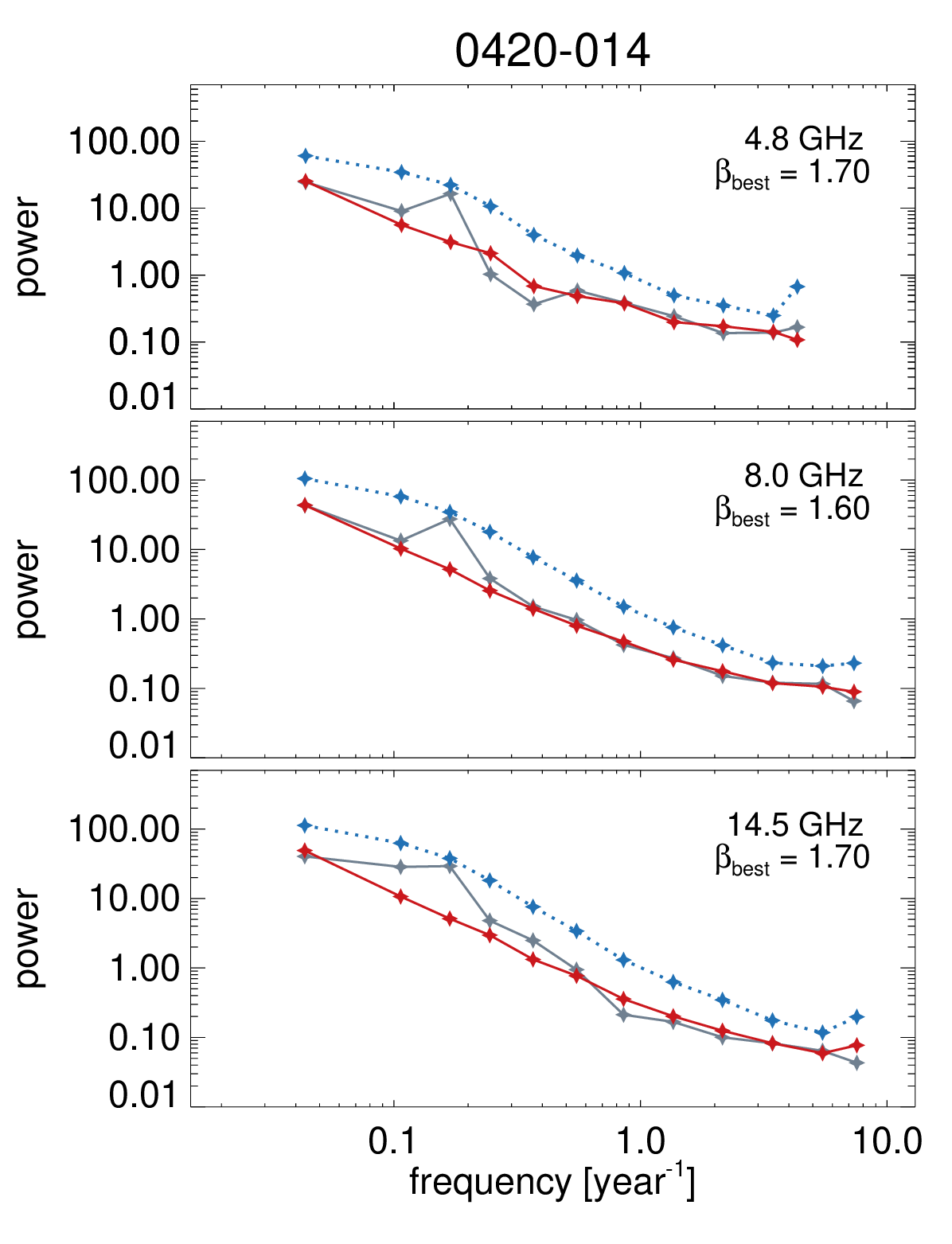}
\includegraphics[trim=2mm 0mm 3mm 0mm, clip, width = 57mm]{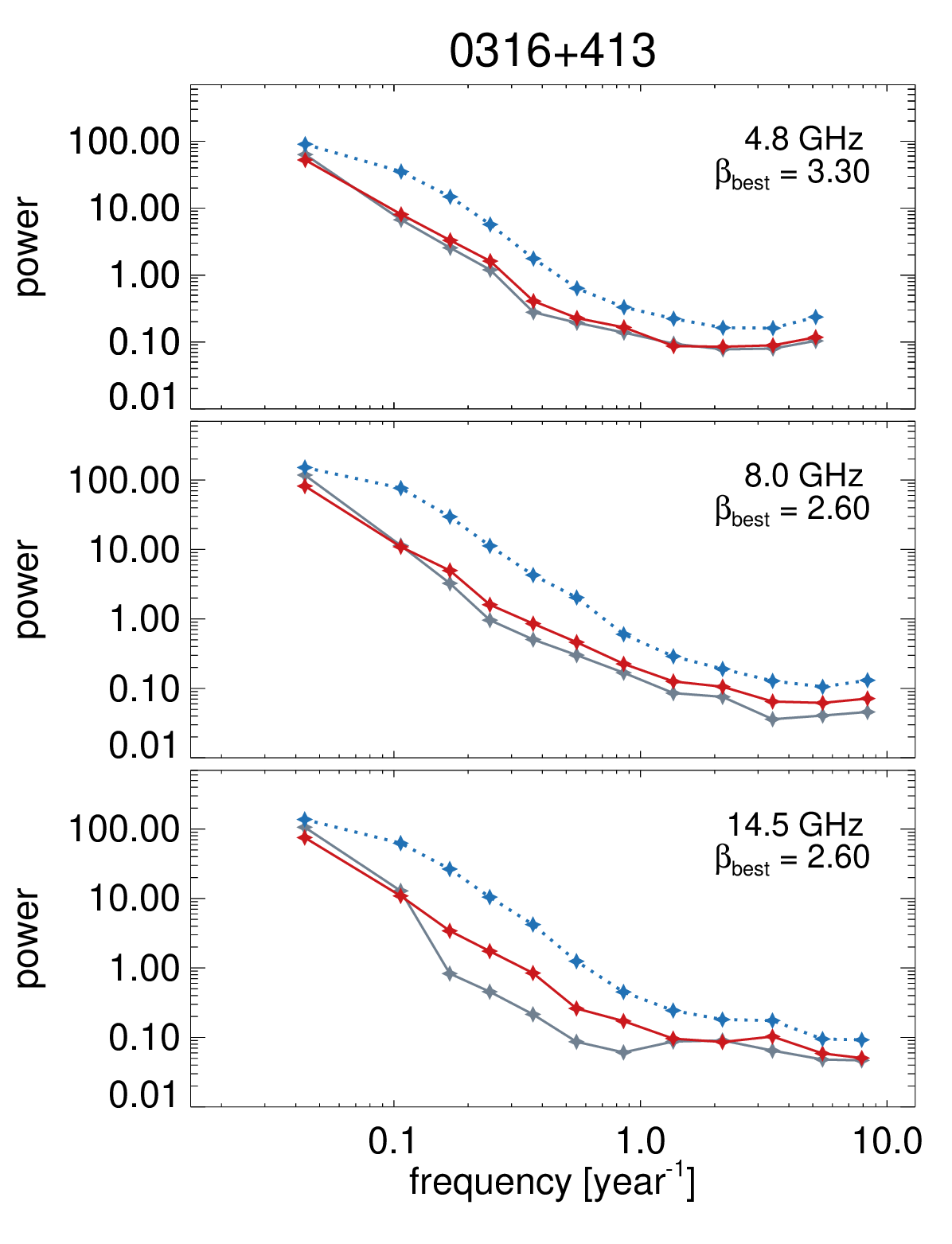}\\
\includegraphics[trim=2mm 0mm 3mm 0mm, clip, width = 57mm]{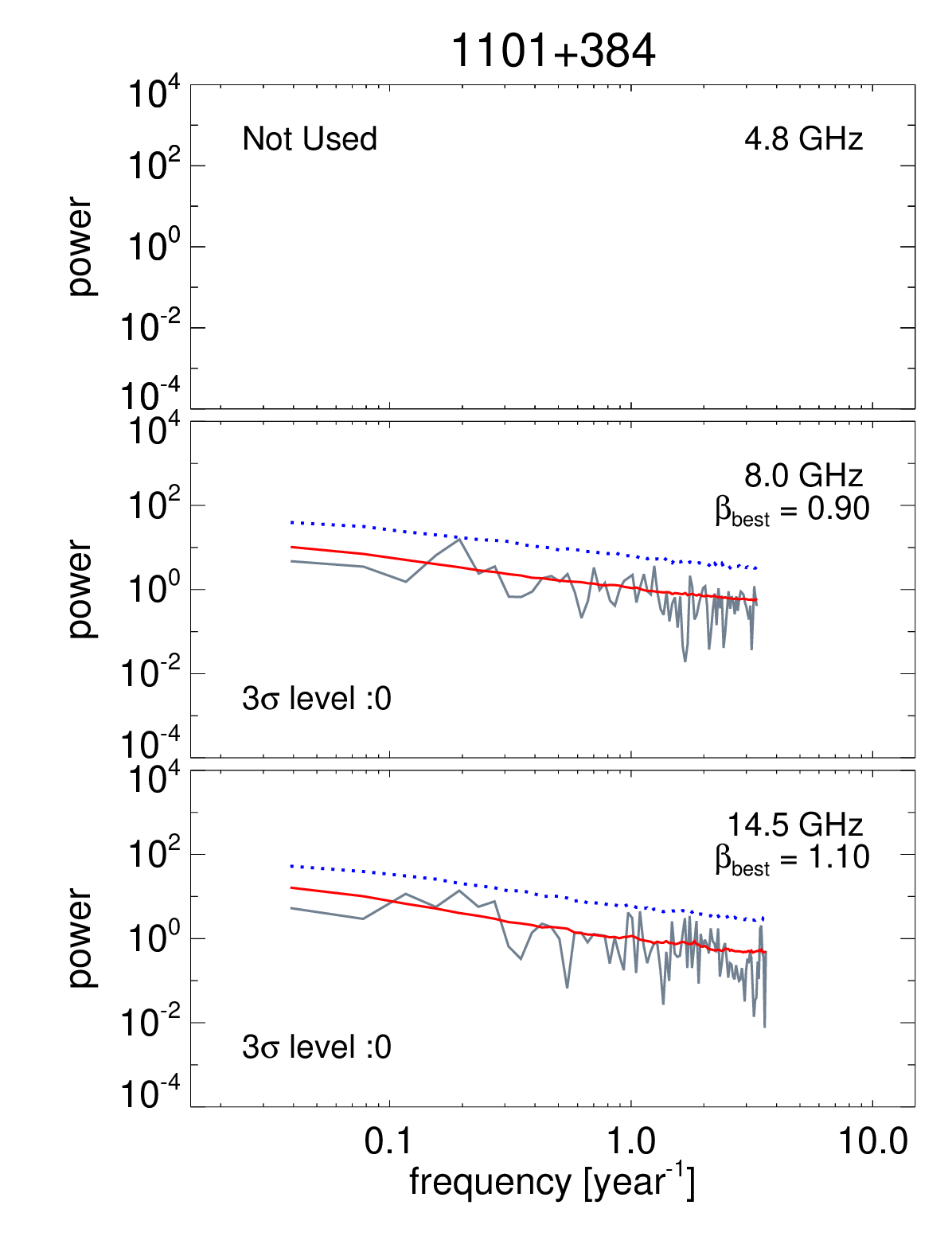}
\includegraphics[trim=2mm 0mm 3mm 0mm, clip, width = 57mm]{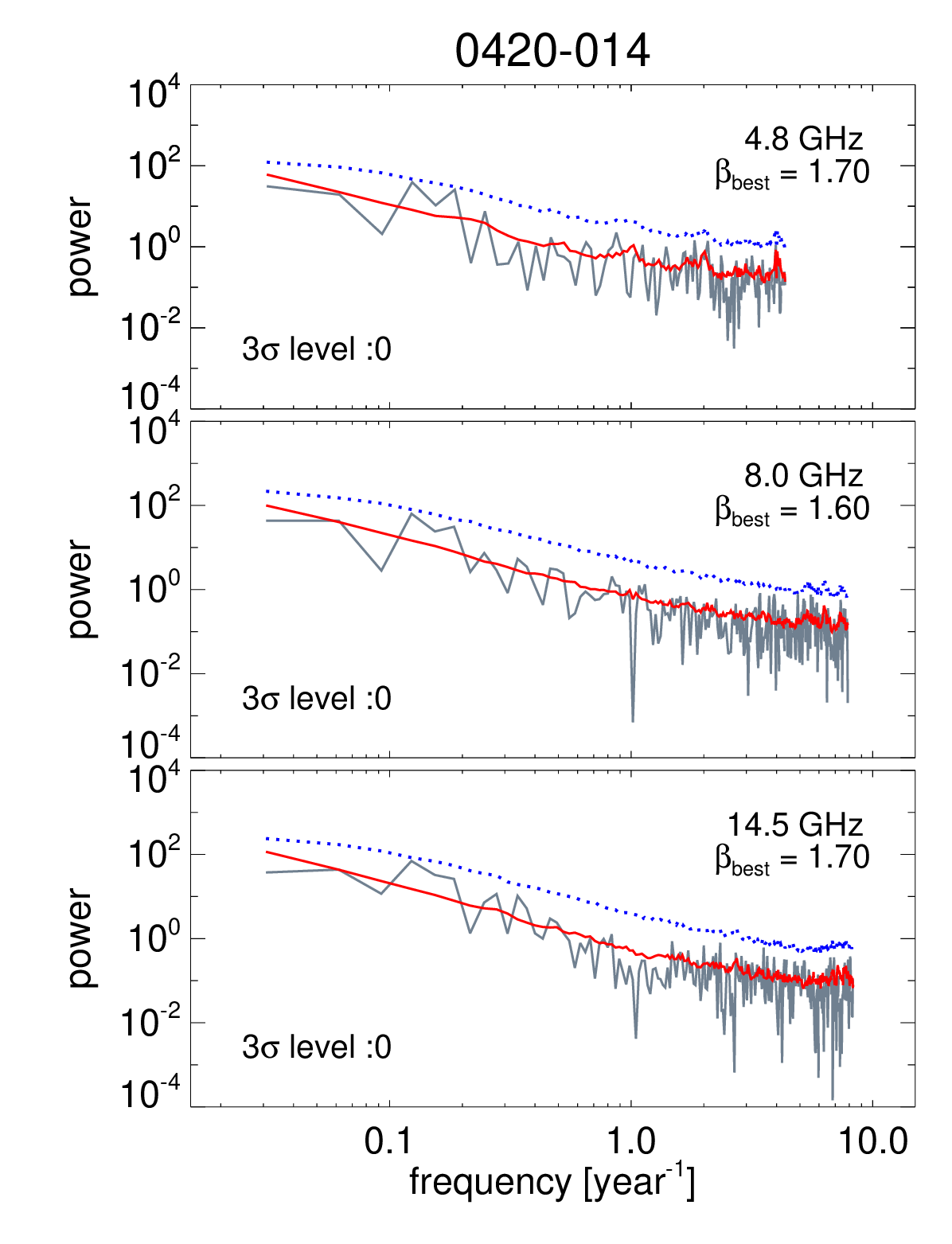}
\includegraphics[trim=2mm 0mm 3mm 0mm, clip, width = 57mm]{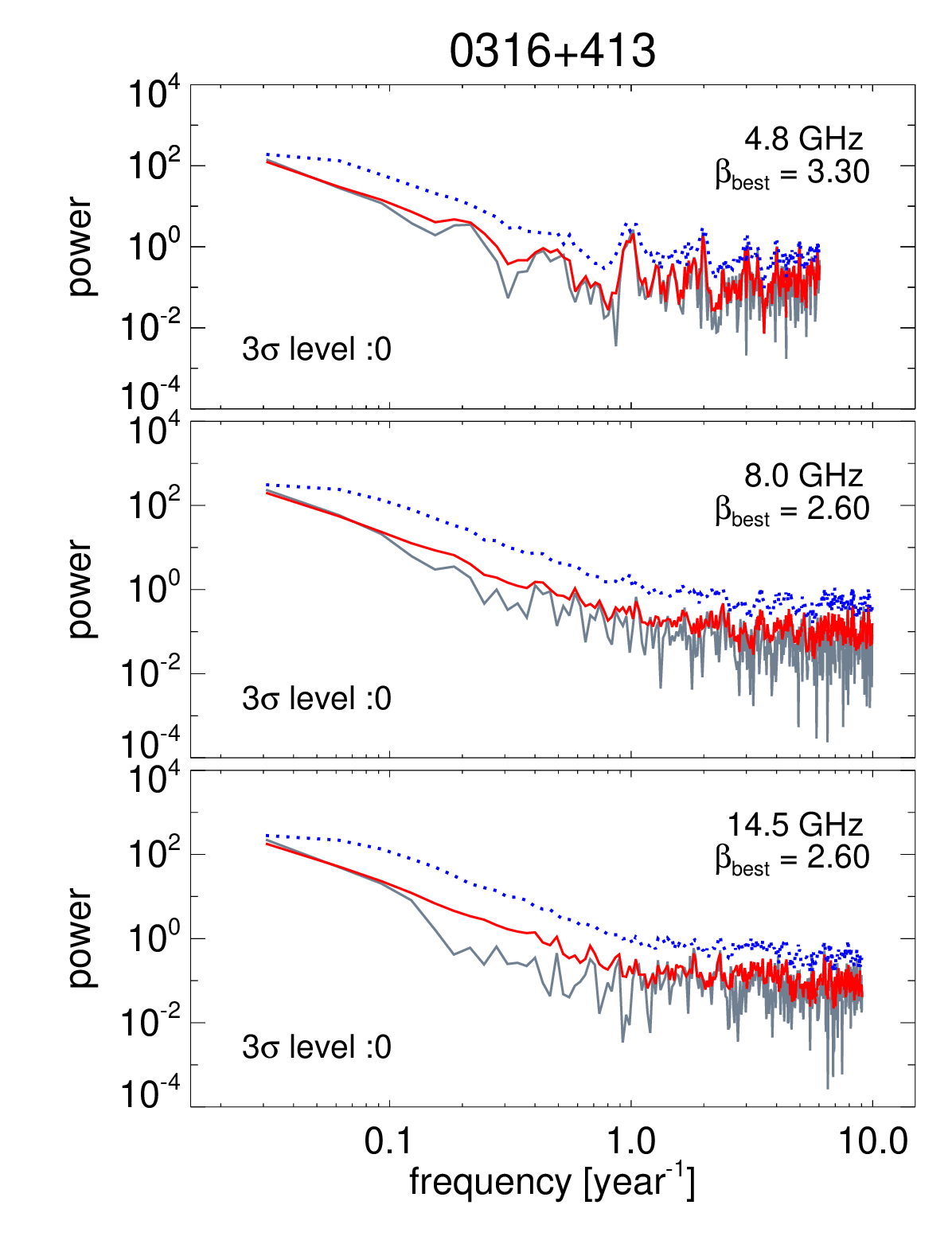}
\caption{\emph{Top row}: Binned logarithmic periodograms (black solid lines), best-fit simple power law models (red solid lines), and  $3\sigma$ significance levels (blue dotted lines) for the three sources presented in Figure~\ref{chi}. In each diagram, $\beta_{\rm best}$ is noted. \emph{Bottom row}: Same as the upper row, for un-binned power spectra. In each diagram the number of data points exceeding the significance threshold ('$3\sigma$ level : ...') is noted. \label{pd}}
\end{center}
\end{figure*}

\emph{Simulated power spectra.}  Artificial lightcurves can be computed by simulating complex spectra using the algorithm of \cite{Timmer1995} and Fourier transforming these spectra. We began with artificial lightcurves that covered a tenfold longer timeline than the observed ones and cut out segments of appropriate length; this procedure reproduces the effects of red-noise leak \citep{Uttley2002}. Even after binning, the observed lightcurves still show gaps, and thus suffer from uneven sampling, because data gaps can be much longer than the length of one bin. Therefore, we mapped the sampling pattern of the observed lightcurves into the simulated ones to take the effects of aliasing into account. We noted the importance of this process already in \cite{Park2014}. We added Gaussian noise to each lightcurve; for each data point, we used the observational measurement error, multiplied by the ratio of the standard deviations of the simulated and the observed lightcurves. We note that our simulation process does not use interpolation of data.

\emph{Model fitting.}  For any given lightcurve, we simulated 5\,000 power spectra for a range of $\beta$ from 1 to 3 (from 1 to 4 for 0316+413) in steps of 0.1. At this stage, using a weighted least-squares `goodness-of-fit' test is not possible because red-noise power spectra follow a non-Gaussian distribution. Therefore, we binned both the observed and the simulated power spectra logarithmically by a factor of 1.6 in frequency, as suggested by \cite{Papadakis1993}. We include at least two data points into each bin. To obtain the best-fit models of the observed power spectra, we calculated the standard goodness-of-fit parameter
\begin{equation}
\chi^2 = \sum_{i = 1}^N\frac{\left[\overline{\log P(f_i)} - \langle\overline{\log P_s(f_i)}\rangle\right]^2}{\sigma^2_{\overline{\log P_s(f_i)}}}, 
\label{eq1}
\end{equation}
where $\overline{\log P(f_i)}$ is the $i$th value of the binned logarithmic periodogram of the observed lightcurve, and $\langle\overline{\log P_s(f_i)}\rangle$ and $\sigma^2_{\overline{\log P_s(f_i)}}$ are the average and the variance of the power spectra of the simulated lightcurves, respectively. Theoretically, covariance of power spectrum bins should be taken into account in addition to the variance. In principle, the spectral power at different bins is uncorrelated when using the sampling frequencies prescribed for a Scargle periodogram \citep{Scargle1982}. Actually however, the observed power spectra are affected by a convolution of the true spectra with a bias function, the `Fejer kernel' \citep{Priestley1981}, which comes from complex red-noise leak and aliasing and introduces correlations between the spectral power values at different sampling frequencies. In our work, this effect is already accounted for in our Monte-Carlo simulations -- which justifies using the variance instead of the full variance-covariance matrix.

We determined the $\beta$ value ($\beta_{\rm best}$) for which $\chi^2$ is minimized ($\chi^2_{\rm min}$) and obtained the errors of $\beta_{\rm best}$ from the boundaries of the interval in $\beta$ where $\chi^2$ becomes $\chi^2_{\rm min} + 1$. Since our simulation is limited to a resolution of 0.1 in $\beta$, we added an additional binning error of 0.05 in squares. We note that \cite{Isobe2015} obtained the best-fit model power spectra of their \emph{Monitor of All Sky X-ray Image} (MAXI) lightcurves of Mrk 421 using this method. We illustrate the typical behavior of $\chi^2 / \rm d.o.f.$ as function of $\beta$ for three sources in Figure~\ref{chi}; these three sources are representative of sources showing fast, moderate, and slow flux variability, respectively. Table~\ref{Information} shows the best-fit values of $\beta$ for our sources plus their errors.

\emph{Significance levels.} We determined a $3\sigma$ (99.7\%) significance level for each sampling frequency from the set of 5\,000 simulated periodograms with $\beta_{\rm best}$ as we did in \cite{Park2014}. A spectral power that exceeds the significance level at a certain sampling frequency might indicate the presence of a (quasi-)periodic signal. In Figure~\ref{pd}, we show the observed power spectra, the expected distributions resulting from averaging over 5\,000 simulated power spectra with $\beta_{\rm best}$, and the significance levels of the three sources presented in Figure~\ref{chi}.

\begin{figure*}[!t]
\begin{center}
\includegraphics[trim=4mm 0mm 4mm 0mm, clip, width = 59mm]{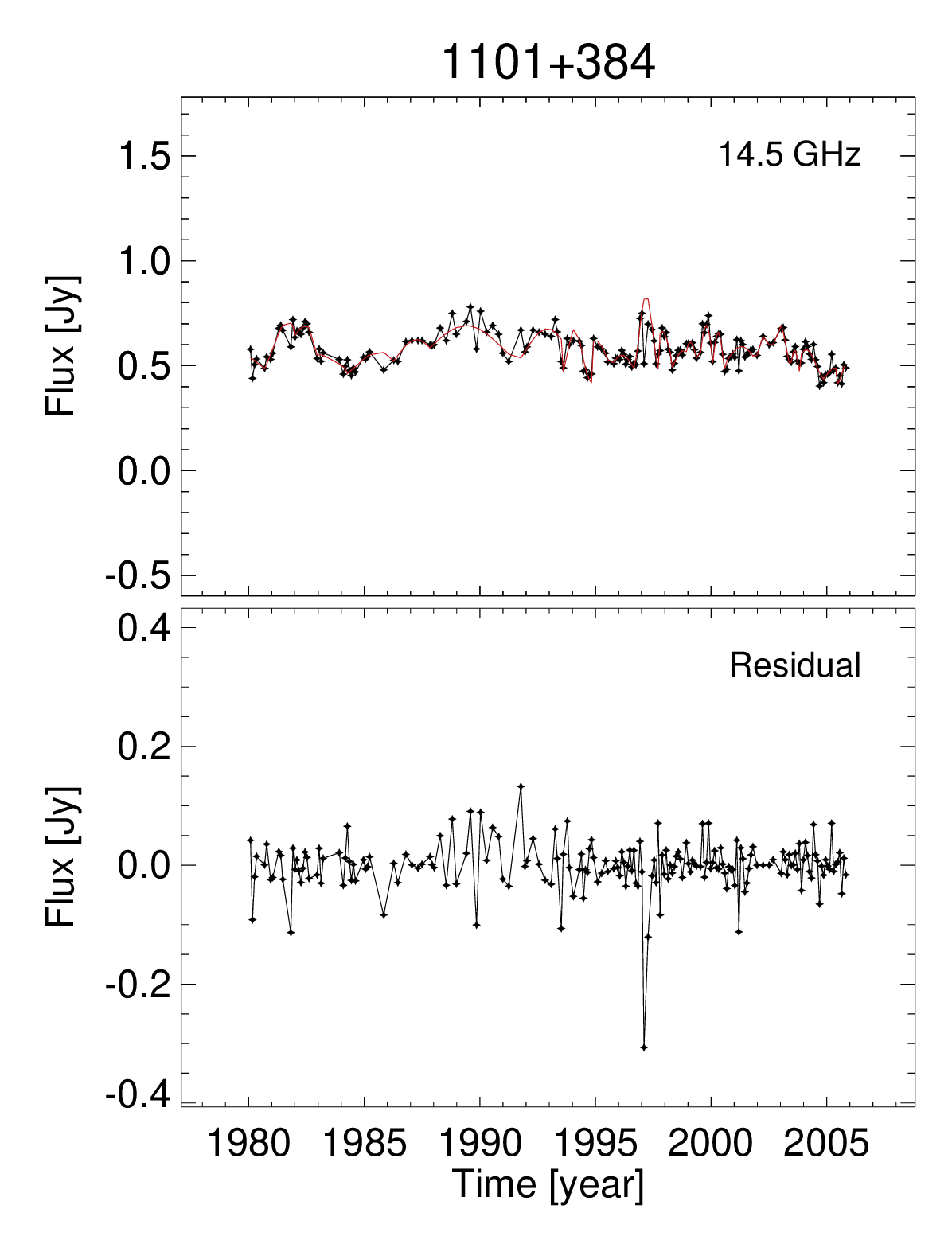}
\includegraphics[trim=4mm 0mm 4mm 0mm, clip, width = 59mm]{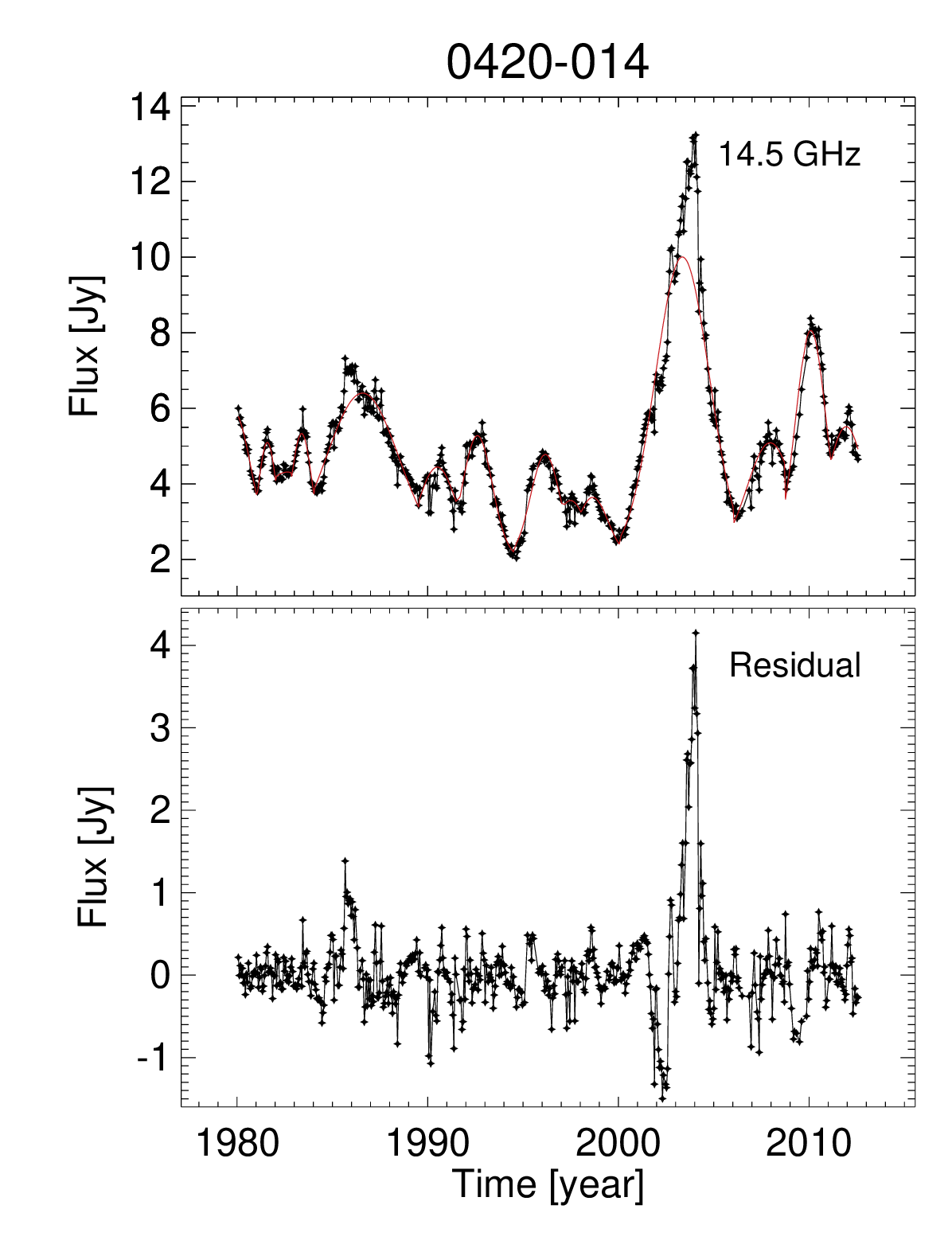}
\includegraphics[trim=4mm 0mm 4mm 0mm, clip, width = 59mm]{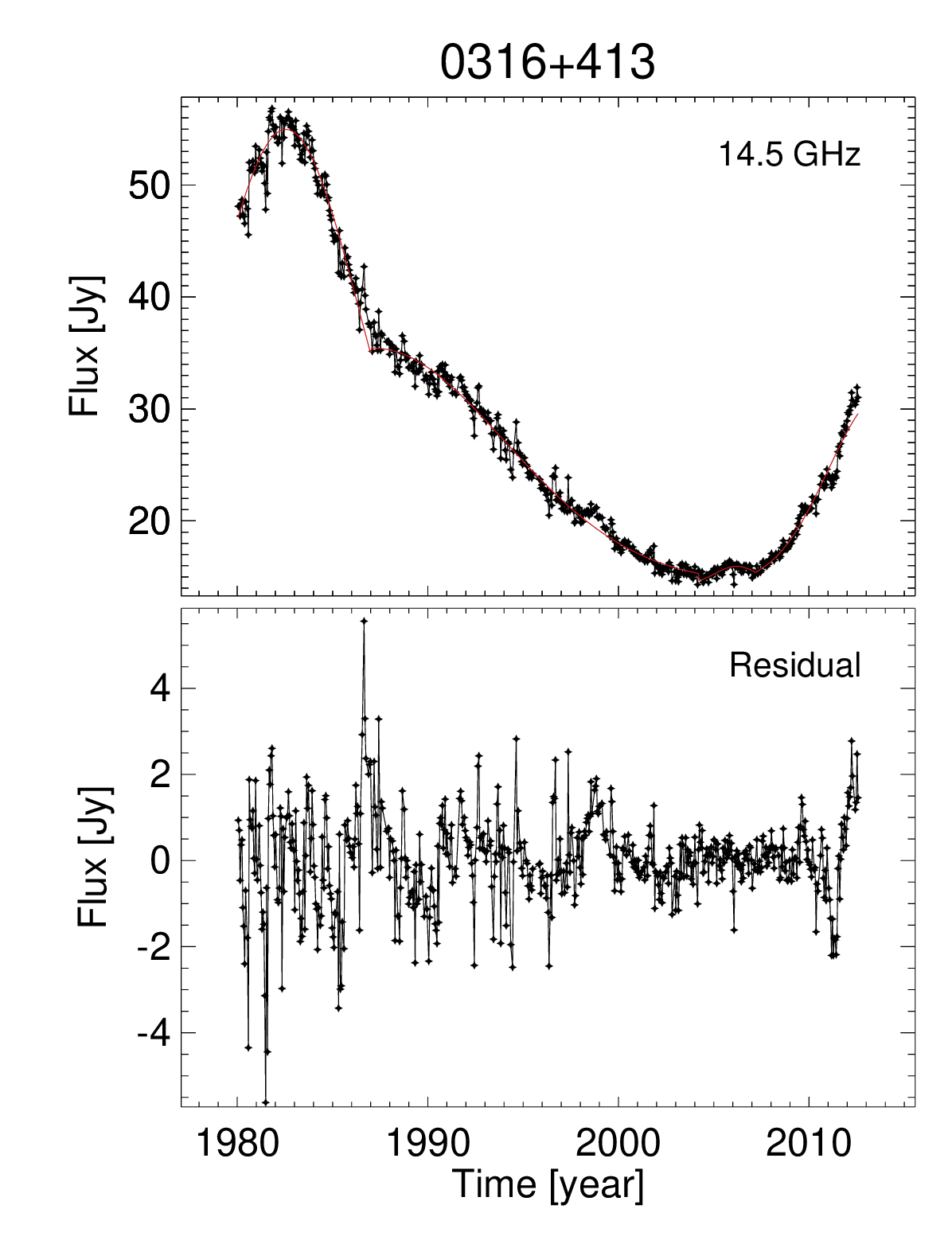}
\caption{\emph{Top panels}: Lightcurves (black solid lines) and model lines composed of Gaussian flares (red solid lines, see Section~\ref{ssect_fit}) of the three sources presented in Figure~\ref{chi} at 14.5 GHz. \emph{Bottom panels}: Residuals between data and models. \label{fig_fit}}
\end{center}
\end{figure*}

\subsection{Fractal Dimension\label{ssect_fractal}}

The variability of a lightcurve can also be quantified via its fractal dimension (see e.g., \citealt{Falconer1990} for an exhaustive review). Basically, this quantity describes how much a given plane -- flux density vs. time in our case -- is filled by the graph of a given function. If small (large) scale fluctuations dominate, corresponding to smaller (larger) values of $\beta$ in periodograms, the lightcurve fills a larger (smaller) fraction of the flux--time plane. The fractal dimension has been used to estimate the strength of spatial clustering of gas or stars and the effects of projection onto the sky plane (see e.g., \citealt{Sanchez2005, Sanchez2010} and references therein). We specifically used the box-counting dimension defined by
\begin{equation}
d_f = -\lim_{\epsilon \to 0}\frac{\log N(\epsilon)}{\log \epsilon},
\end{equation}
where $N(\epsilon)$ is the number of cells of (dimensionless) size $\epsilon$ occupied by the lightcurve. In practice, $\epsilon$ is limited by the sampling of the lightcurve. For each lightcurve, we normalized the time axis to the interval from 0 to 1 and the flux density to zero mean and unity standard deviation. We divided the normalized time axis into $n$ sections with binsize $\epsilon$ and the normalized flux axis into $n$ sections of size $10 / n$ (because the flux densities happen to lie in the range of $-5$ to $5$ for all lightcurves). In Figure~\ref{fig_fractal}, we show the normalized lightcurves and corresponding filled grid cells for the three sources presented in Figure~\ref{chi}.

\subsection{Fitting Lightcurves Piecewise with Gaussian Peaks\label{ssect_fit}}

Radio lightcurves of radio bright AGNs (mostly blazars) are characterized by multiple flaring events. Many studies have described flares either as exponentially rising and decaying (e.g., \citealt{Valtaoja1999, Chatterjee2008, Chatterjee2012, Abdo2010}) or Gaussian (e.g., \citealt{Pyatunina2006, Pyatunina2007, Mohan2015}) outbursts of radiation. As already noted by \cite{Valtaoja1999}, the decomposition (or deconvolution) of lightcurves into several flares (specifically, the one-dimensional CLEAN method) do not work well at observing frequencies below  22 GHz where the overlap of individual flares is very strong because of the rather long evolutionary timescales of the outbursts. Thus, we divided the lightcurves of our sources into several pieces by visual inspection and fitted a single Gaussian function to each piece. We note that this process is different from the aforementioned deconvolution because we only analysed discrete (non-overlapping) segments of the lightcurves. In this case, the \emph{amplitude} of the model flares can be substantially overestimated. However, our primary aim is to obtain the \emph{duration} of the flares, for which our procedure is sufficient. In Figure~\ref{fig_fit}, we show the observed lightcurves, the model lightcurves generated by combining the individual best-fit Gaussians, and the residuals between the data and the models of the three sources presented in Figure~\ref{chi} at 14.5 GHz as an example. The model lightcurves represent the data very well in general, with the exception of some narrow spikes that are not caught by the smooth Gaussian profiles. Table~\ref{Information} shows the median duration of flares for each lightcurve. Here, $\sigma$, the duration of the flare, refers to the Gaussian width, i.e., $f(t) \propto \exp[-(t-t_0)^2/2\sigma^2]$.

\subsection{Derivatives of Lightcurves\label{ssect_der}}

\begin{figure*}[!t]
\begin{center}
\includegraphics[trim=4mm 0mm 4mm 0mm, clip, width = 59mm]{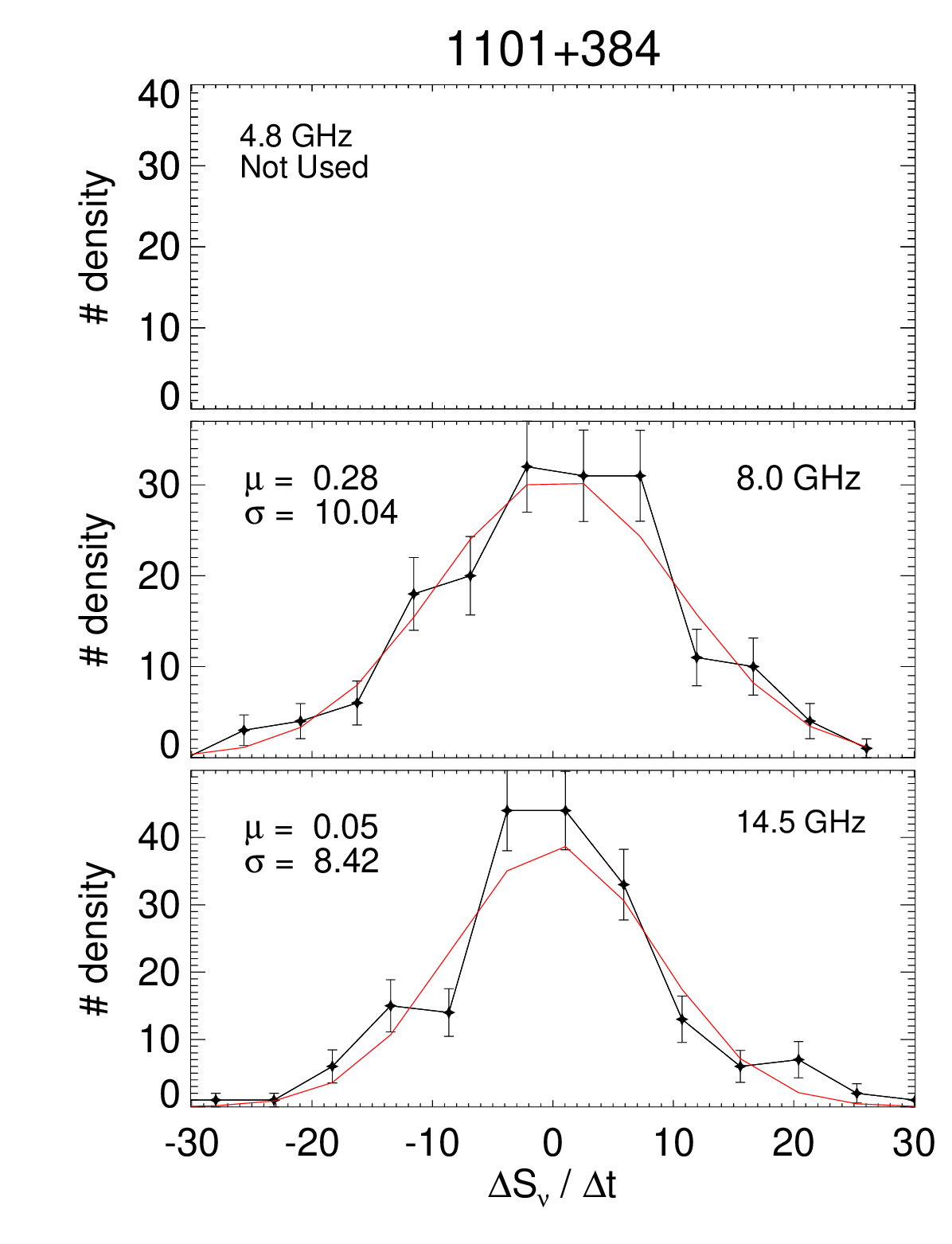}
\includegraphics[trim=4mm 0mm 4mm 0mm, clip, width = 59mm]{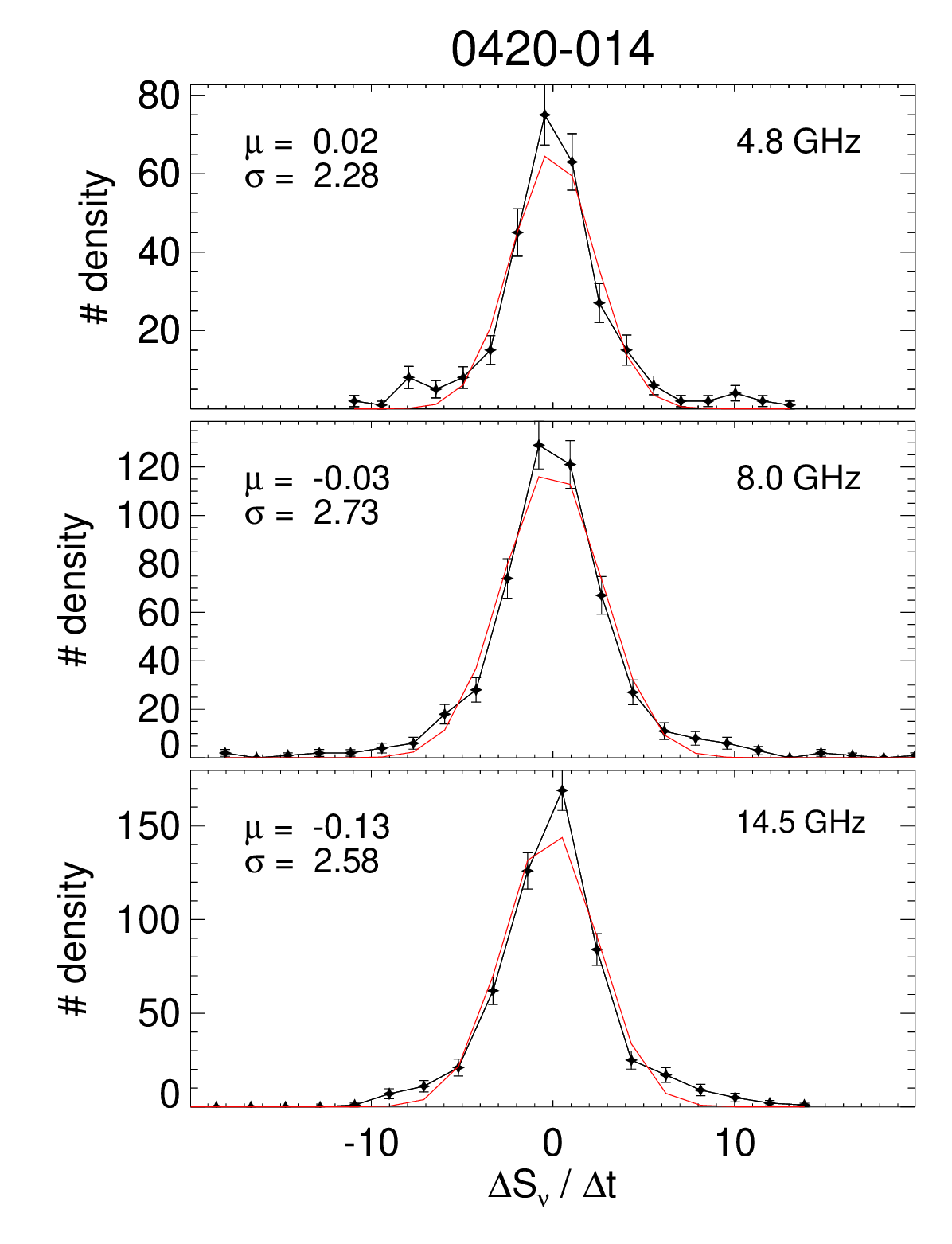}
\includegraphics[trim=4mm 0mm 4mm 0mm, clip, width = 59mm]{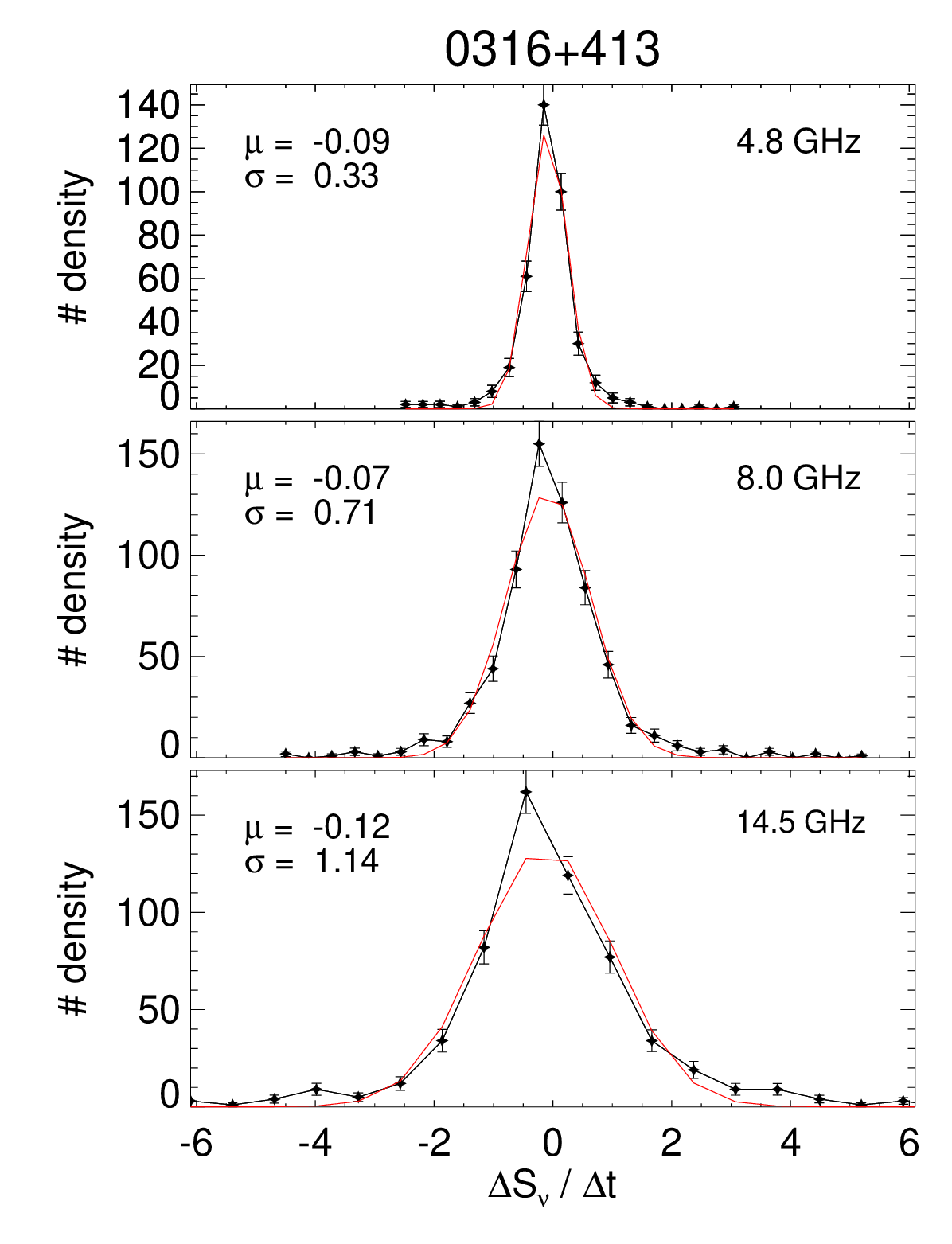}
\caption{Distributions of the derivatives of the normalized lightcurves (black solid lines) and the best-fit Gaussian functions (red solid lines, see Section~\ref{ssect_der}) of the three sources presented in Figure~\ref{chi}. Errors were estimated via bootstrapping with 1\,000 random resamplings of a given data set. The means and the standard deviations of the best-fit Gaussians are $\mu$ and $\sigma$, respectively. \label{Diff}}
\end{center}
\end{figure*}

We devised a new method to obtain variability timescales free from any \emph{a priori} assumption on the functional form of flux variations (as used in Section~\ref{ssect_fit}). The main idea is to take the derivative of a lightcurve as function of time, $\Delta S_{\nu} / \Delta t$ with $\Delta S_{\nu}$ and $\Delta t$ being the difference in flux density and time between adjacent data points, respectively, and to obtain the distribution function of the derivative values. Before taking the derivative, we normalized the lightcurves to zero mean and unity standard deviation. We used bootstrapping for estimating the errors of the distributions, which turned out to be close to binomial errors. We fitted a single Gaussian function to each distribution function, which was usually a good representation. We obtained the standard deviations of the best-fit Gaussians, $\sigma_{\mathrm{der}}$. Smaller values of $\sigma_{\mathrm{der}}$ mean that more time is necessary to make a certain amount of change in flux density. Accordingly, the inverse of $\sigma_{\mathrm{der}}$ provides an effective variability timescale; in our case, the unit of $\sigma_{\mathrm{der}}$ is yr$^{-1}$. In Figure~\ref{Diff}, we show the distribution functions of the derivatives and the fitted Gaussian functions of the three sources shown in Figure~\ref{chi}. We provide the $\sigma_{\mathrm{der}}$ values for our sources in Table~\ref{Information}.

\subsection{Black Hole Masses and Accretion Rates\label{ssect_BH}}

In order to examine if there is any correlation between variability timescale and black hole mass or accretion rate (and thus Eddington ratio), we searched the literature for the black hole masses $M_{\rm BH}$ and the disk luminosities $L_{\rm disk}$ of our sources. There has been significant progress in measuring the black hole mass of AGNs with various methods, including stellar dynamics (e.g., \citealt{Kormendy1995, Ferrarese2005}), gas dynamics (e.g., \citealt{Macchetto1997}), the black hole mass--bulge luminosity relation ($M_{\rm BH}$-$L_{\rm bulge}$ relation; e.g., \citealt{McLure2001}), single-epoch spectroscopy using the size--luminosity relation for AGN broad line regions (BLR) derived from reverberation mapping (e.g., \citealt{Kaspi2000, McLure2002, Kaspi2005}), and the relation between the black hole mass and the velocity dispersion $\sigma$ of the stellar system around the black hole, i.e., the $M_{\rm BH}$-$\sigma$ relation (e.g., \citealt{Gebhardt2000, Merritt2001, Tremaine2002}). Using these empirical relations is a reasonable choice in our case; direct estimates of the black hole masses via, e.g., reverberation mapping, would require a dedicated long-term monitoring program which is beyond the scope of our work. A few studies have presented the black hole masses for radio bright AGNs, including many of our sources, with various methods \citep{Gu2001, Woo2002, Falomo2002, Falomo2003a, Falomo2003b, Barth2003, Wang2004, Liu2006}. However, some authors did not consider the contribution of non-thermal continuum emission of jets to the observed optical continuum luminosity in their single epoch spectroscopic mass measurements. This leads to overestimates of the sizes of BLRs and thus of the black hole masses. \cite{Liu2006} showed that the non-thermal contribution is indeed significant for their sample of radio loud AGNs. Therefore, we had to recalculate the black hole masses given in the abovementioned works. Our calculations use (a) emission line luminosities, which are almost not affected by the non-thermal continuum, and (b) the relation between the continuum luminosity and the emission line luminosity of radio quiet AGNs \citep{Liu2006}.

We used the data for three emission lines, H$\beta$, Mg\,II, and C\,IV (mainly from \citealt{Wang2004, Liu2006}; and \citealt{Torrealba2012}) for estimating black hole masses via single epoch spectroscopy. We used the relation between black hole mass, full width at half maximum (FWHM), and luminosity of H$\beta$ line of \cite{Vestergaard2006},
\begin{equation}
\frac{M_{\rm BH}}{M_{\odot}} = 10^{6.67}\left[\frac{\rm FWHM(H\beta)}{\rm 1000\ km\ s^{-1}}\right]^2\left[\frac{L(\rm H\beta)}{\rm 10^{42}\ ergs\ s^{-1}}\right]^{0.63}
\end{equation}
and the corresponding relation of \cite{Vestergaard2009} for the Mg\,II line,
\begin{equation}
\frac{M_{\rm BH}}{M_{\odot}} = 10^{6.96}\left[\frac{\rm FWHM(Mg\,II)}{\rm 1000\ km\ s^{-1}}\right]^2\left[\frac{L_{5100{\textrm\AA}}}{\rm 10^{44}\ ergs\ s^{-1}}\right]^{0.5}
\end{equation}
where $L_{5100{\textrm\AA}}$ is the monochromatic luminosity at 5100\,\AA. For using the latter relation, we first converted the Mg\,II luminosity to H$\beta$ luminosity following \cite{Francis1991} who found the ratio of the luminosities between these emission lines to be $L({\rm H}\beta):L({\rm Mg\, II}) = 22:34$. Then, we obtained $L_{5100\textrm\AA}$ from the relation between the monochromatic luminosity at $5100{\textrm\AA}$ and the H$\beta$ luminosity for radio-quiet AGNs presented in \cite{Liu2006}, assuming the same relation holds for radio-loud AGNs:
\begin{equation}
L_{5100{\textrm\AA}} = 0.843\times10^2L_{\rm H\beta}^{0.998}.
\end{equation}
For the C\,IV lines, we followed \cite{Liu2006} who assumed that the radius of C\,IV emitting BLRs is about half of that of H$\beta$ emitting BLRs (see references therein) and used the BLR size--luminosity relation of \cite{Kaspi2005} for H$\beta$. We recalculated the black hole masses presented in \cite{Wang2004} obtained via single epoch spectroscopy, using the aforementioned methods, scaling relations, and our adopted cosmological parameters that are slightly different from those in their work. We used the black hole masses from \cite{Liu2006} without any modification because they already took the contribution by non-thermal emission into account. We also made use of the optical spectroscopic atlas of the MOJAVE / 2~cm AGN sample of \cite{Torrealba2012} and estimated black hole masses from their velocity dispersions and line luminosities.

We also included black hole masses derived via the $M_{\rm BH}$-$\sigma$ relation \citep{Falomo2003a, Falomo2003b, Barth2003}, the rotation velocity of H$_2$ gas around the black hole \citep{Wilman2005}, and the $M_{\rm BH}$-$L_{\rm bulge}$ relation \citep{Bettoni2003} from the literature. We unified the black hole masses derived from various $M_{\rm BH}$-$\sigma$ relations into that of \cite{Tremaine2002}. We summarized all black hole masses we obtained in Table~\ref{table_BH}. The absolute magnitudes of host galaxies, $M_R$, shown in Table~\ref{table_BH} were obtained by using cosmological parameter values ($H_0 = \rm 50\ km\ s^{-1}\ Mpc^{-1}$ and $\Omega_0 = 0$; see \citealt{Falomo2003a}) different from ours. We did not modify them because the same parameter values were used to derive the $M_{\rm BH}$-$L_{\rm bulge}$ relation in \cite{Bettoni2003}. We averaged the black hole masses for each source (if there was more than one measurement)\footnote{We used the geometric mean for the averaging of the black hole masses. We note that using the geometric and the arithmetic mean in linear space led to almost the same result.}. 

The uncertainty of a given black hole mass is hard to quantify because of different geometries and kinematics of BLRs that give rise to errors in single epoch spectroscopic mass measurements (e.g., \citealt{Vestergaard2006, Park2012}), intrinsic scatter in the $M_{\rm BH}$-$\sigma$ relation (e.g., \citealt{Kormendy2013}), insufficient bolometric corrections of monochromatic continuum luminosities \citep{Trippe2015}, and potentially further effects. We adopt an error of 0.3 dex if the averaged black hole mass involves the $M_{\rm BH}$-$\sigma$ relation or single epoch spectroscopy with H$\beta$ and Mg II lines ($M_{\rm BH}(\rm{H}\beta)$, $M_{\rm BH}(\rm{Mg\ II})$), and of 0.4 dex if only $M_{\rm BH}(\rm{C\ IV})$ or a black hole mass obtained with the $M_{\rm BH}$-$L_{\rm bulge}$ relation was available. The above values were adopted based on the discussion on errors in black hole mass estimation with single epoch spectroscopy of \cite{Ho2012}, the intrinsic scatter in the $M_{\rm BH}$-$\sigma$ relation shown in \cite{Kormendy2013}, and the scatter found in the $M_{\rm BH}$-$L_{\rm bulge}$ relation of \cite{Bettoni2003}.

\begin{figure}[!t]
\begin{center}
\includegraphics[trim=10mm 0mm 0mm 0mm, clip, width = 89mm]{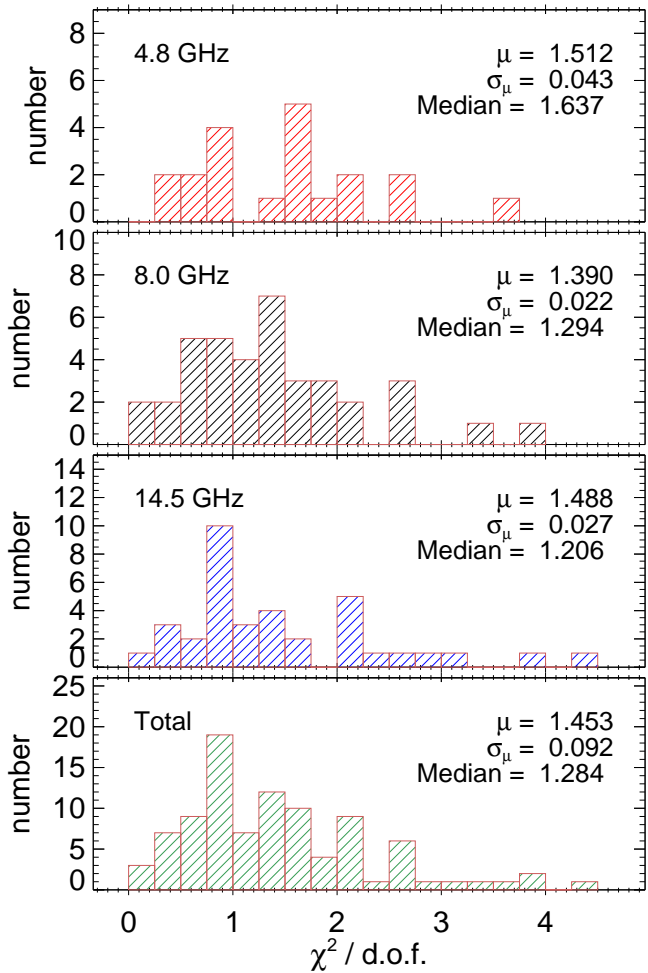}
\caption{Distribution of $\chi^2 / \rm d.o.f.$ for best-fit model periodograms calculated as outlined in Section~\ref{ssect_lc}. From top to bottom, panels show the results for 4.8 GHz, 8 GHz, 14.5 GHz, and all lightcurves combined, respectively. The mean ($\mu$), the standard error of mean ($\sigma_{\mu}$), and the median of the distribution are noted in each panel. \label{minchi}}
\end{center}
\end{figure}

We obtained the disk luminosities in Eddington units using the adopted black hole masses and the relation $L_{\rm Edd} \approx 1.5\times 10^{38}(M_{\rm BH}/M_{\odot})\rm\ erg\ s^{-1}$ \citep[cf., e.g.,][]{Netzer2013}, which can be used to obtain accretion rates and Eddington ratios when employing certain reasonable assumptions (see Section~\ref{ssect_bb}). The disk luminosity was calculated by assuming $L_{\rm disk} \approx 10L_{\rm BLR}$ according to \cite{Ghisellini2011}, where $L_{\rm BLR}$ is the BLR luminosity and was obtained following \cite{Celotti1997} who showed that $L_{\rm BLR} / L_{\rm Ly \alpha} = 5.56$, based on, e.g., \cite{Francis1991}. We averaged the disk luminosities for each source if multiple values from different line luminosities were available (mostly from \citealt{Wang2004, Liu2006}, and \citealt{Torrealba2012}). We obtained measurement uncertainties for the disk luminosities from those sources with $L_{\rm disk}$ derived from various emission lines, which allowed us to use the standard deviation of the luminosities as error. Since this was possible for only some of our sources, we adopt the mean values of their errors as typical errors for the other sources. We note that the estimated errors are governed by the assumption of constant line ratios rather than measurement errors in the luminosity of each line.

\section{Results and Discussion}
\label{sect4}

\subsection{General Features of Power Spectra \label{ssect_beta}}

Using the procedure outlined in Section~\ref{ssect_lc}, we obtained best-fit $\beta$ values for our sources ranging from $\approx$1 to $\approx$3. The observed power spectra are in general well described by simple powerlaw models. In Figure~\ref{minchi}, we show the distribution of $\chi^2 / \rm d.o.f.$ of our best-fit models, which is concentrated around unity, though with notable scatter. We deal with a few sources with large $\chi^2 / \rm d.o.f.$ in Section~\ref{ssect_br}. We found the timescales $\tau_{\rm med}$ to range from $\approx$0.3 to $\approx$6.5 years and $\sigma_{\mathrm{der}}$ to range from $\approx$0.3 to $\approx$10 $\rm yr^{-1}$ over all sources. We note that our best-fit model periodograms reproduce the local peaks seen in the observed power spectra for some sources (e.g.,1253$-$055 in Figure~\ref{ex} and 0316+413 in Figure~\ref{pd}). This result indicates that such patterns are introduced by the sampling of the lightcurves, not by source-intrinsic variability. Accordingly, interpreting any local peak in a power spectrum as an indication for quasi-periodic oscillations requires careful modelling of the power spectrum in order to prevent false positives.

\begin{figure}[!t]
\begin{center}
\includegraphics[trim=15mm 7mm 7mm 0mm, clip, width = 89mm]{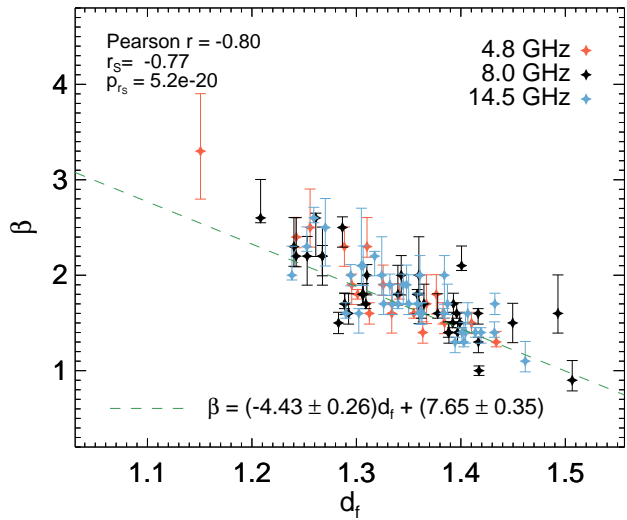}
\caption{Power spectral index $\beta$ versus fractal dimension $d_f$. Red, black, and blue points are 4.8, 8, and 14.5 GHz data, respectively. The green dashed line is the best-fit line to the data; the best-fit parameters are shown below at the bottom of the plot. The Pearson correlation coefficient $r$, the Spearman rank correlation coefficient $r_s$, and the statistical false-alarm probability of $r_s$, $p_{r_s}$, are noted in the top left of the diagram. \label{fig_Tfrac}}
\end{center}
\end{figure}

\begin{figure*}[!t]
\begin{center}
\includegraphics[trim=15mm 7mm 7mm 10mm, clip, width = 89mm]{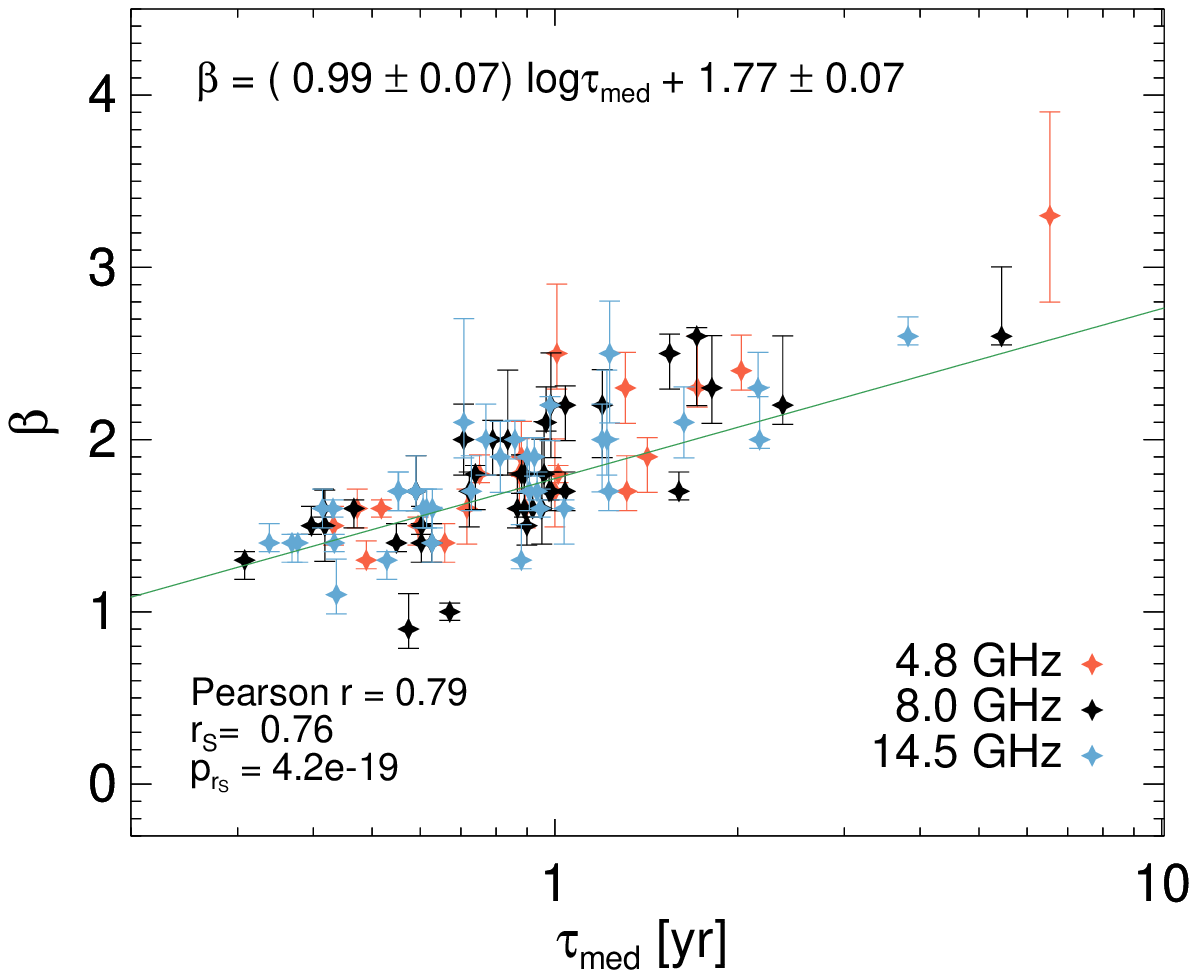}
\includegraphics[trim=12mm 7mm 10mm 10mm, clip, width = 89mm]{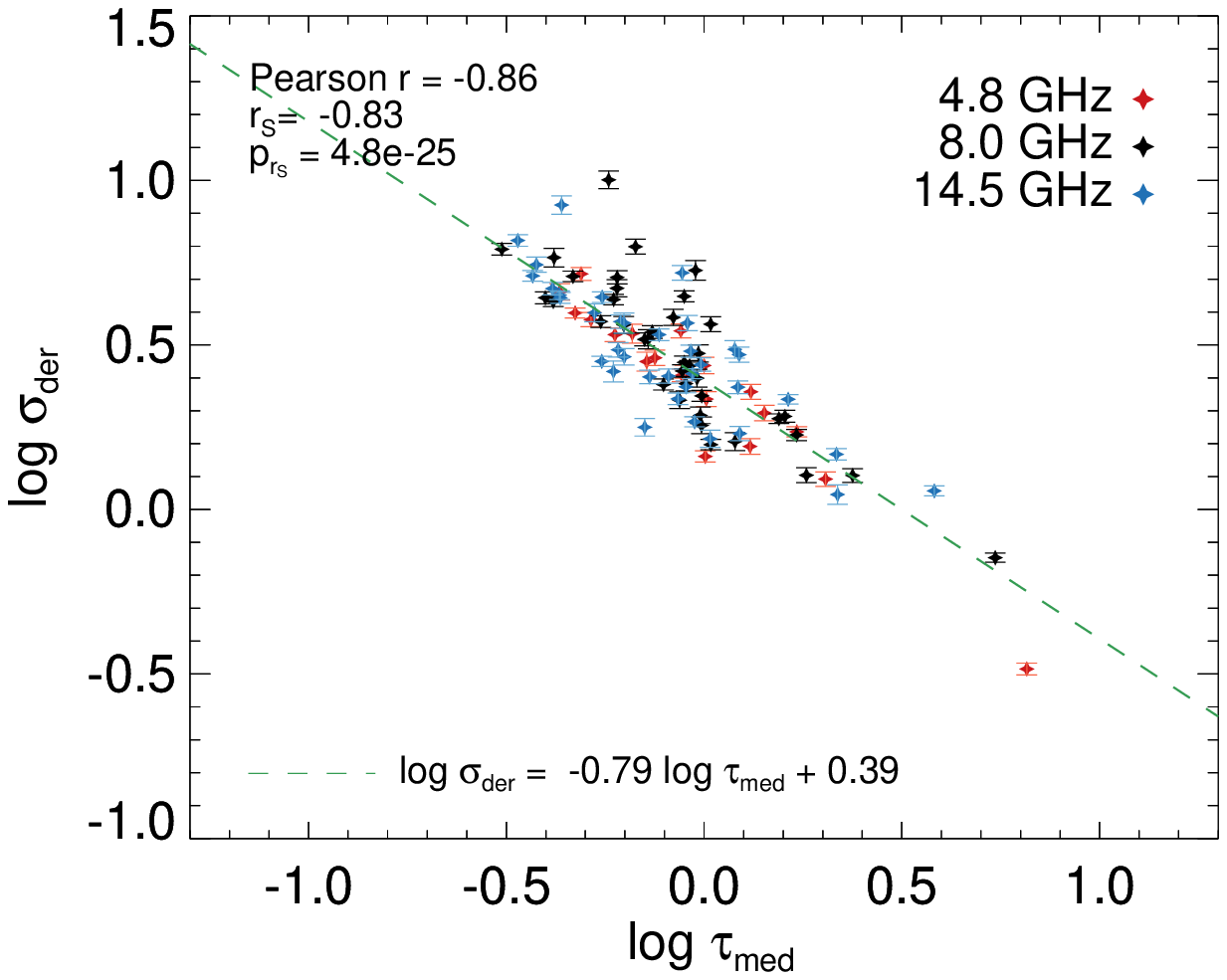}
\caption{\emph{Left}: Power law index $\beta$ as function of median duration of flares, $\tau_{\rm med}$. Red, black, and blue are 4.8, 8, and 14.5 GHz data, respectively. The green solid line is the best-fit line, the corresponding parameters are shown at the top of the plot. The Pearson correlation coefficient $r$, the Spearman rank correlation coefficient $r_s$, and the false-alarm probability of $r_s$, $p_{r_s}$, are given at the bottom left. \emph{Right}: Logarithm of the width of the distribution of the derivatives of lightcurves, $\log\sigma_{\mathrm{der}}$, as function of $\log\tau_{\rm med}$. The green dashed line is the best-fit line, the best-fit parameters are provided at the bottom of the plot.\label{Deconv}}
\end{center}
\end{figure*}

\subsection{Distributions of Fractal Dimension\label{ssect_fracres}}

We show the relation between $\beta$ and the fractal (box-counting) dimension of the lightcurves (cf. Section~\ref{ssect_fractal}), $d_f$, in Figure~\ref{fig_Tfrac}. We find a strong anti-correlation with correlation coefficients (Pearson and Spearman rank) around $-0.8$.\footnote{Obtaining meaningful correlation coefficients requires that the data under study are uncorrelated. This is not always strictly the case in our analysis because we include data originating from different lightcurves (at two or three frequencies) from the same source. However, the lightcurves at different frequencies are quite different in general and show different sampling patterns. Therefore, we do not average data in frequency except when studying parameters (such as black hole mass) that cannot depend on frequency.} At least qualitatively, this seems rather obvious because a larger fractal dimension means that a light curve fills more grid cells. This in turn implies a more strongly fluctuating lightcurve which comes with a smaller value of $\beta$. Even though, we present here for the first time the quantitative relation between $\beta$ and $d_f$,
\begin{equation}
\beta = -(4.43 \pm 0.26) d_f + (7.65 \pm 0.35).
\end{equation}
This relation holds over a wide range of $\beta$ values from $\approx$1 to $\approx$3 within errors with no notable dependency on observing frequency. This result provides a good independent check of our methodology.

\subsection{$\beta$ as an Indicator of Variability Timescale\label{ssect_bt}}

 We find a strong correlation between the power spectral index $\beta$ and the logarithm of the median duration of the flares obtained in Section~\ref{ssect_fit} (left panel of Figure~\ref{Deconv}). The best-fit linear relation is $\beta \propto 0.99\log\tau_{\rm med},$\footnote{Actually, $\chi^2$ fitting assumes that errors are symmetric, whereas we obtained asymmetric errors for $\beta$. To be conservative, we used the larger error for fitting.} where $\tau_{\rm med}$ is the median duration of flares. This result implies that the longer the overall duration of the flares of radio bright AGNs, the steeper the power spectra. Technically, such behavior is straightforward to understand: if a source shows flares with long duration, its lightcurve is dominated by long-term variability, which leads to steeper power spectra. However, the duration of flares is arguably related to fundamental physical processes in the AGN (like shocks in jets), and thus to various physical parameters (e.g., \citealt{Marscher1985, Fromm2011}). In turn, this indicates that $\beta$ can be used to derive some physical parameters of AGNs (to be specified below).

Else than AGN variability at high observing frequencies (especially X-rays), radio variability has not received much attention because of the difficulties in quantifying the properties of the variability. The power spectra of AGNs at radio wavelengths usually show featureless simple power-law noise as seen in Figure~\ref{pd} but no characteristic break frequencies as found in X-rays. From the relation between $\beta$ and the median duration of flares, we conclude that the slope of the power spectra represents the variability timescales of radio bright AGNs. This implies that measurements of $\beta$ are able to reveal the complex accretion and jet physics of AGNs.

The left panel of Figure~\ref{Deconv} shows that the data points at large $\tau_{\rm med}$ tend to lie above the best fit line. This is mainly because the number of flares becomes very small (three to five) for sources with large flare durations, thus making the use of the median problematic. In addition, the approach used in Section~\ref{ssect_fit} makes a strong assumption, namely that all lightcurves can be described as sequences of flares with Gaussian profiles -- an assumption that may or may not be generally valid.

In order to arrive at a more robust estimate, we focus on the parameter $\sigma_{\mathrm{der}}$, the width of the distribution of the derivatives of a lightcurve obtained in Section~\ref{ssect_der}, instead of $\tau_{\rm med}$ in the following. The parameter $\sigma_{\mathrm{der}}$ is inversely proportional to the effective variability timescale of a given lightcurve. It uses all data in a lightcurve, making its use statistically more rigorous than using the median duration of flares. In addition, the (statistical) errors of $\sigma_{\mathrm{der}}$ are known. As shown in the right panel of Figure~\ref{Deconv}, $\log \sigma_{\mathrm{der}}$ and $\log \tau_{\rm med}$ indeed show a strong anti-correlation. The scatter around the best-fit line in that figure and the fact that the two quantities do not show a one-to-one relation demonstrate the limited accuracy of the median duration of flares as a proxy for an effective variability timescale. Indeed, $\sigma_{\mathrm{der}}$ does contain information on flaring activity, especially the duration of flares: if the variability is dominated by flares with longer duration, this will lead to smaller $\sigma_{\mathrm{der}}$ regardless of our choice of models for fitting the lightcurves (compare Figure~\ref{fig_fit} and~\ref{Diff}).

\begin{figure*}[!t]
\begin{center}
\includegraphics[trim=15mm 7mm 7mm 10mm, clip, width = 89mm]{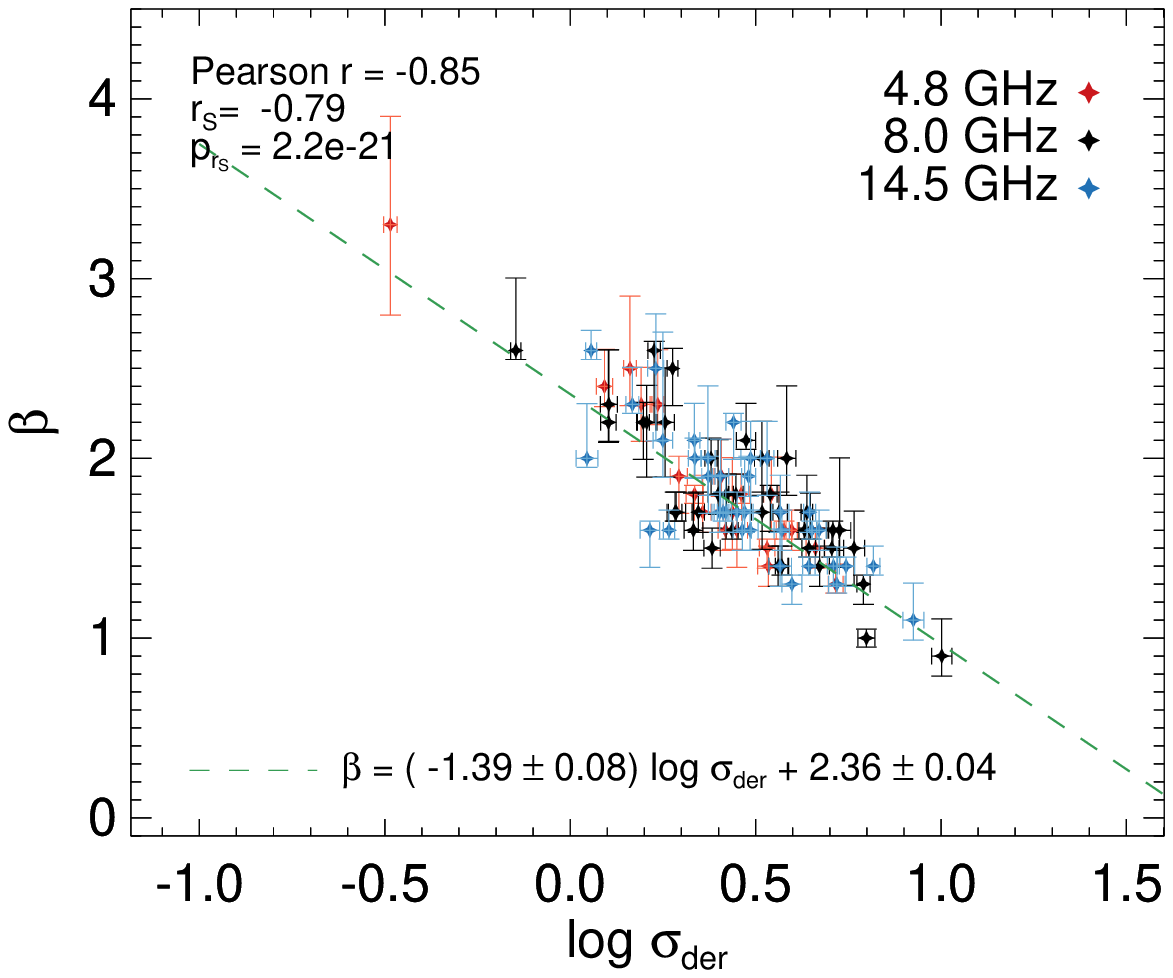}
\includegraphics[trim=12mm 7mm 10mm 10mm, clip, width = 89mm]{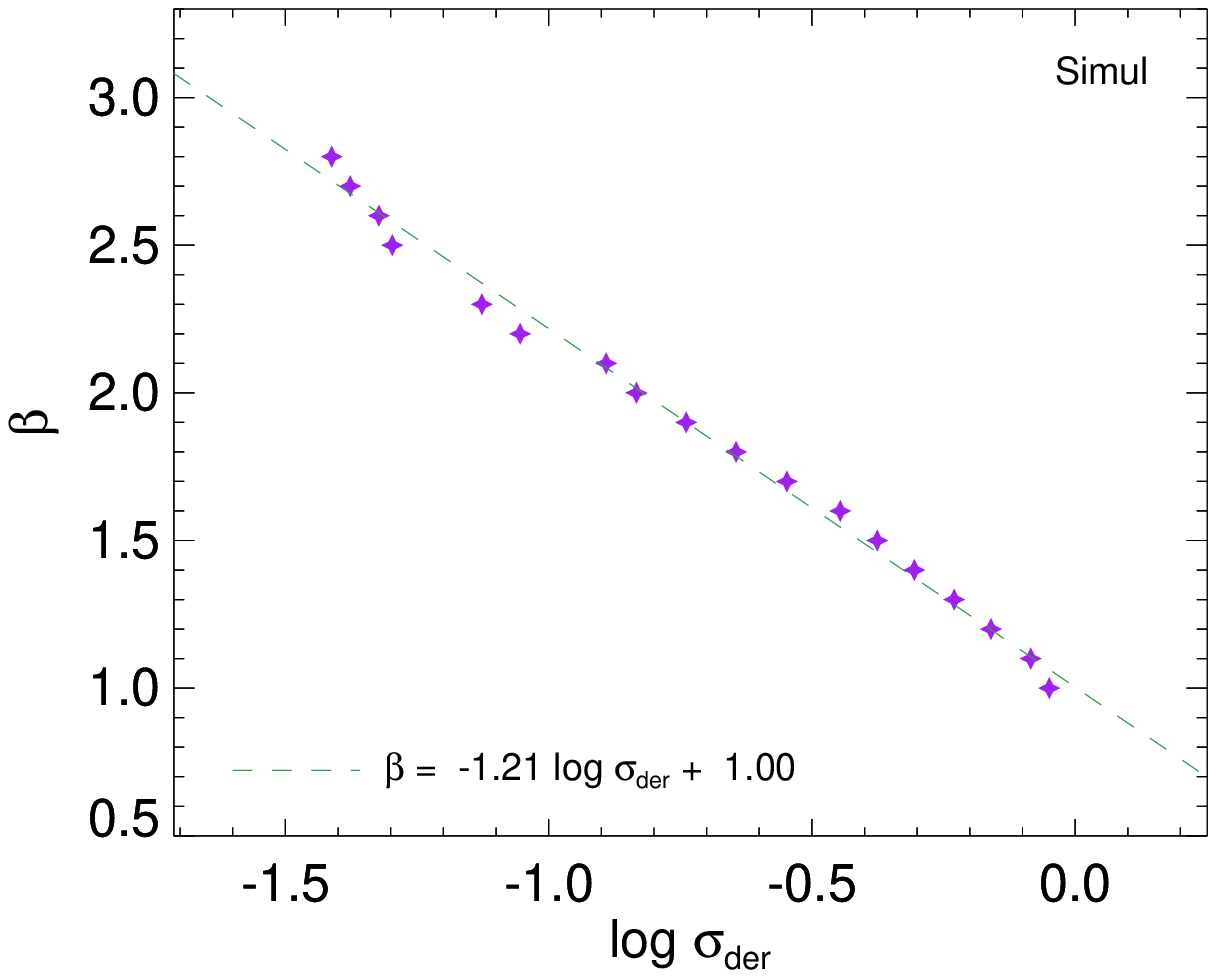}
\caption{\emph{Left}: Power law index $\beta$ as function of $\log\sigma_{\mathrm{der}}$ (which is defined in Section~\ref{ssect_der}, in units of yr$^{-1}$). Red, black, and blue colours indicate 4.8, 8, and 14.5 GHz data, respectively. The green dashed line shows the best-fit line (using errors on both axes with the FITEXY estimator; cf. Section~\ref{ssect_bt}). The error bars of $\log\sigma_{\mathrm{der}}$ were obtained by using the standard error propagation of the errors in $\sigma_{\mathrm{der}}$. The Pearson correlation coefficient $r$, the Spearman rank correlation coefficient $r_s$, the false-alarm probability of $r_s$, $p_{r_s}$, and the result of the linear regression are noted in the diagram. \emph{Right:} Same as the left panel, for \emph{simulated} red-noise lightcurves (see Section~\ref{ssect_bt} for details). \label{b_diff}}
\end{center}
\end{figure*}

We analyzed the relation between $\beta$ and $\log \sigma_{\mathrm{der}}$. A linear regression returns $\beta \propto -(1.39\pm0.08)\log\sigma_{\mathrm{der}}$ (see the left panel of Figure~\ref{b_diff}). We employed the FITEXY estimator \citep{Press1992} for a linear fit to data with errors on both axes.\footnote{We refer the reader to \cite{Tremaine2002} who discuss the advantages and disadvantages of this method over other fitting algorithms and to \cite{Kelly2007} who deal with more complicated situations.} We checked whether the observed relation between $\beta$ and $\log\tau_{\rm med}$ also appears in simulated data using red-noise only lightcurves. We generated 100 artificial lightcurves using the method of \cite{Timmer1995} with $\beta$ ranging from 1 to 2.8. The simulated lightcurves were sampled at equal intervals. We added Gaussian noise amounting to 2\% of the standard deviation of a given lightcurve to take the effect of measurement noise into consideration. We obtained $\sigma_{\mathrm{der}}$ from the distribution of the derivatives of the normalized simulated lightcurves as we did for the observed lightcurves. The relation between $\beta$ and $\log\sigma_{\mathrm{der}}$ for the simulated data is shown in the right panel of Figure~\ref{b_diff}. Overall, the data points are described well by a power-law function, the slope is (within $\sim$2$\sigma$) consistent with the observed one. We note that the value of the constant term is arbitrary because the unit of time is arbitrary.

The consistency between observed and simulated $\beta-\log \sigma_{\mathrm{der}}$ relations indicates that the observed relation is actually a generic feature of red noise lightcurves. One might ask if this conclusion is consistent with the presence of distinct flares in the radio lightcurves of AGNs -- flares are deterministic, whereas red-noise time series are stochastic by nature. When a flare begins, the flux density increases during the Compton and synchrotron stages (see e.g., \citealt{Marscher1985, Valtaoja1992, Fromm2011}) but decays eventually. Accordingly, we have to conclude that the aperiodic occurrence of flares makes an AGN lightcurve a red-noise time series. The duration of flares changes with time and flares occur at random times, leading to frequent superposition of flares (see, e.g., Figure~\ref{fig_fit}). This behaviour might originate from underlying physical processes such as a long-term, red-noise like variability of accretion rate and/or particle injection rate into jets that is physically correlated over a time comparable to the observation period. In this case, we expect to find power spectra with a single slope over a large range of sampling frequencies. However, as we will see below, a few sources show a flattening of their power spectra at low sampling frequencies; this implies that the timescales over which an emission process is correlated can be shorter than the observation period, roughly on the order of the typical duration of flares. We will discuss the possible origin of featureless red-noise power spectra in more detail in Section~\ref{ssect_br}.

\subsection{Relation between $\beta$ and the Accretion Rate\label{ssect_bb}}

The reason why different radio bright AGNs show different variability patterns, specifically different $\beta$, is not well-studied. In the case of radio-quiet, optical/X-ray bright AGNs and GBHs, a well-known scaling relation between the timescales that correspond to the break frequencies in their power spectra and the black hole mass indicates that their variability timescales are determined by the size of the emitting region (e.g., \citealt{McHardy2004, UM2005, Kelly2009, Kelly2011}). This size would be proportional to the Schwarzschild radius which in turn scales linearly with the black hole mass. However, it is not clear if a similar scaling relation also holds for radio bright AGNs. Radio bright AGNs emit their flux from relativistic jets instead of accretion disks, and emit synchrotron radiation instead of thermal radiation. Therefore, one first needs to find a physical mechanism that determines the duration of radio flares.

As noted in Section~\ref{ssect_fit}, our lightcurves can be described as sequences of Gaussian -- and thus symmetric in time -- flux peaks. Time-symmetric flares from blazars have been observed at multiple observing frequencies (see, e.g., \citealt{Hovatta2008} for radio, \citealt{Chatterjee2012} for optical and $\gamma$-ray, and \citealt{Abdo2010} for $\gamma$-ray observations). The symmetry in time has been interpreted as the result of rise and decay timescales being determined by the crossing time of radiation (or particles) through the emission region. \cite{Jorstad2005} showed that the radiative cooling time is shorter than the cooling timescale of adiabatic expansion for almost all jet components in their VLBI blazar sample. Therefore, one may expect the duration of flares to be given by the sizes of emission regions.

In most cases, radio flares of AGNs are associated with the inner regions of jets (often identified with compact VLBI cores). Especially, there is growing evidence for interaction of moving jets with (probably) stationary cores, leading to strong flares at high energies (from optical to $\gamma$-rays) and, likewise, at cm/mm wavelengths \citep{Savolainen2002, Leon-Tavares2010, Arshakian2010}. The cm/mm flares show much broader flare widths and time delays relative to the high energy flares, likely due to relatively long cooling timescales and high optical depths (e.g., \citealt{Savolainen2006, Jorstad2010, Marscher2008, Marscher2010, Marscher2011, Marscher2013}). If the core is a conical, standing shock (commonly assumed to be a recollimation shock, see e.g., \citealt{Cawthorne2006, Cawthorne2013, Marscher2006, Marscher2014}), then the duration of flares would be determined by the crossing time of jet material through the shock. The core might actually consist of multiple stationary shocks; stationary knots in addition to the cores have been discovered by VLBI for many, usually nearby, sources (e.g., \citealt{Jorstad2005, Jorstad2010, Cohen2014}).

\begin{figure*}[!t]
\begin{center}
\includegraphics[trim=12mm 7mm 10mm 10mm, clip, width = 89mm]{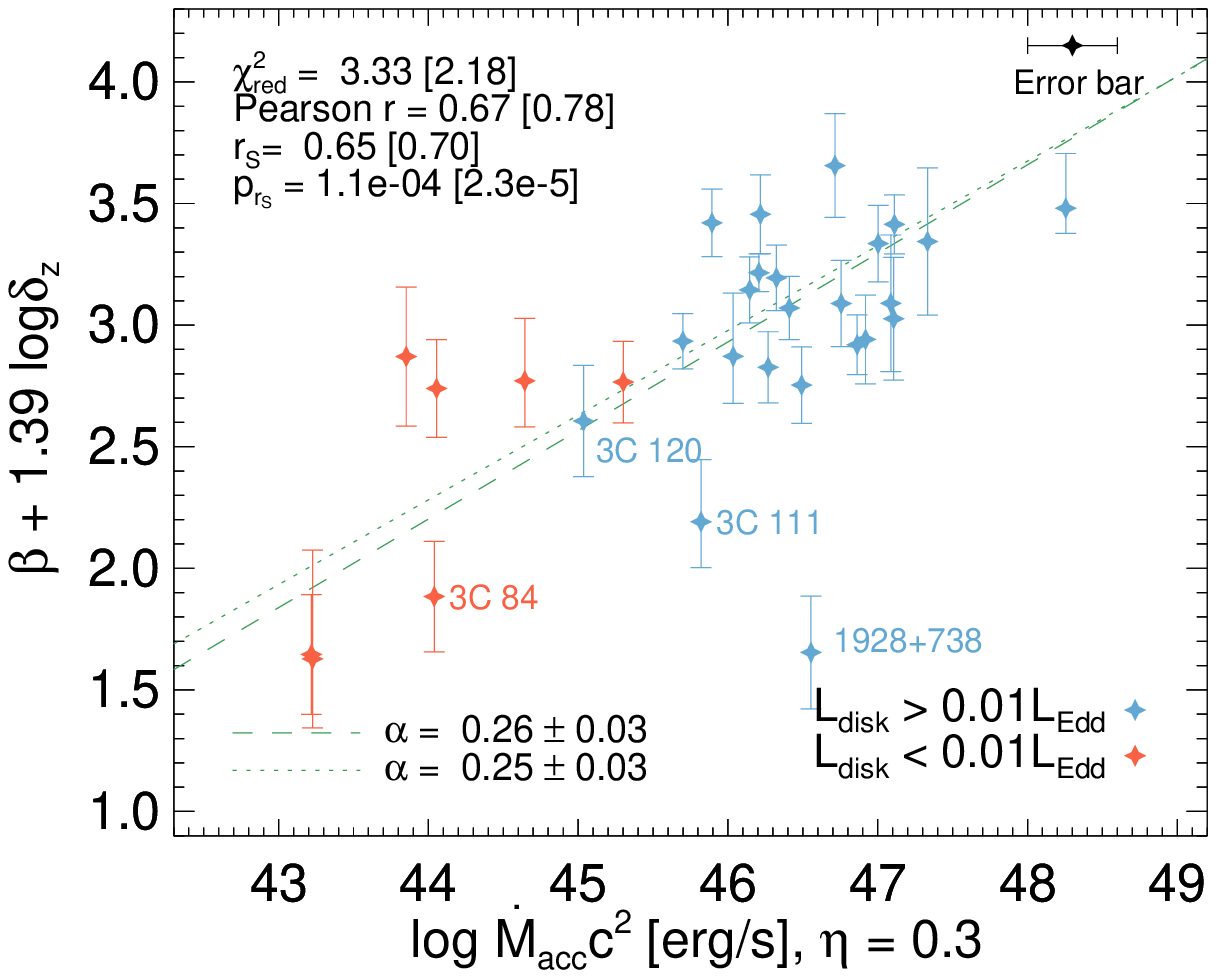}
\includegraphics[trim=12mm 7mm 10mm 10mm, clip, width = 89mm]{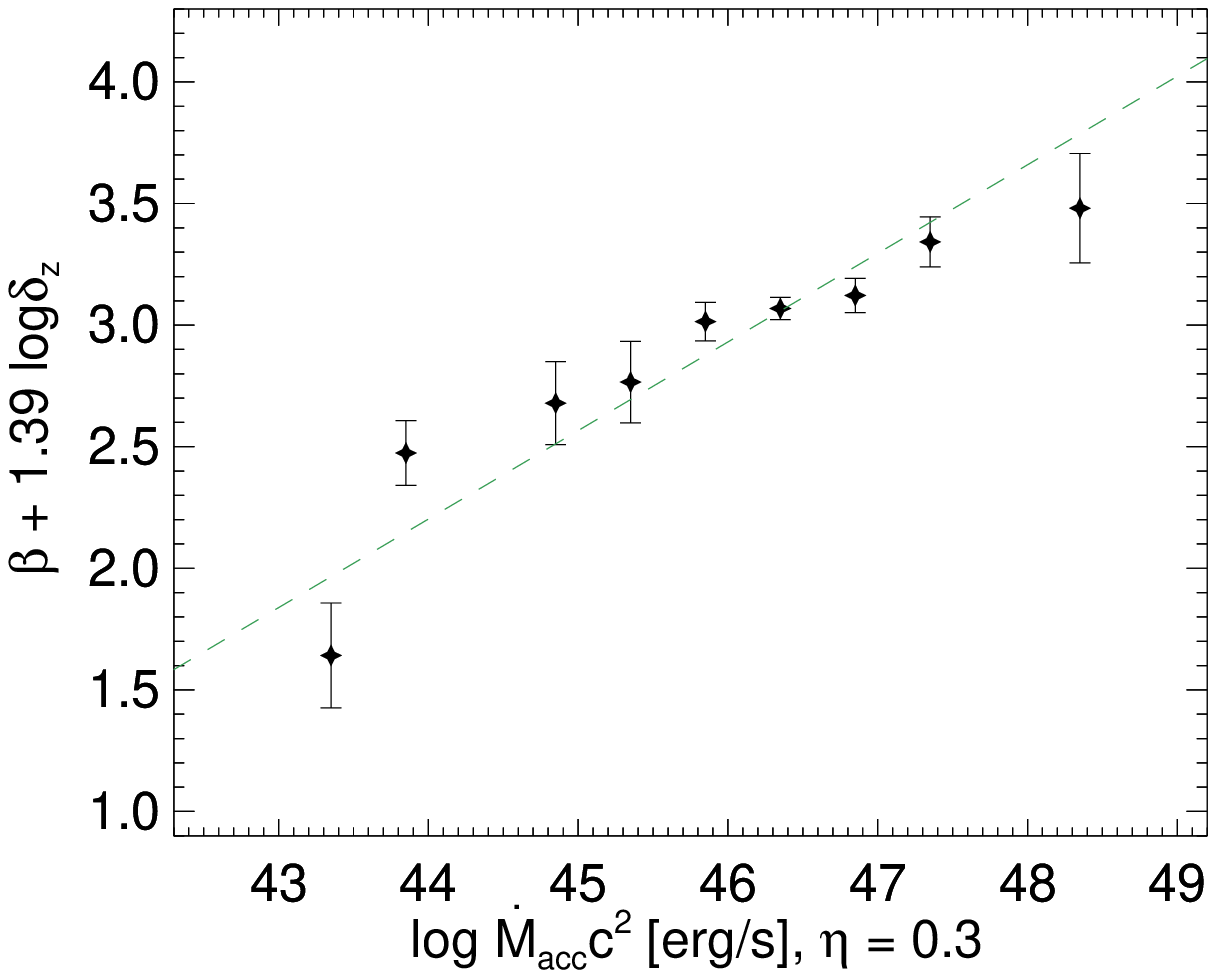}
\caption{\emph{Left}: The parameter $\beta + 1.39\log\delta_z$ as function of accretion disk luminosity. The sources are divided into two groups based on their disk luminosities in Eddington units (see Section~\ref{ssect_bb} for details): sources with disk luminosities $>0.01L_{\rm Edd}$ in blue, those with $<0.01L_{\rm Edd}$ in red. The names of the radio galaxies (3C 84, 3C 111, 3C 120) are noted. We used the weighted average of $\beta$ for a given source if values from two or three frequencies were available. Errors along the ordinate are obtained by standard propagation of errors in $\beta$ and $\log\delta_z$. The typical error along the abscissa, 0.29 dex, is illustrated by the black data point with error bars in the top right. The green dashed (dotted) line illustrates a linear regression with (without) including the outlier 1928+738, using the errors along both axes. The best-fit slopes divided by 1.39, denoted $\alpha$, and their statistical error are provided at the bottom. The $\chi^2 / \rm d.o.f$ of the best-fit model, the Pearson correlation coefficient $r$, the Spearman rank correlation coefficient $r_s$, and the false-alarm probability of $r_s$, $p_{r_s}$, are noted. The values in the bracket are obtained excluding the outlier 1928+738 in the calculation. \emph{Right}: Same as the left panel but with all the data after binning logarithmically in accretion power with a binsize of 0.5 dex. The green dashed line, the same one as in the left panel, is drawn to show that the binned data also follows the best fit line.\label{fig_BH}}
\end{center}
\end{figure*}

In this scenario, higher rates of matter injection into AGN jets would lead to longer flare durations, or variability timescales, if the particle densities and bulk velocities of inner jets are similar across our sample. These assumptions are supported by observations of the particle densities in the jets of several blazars \citep{OG2009} and the fact that the location of standing shocks is expected to be at the end of the acceleration and collimation zone of the jet (e.g., \citealt{Marscher2008, Marscher2014}). The rate of matter injection into the jet, $\dot{M}_{\rm jet}$, would (largely) determine variability timescales in radio bright AGNs. In a given time interval an AGN with higher $\dot{M}_{\rm jet}$ would show, say, one major flare while those with smaller $\dot{M}_{\rm jet}$ would show multiple minor flares. Recent theoretical studies actually show that the rate of electron injection into jets can play an important role in determining the slope of power spectra \citep{Finke2014, Finke2015}. 

\cite{Ghisellini2014} found a correlation between the jet power $P_{\rm jet}$ and the accretion power of blazars of the form $P_{\rm jet} \propto \dot{M}_{\rm acc}$, where $\dot{M}_{\rm acc}$ is the accretion rate. The jet power $P_{\rm jet}$ is given by the kinetic energy per time, i.e., $P_{\rm jet} \approx (\Gamma - 1)\dot{M}_{\rm jet}c^2$, where $\Gamma$ is the Lorentz factor. Since our sources are luminous blazars and usually show superluminal proper motions, their Lorentz factors are located in a rather narrow range \citep[cf.][]{Ghisellini2014}. Thus, we have $P_{\rm jet} \propto \dot{M}_{\rm jet} \propto \dot{M}_{\rm acc}$. Evidence for the proportionality between $\dot{M}_{\rm jet}$ and $\dot{M}_{\rm acc}$ has been provided by \cite{Chatterjee2009, Chatterjee2011} who discovered that significant dips in the X-ray light curves of the radio galaxies 3C 111 and 3C 120 are followed by ejections of new superluminal jet components. This indicates that X-ray emitting matter in hot coronae and/or the innermost accretion disks is ejected in a jet outflows. Combining the various arguments, we examined if radio bright AGNs indeed show a scaling relation between the variability timescales and the accretion rates.

When comparing the observed variability timescales with other parameters we need to correct for the effects of Doppler boosting. The observed variability timescale is decreased relative to the intrinsic one, $\tau_{\rm var}$, by the Doppler factor $\delta = 1 / \Gamma(1-\beta\cos\theta)$, where $\Gamma$ is the Lorentz factor, $\beta$ is the jet speed in units of speed of light, and $\theta$ is the angle between the jet axis and the line of sight. If cosmological redshift is non-negligible, the total Doppler factor is $\delta_z = \delta / (1+z)$. If we assume that the variability timescales of our sources scale with the accretion rate to a power $\alpha$, i.e., $\tau_{\rm var} \propto \dot{M}_{\rm acc}^\alpha / \delta_z$, then we find from the relation $\beta \propto 1.39\log\tau_{\rm var}$ (see Section~\ref{ssect_bt})
\begin{equation}
\label{eq5}
\beta \propto 1.39\alpha\log \dot{M}_{\rm acc} - 1.39\log\delta_z.
\end{equation}

The Doppler factor is difficult to measure directly because the two parameters involved, intrinsic jet speed and viewing angle, are hard to disentangle in many cases. Nevertheless, \cite{Hovatta2009} obtained the Doppler factors of many of our sources. They decomposed their lightcurves obtained at 22 and 37 GHz into exponentially rising and decaying flares. They assumed that the brightness temperature derived from the flux variability differs from the radiating particle--magnetic field energy equipartition temperature by the Doppler factor (\citealt{Readhead1994}, see also \citealt{LV1999, Lahteenmaki1999, Savolainen2010}). In addition, \cite{Jorstad2005} obtained the Doppler factors for some of our sources assuming that the observed variability timescales differ from the light crossing time across the emitting region because of Doppler boosting affecting observed jet components. For 1101+384 and 1652+398, we took the values from \cite{Lico2012} and \cite{Tavecchio1998}, respectively. We note that the Doppler factor of 1101+384 measured by \cite{Lico2012} is somewhat different from that of \cite{Tavecchio1998}. However, we adopted the argument of the former that different Doppler factors for radio and high energy photons are necessary for this source. We refer the readers to \cite{Katarzynski2001} who obtained $\delta = 7-14$ for 1652+398, depending on their models for the observed spectral energy distribution. This result is consistent with that of \cite{Tavecchio1998} and we adopted their value, $\delta = 10$. We note, however, that this value could be biased because it was derived from modelling the spectral energy distribution of higher energy photons, which might originate from a different emission region. In total, we were able to retrieve the Doppler factors for 39 out of 43 sources; the values are shown in Table~\ref{Information}. Where Doppler factors from both \cite{Hovatta2009} and \cite{Jorstad2005} were available, we first took the average of all values of the latter because they provided individual Doppler factors for each jet component of a given source. Then, we took the average of the Doppler factor of \cite{Hovatta2009} and the averaged one of \cite{Jorstad2005}. We used the standard deviation of the logarithms of the Doppler factors of \cite{Jorstad2005} as the error of $\log\delta$, i.e., $\sigma_{\log\delta}$, for each source. Variations in the values for different jet components might originate from intrinsic variability of the Doppler factors and/or measurement errors. For some sources, only the values of \cite{Hovatta2009} were available; in those cases, we assigned the average $\sigma_{\log\delta}$ from sources for which we could actually estimate the error ($\approx 0.147$ dex) as ``typical'' error. We note that this value is consistent with an independent estimate of the mean uncertainty of variability Doppler factors, which is $\approx30\%$ \citep{Liodakis2015}.

\begin{figure*}[!t]
\begin{center}
\includegraphics[trim=12mm 7mm 10mm 10mm, clip, width = 89mm]{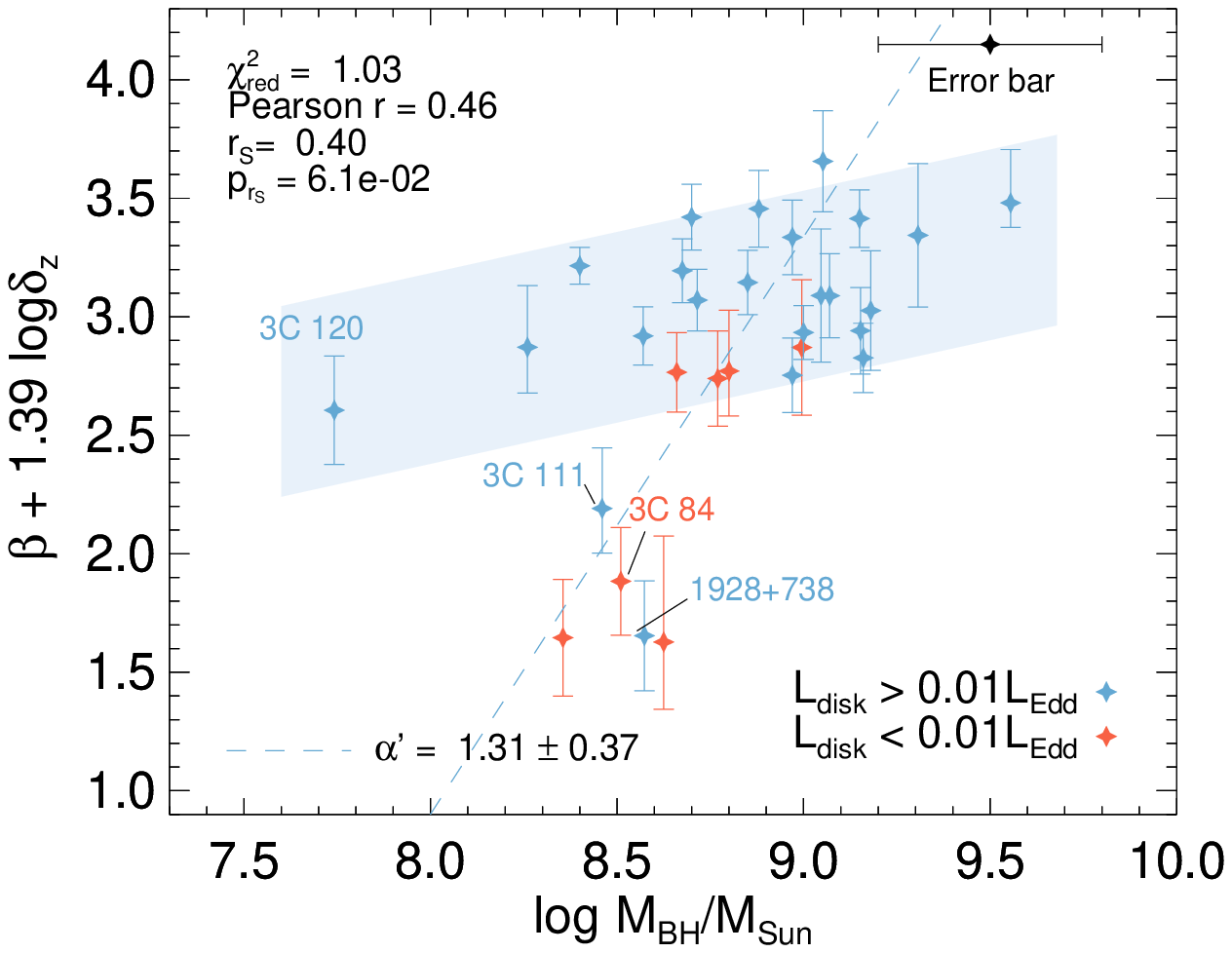}
\includegraphics[trim=12mm 7mm 10mm 10mm, clip, width = 89mm]{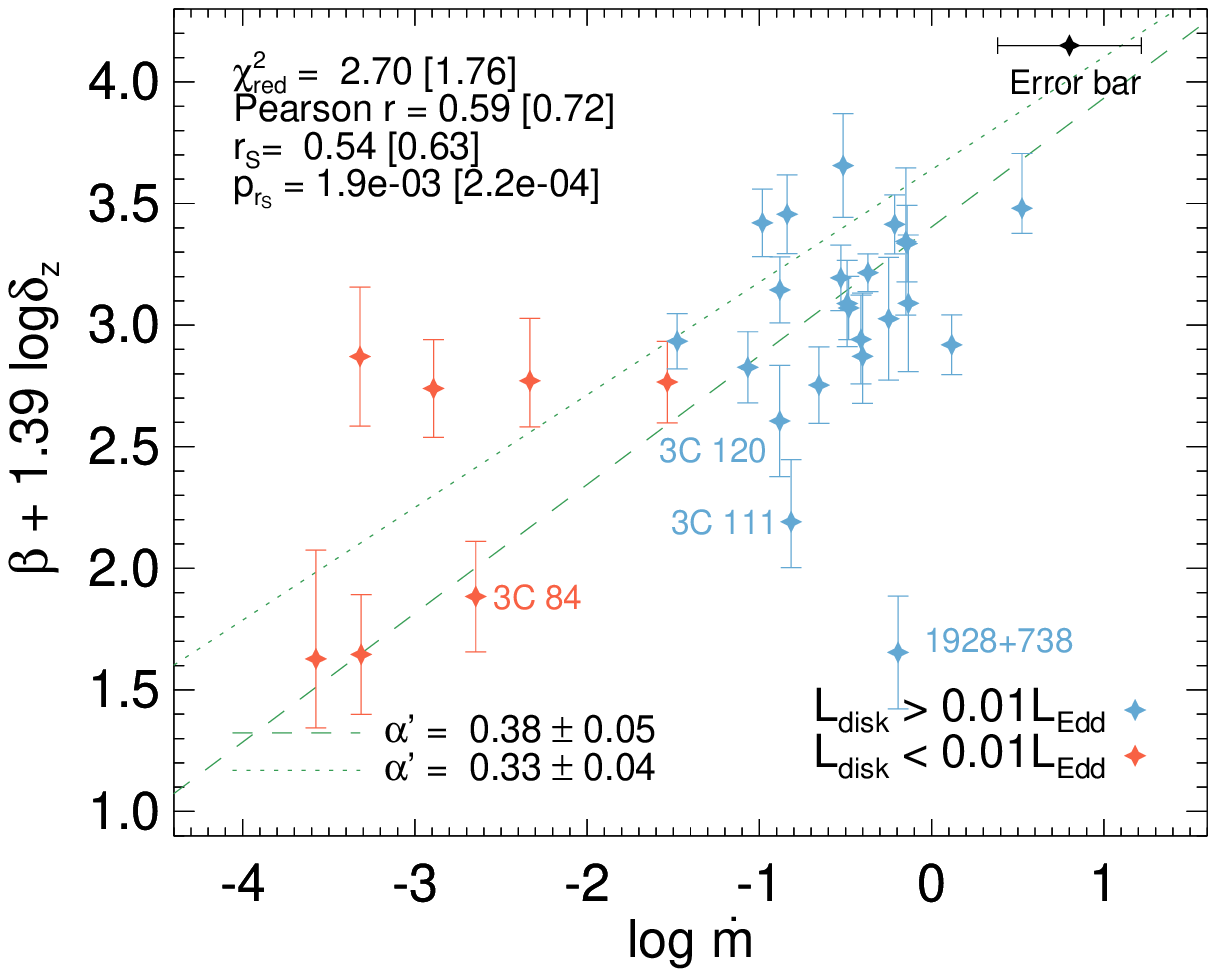}
\caption{\emph{Left}: Same as the left panel of Figure~\ref{fig_BH} but with the black hole mass on the abscissa. The typical error along the abscissa, 0.3 dex, is illustrated by the black data point with error bars in the top right. The blue dashed line illustrates a linear regression, using the errors along both axes. The blue shaded region indicates a relation with slope $\alpha = 1/4$ (drawn manually). The best-fit slope divided by 1.39, i.e., $\alpha$, and its statistical error are provided at the bottom. The $\chi^2 / \rm d.o.f$ of the best-fit model, the Pearson correlation coefficient $r$, the Spearman rank correlation coefficient $r_s$, and the false-alarm probability of $r_s$, $p_{r_s}$, are noted. \emph{Right}: Same as the left panel in Figure~\ref{fig_BH} but with the Eddington ratio on the abscissa. The typical error along the abscissa, 0.42 dex, obtained via standard error propagation of the errors in $\log \dot{M}_{\rm acc}$ and $\log M_{\rm BH}$, is illustrated by the black data point with error bars in the top right.  \label{fig_BH2}}
\end{center}
\end{figure*}

The disk luminosity, in units of Eddington luminosity, is an indicator of accretion rate because the normalized accretion rate is given by $\dot{m} \equiv \dot{M}_{\rm acc} / \dot{M}_{\rm Edd} = L_{\rm disk} / \eta L_{\rm Edd}$, where $\eta$ is the radiative efficiency of accretion. According to the standard, geometrically thin accretion disk theory \citep{Shakura1973}, $\eta$ depends on the location of the innermost stable orbit of the disk and thus on the spin of the black hole. \cite{Ghisellini2014} showed that jet launching and acceleration must be extremely efficient for blazars to explain the excess of jet power over accretion power. This requires almost maximally rotating black holes (See also e.g., \citealt{Tchekhovskoy2011} and \citealt{Zamaninasab2014}). Therefore, our sources likely have $\eta$ values close to the limiting case $\eta \approx 0.3$ chosen by \cite{Ghisellini2014}. In summary, we compared the intrinsic variability timescales (Doppler-corrected) with the accretion power, $\dot{M}_{\rm acc}c^2$, where the accretion rate is derived from the disk luminosity assuming $\eta = 0.3$, in the left panel of Figure~\ref{fig_BH}. We note that we rearranged Equation~\ref{eq5} in order to avoid displaying large errors along one axis.

Despite some scatter, we find a strong correlation. We note that the correlation coefficients become significantly larger when we exclude the FSRQ 1928+738 from the calculation which is arguably an outlier. We suspect that the Doppler factor of this source is systematically underestimated ($\delta_z = 1.5$), even though it shows quite fast superluminal motion with a maximum jet speed of 8.16 times the speed of light without showing any indication of counter jet emission \citep{Lister2013}. The Spearman rank correlation coefficient -- which is less sensitive to outliers -- shows that the positive correlation between $\log \dot{M}_{\rm acc}c^2$ and $\beta + 1.39\log\delta_z$ is statistically significant, with the false alarm probability $p_{\rm r_{s}}$ being about 0.01\%. We performed a linear regression using the errors on both axes with the FITEXY estimator (Section~\ref{ssect_bt}) and obtained a slope of $0.36 \pm 0.04$. This value translates into $\alpha = 0.26 \pm 0.03$ according to Equation~\ref{eq5}. The value of $\chi_{\rm red}^2 = \chi^2 / \rm d.o.f.$ is close to one, especially when (the value given in the bracket) 1928+738 is excluded; this indicates a good agreement of model and data over five orders of magnitude in accretion power. In the right panel of Figure~\ref{fig_BH}, we binned the data in the left panel logarithmically in accretion power with a binsize of 0.5 dex and the best fit line from the un-binned data is shown together.

We investigated possible differences in the scaling relation for the two classes of radio bright AGNs, i.e., BLOs and FSRQs. We divided our sources into those with disk luminosities above and those with disk luminosities below 1\% of the Eddington luminosity, which corresponds to FSRQs and BLOs, respectively. This approach is based on \cite{Ghisellini2011} who showed that using the ratio of disk or BLR luminosity and Eddington luminosity is more adequate to distinguish FSRQs and BLOs compared to the classical one using the equivalent width of emission lines. The value of $\approx 1\%$ is known to divide different accretion regimes of AGNs (e.g., \citealt{Ghisellini2011, Heckman2014}) and the same parameter can also be used to distinguish FR 1 and FR 2 radio galaxies (e.g., \citealt{Baum1995}). We counted FR I galaxies, in our case 3C 84, as BLOs and the FR II galaxy 3C 111 and 3C 120 as FSRQ (see Table~\ref{table_BH}). This is in accord with, e.g., \cite{Padovani1992}, \cite{Maraschi1994}, and \cite{Cavaliere2002} who suggested that FR I and FR II radio galaxies are the parent populations of BLOs and FSRQs respectively. However, as seen in the left panel of Figure~\ref{fig_BH}, we do not see any indication of difference in the scaling relation between different classes of radio bright AGNs, although the small number of BLOs and the fact that all BLOs in our sample are among the most radio-loud objects prevent us from drawing strong conclusions.

\begin{figure}[!t]
\begin{center}
\includegraphics[trim=15mm 7mm 7mm 0mm, clip, width = 89mm]{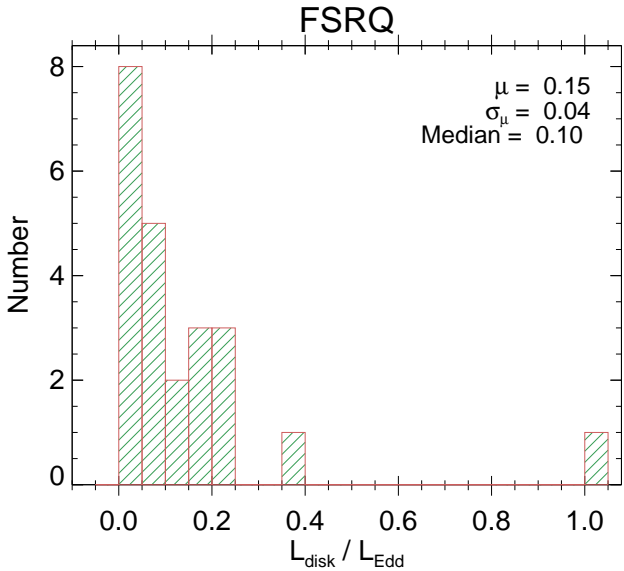}
\caption{Distribution of the disk luminosities, in units of Eddington luminosity, for FSRQs. The mean ($\mu$), the standard error of mean ($\sigma_{\mu}$), and the median of the distribution are noted. \label{fig_Mdot}}
\end{center}
\end{figure}

The fact that all our sources share the same scaling relation regardless of their source types implies that the variability timescales of radio bright AGNs are determined by a relatively simple physical process -- only weakly (if at all) dependent on jet powers (e.g., \citealt{Ghisellini2011}), radiative cooling mechanisms (e.g., \citealt{Ghisellini2009a}), and possible differences in the geometry of magnetic field lines pervading in jets (e.g., \citealt{Marscher2002, LH2005}, see also \citealt{Lyutikov2005}). The clear relation between variability timescales of AGNs at radio wavelengths and accretion rates measured at optical wavelengths comes as a surprise: this behavior indicates that the radio variability of radio-bright AGNs is governed by the accretion process. However, the rather shallow ($\alpha \approx 0.25$) slope in the scaling relation is hard to explain in the frame of a simple conical jet scenario. If (a) flares arise when a conical jet flow passes through a standing shock and (b) jet opening angles do not vary substantially among different AGNs, one arrives at a simple relation between the accretion rate and the length of the jet along the jet direction, $l$, namely: $\dot{M}_{\rm acc}\Delta t \propto \rho l^3$. Here $\Delta t$ is a rest-frame time interval (which is different from the observer frame interval by a factor $(1+z)$) and $\rho$ is the mass density of the jet. If the jet is in a steady state, we can expect $\rho \propto l^{-2}$ which leads to a linear proportionality between $\dot{M}_{\rm acc}$ and $l$ -- thus the intrinsic variability timescale is proportional to the accretion rate. However, the slope we find, 0.25, is quite different from the one expected from this simple scenario. This might be the result of complicated jet geometries, such as localized emission regions (often referred to as ``blobs''), or quasi-spherical emission regions, which have succeeded in explaining the broadband variable emission of blazar jets (e.g., \citealt{BM1996, MK1997, BC2002}, but see also e.g., \citealt{Marscher1985, MT1996, Marscher2006}). In this case, one would expect a proportionality $\tau_{\rm var} \propto r \propto \dot{M}_{\rm acc}^{1/3}$, where $r$ is the size of the blobs, if (a) there is no density gradient in the blobs and (b) the density does not vary substantially from source to source. In addition, recollimation of jets (e.g., \citealt{DM1988}), strong superposition of multiple flares arise in different shock regions and possible time delays at cm wavelengths (e.g., \citealt{Jorstad2010}), and shock-shock interactions in jets (e.g., \citealt{Fromm2011}) might play an important role. The possible effect of superposition of multiple flares might be investigated by using high-frequency data (mm/sub-mm wavelengths). In addition, dedicated numerical simulations would be helpful to investigate the complicated coupling behavior between mass accretion rate and jet structure (e.g., \citealt{Tchekhovskoy2011, Marscher2014}).

As illustrated in Figure~\ref{fig_BH2}, we checked if the intrinsic variability timescale is related to Eddington ratio and black hole mass. In the left panel, the scaling with black hole mass shows a large scatter, with correlation coefficients of $\approx 0.4$, indicating a moderate correlation. This correlation is probably a consequence of the correlation seen in Figure~\ref{fig_BH} because (i) all the FSRQs (except the outlier 1928+738) follow the relation with $\alpha = 1/4$ due to their Eddington ratio being concentrated around $\approx 0.1$ (see Figure~\ref{fig_Mdot}) and (ii) the BLOs lie systematically below the FSRQs with similar black hole masses, which indicates their low accretion rates lead to low variability timescales. In the right panel, we see a correlation of time scale and Eddington ratio with correlation coefficients as high as 0.7 when excluding 1928+738. This, too, is probably a corollary of the $\beta$--$\dot{M}_{\rm acc}$ relation because our sources span only $\approx 1$ dex in black hole mass (as can be seen in the left panel of Figure~\ref{fig_BH}) and more than four orders of magnitude in accretion rate.

\begin{figure*}[!t]
\begin{center}
\includegraphics[trim=8mm 0mm 4mm 0mm, clip, width = 44mm]{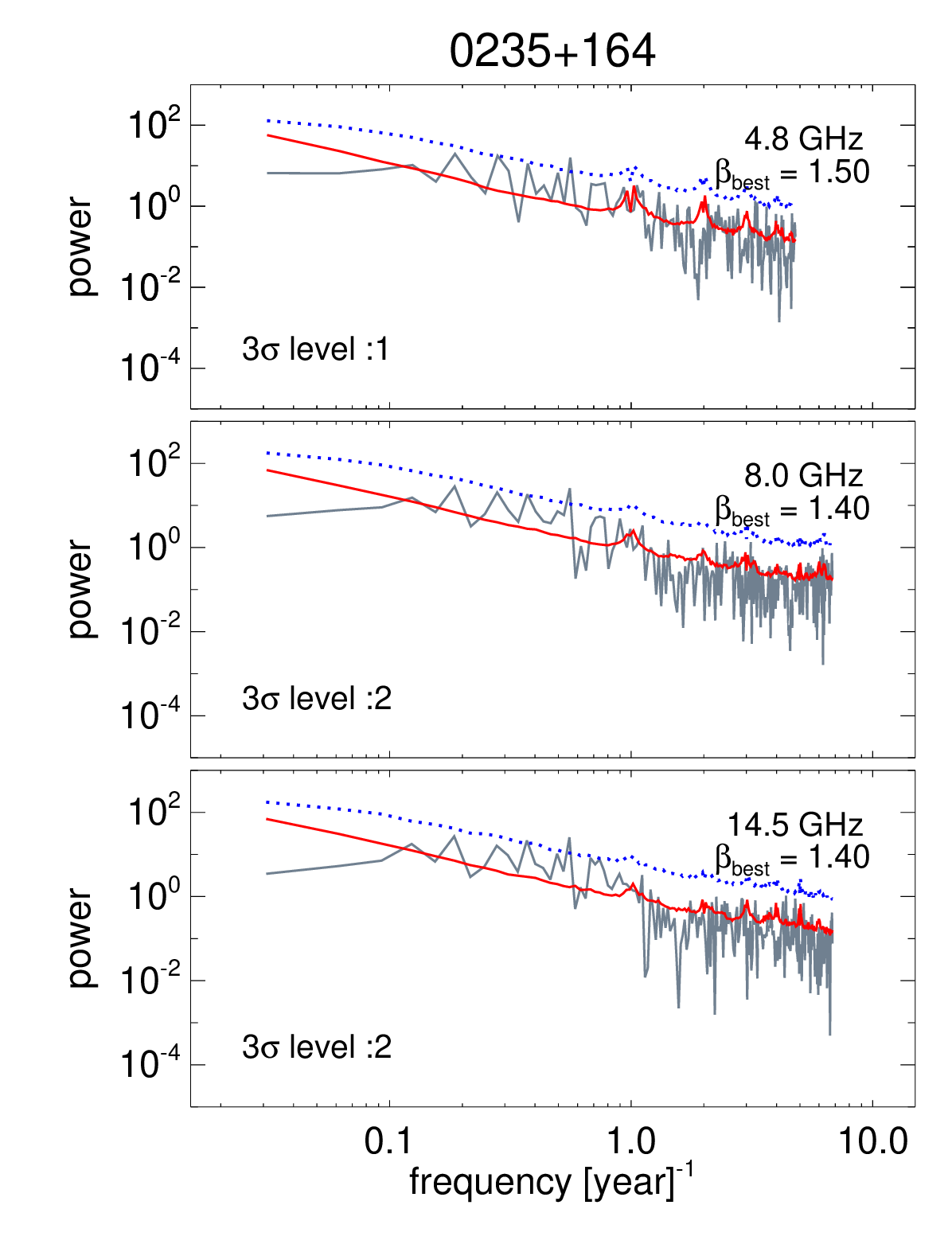}
\includegraphics[trim=8mm 0mm 4mm 0mm, clip, width = 44mm]{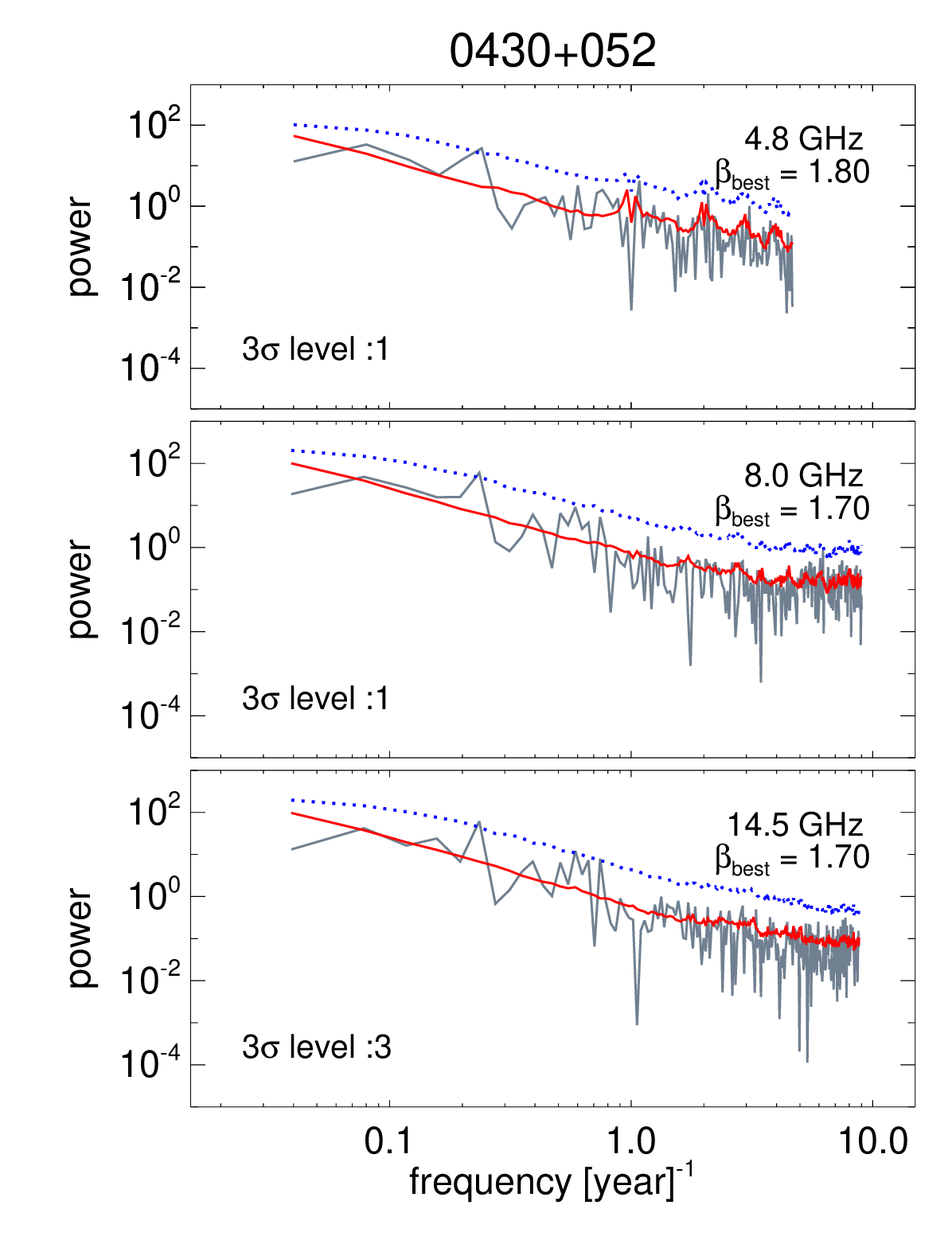}
\includegraphics[trim=8mm 0mm 4mm 0mm, clip, width = 44mm]{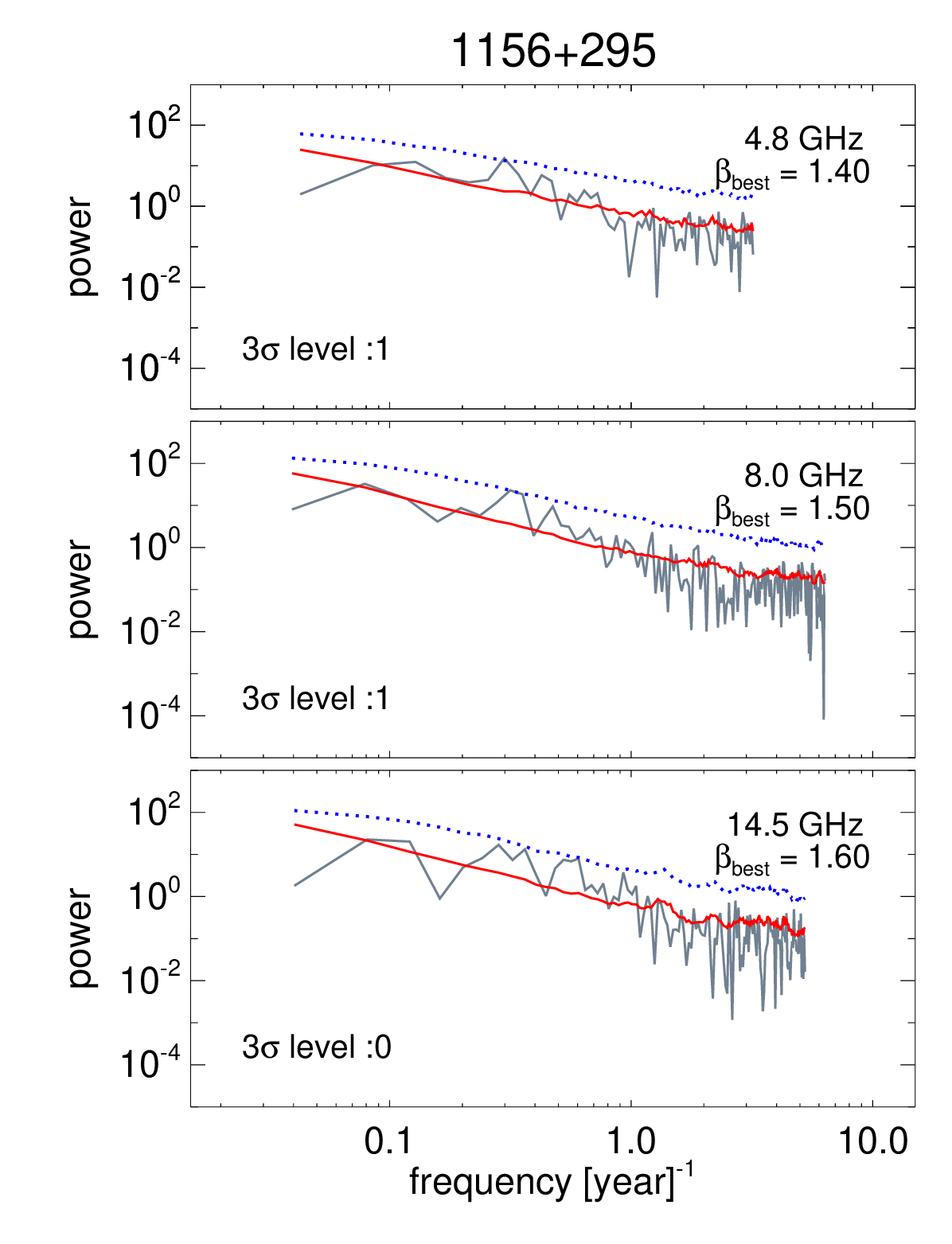}
\includegraphics[trim=8mm 0mm 4mm 0mm, clip, width = 44mm]{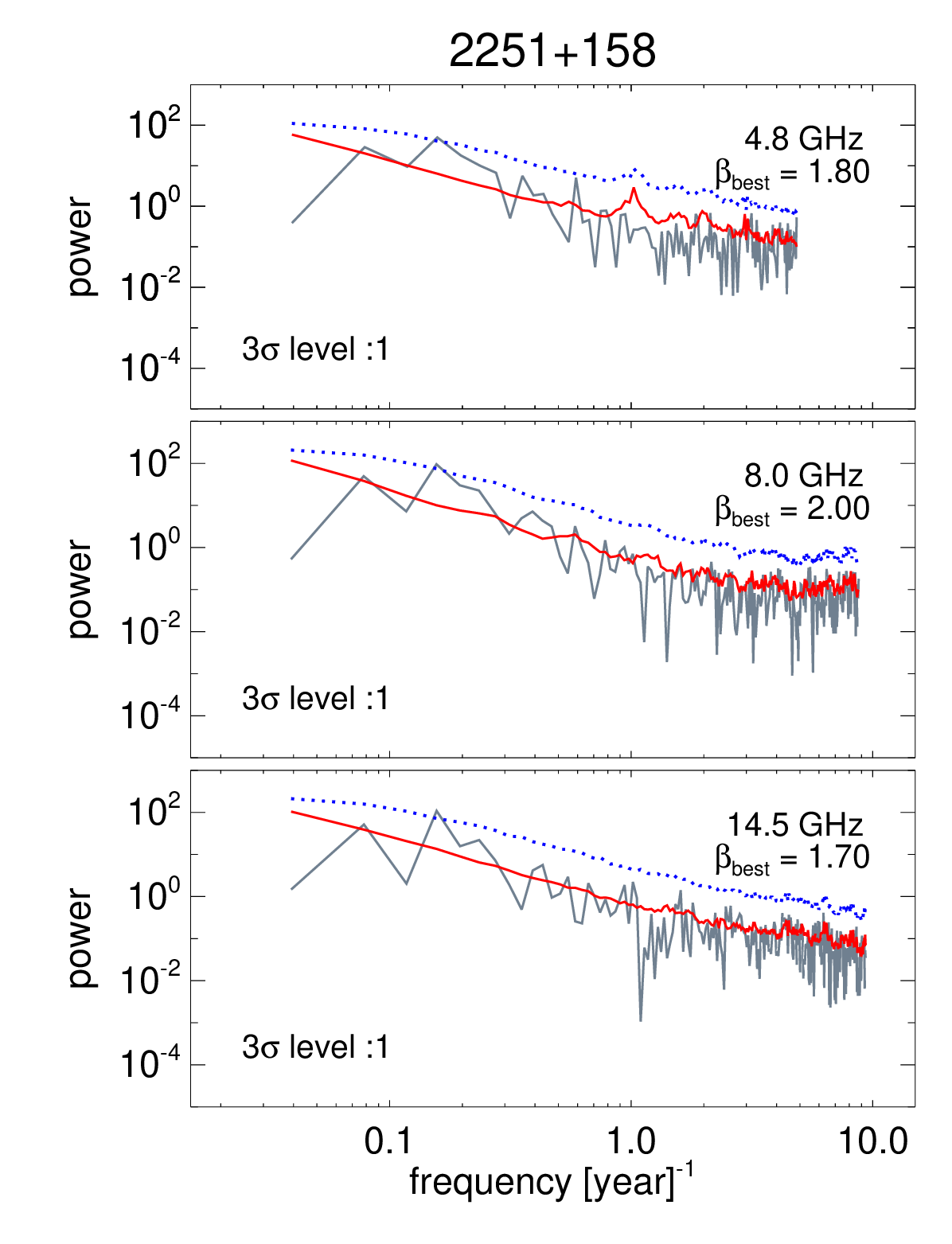}
\caption{Same as the lower panels of Figure~\ref{pd} but for the four sources with significant ($3\sigma$) excess spectral power at two or three observing frequencies (see Section~\ref{ssect_br} for details). \label{fig_QPO}
}
\end{center}
\end{figure*}

\begin{figure*}[!t]
\begin{center}
\includegraphics[trim=8mm 0mm 4mm 0mm, clip, width = 44mm]{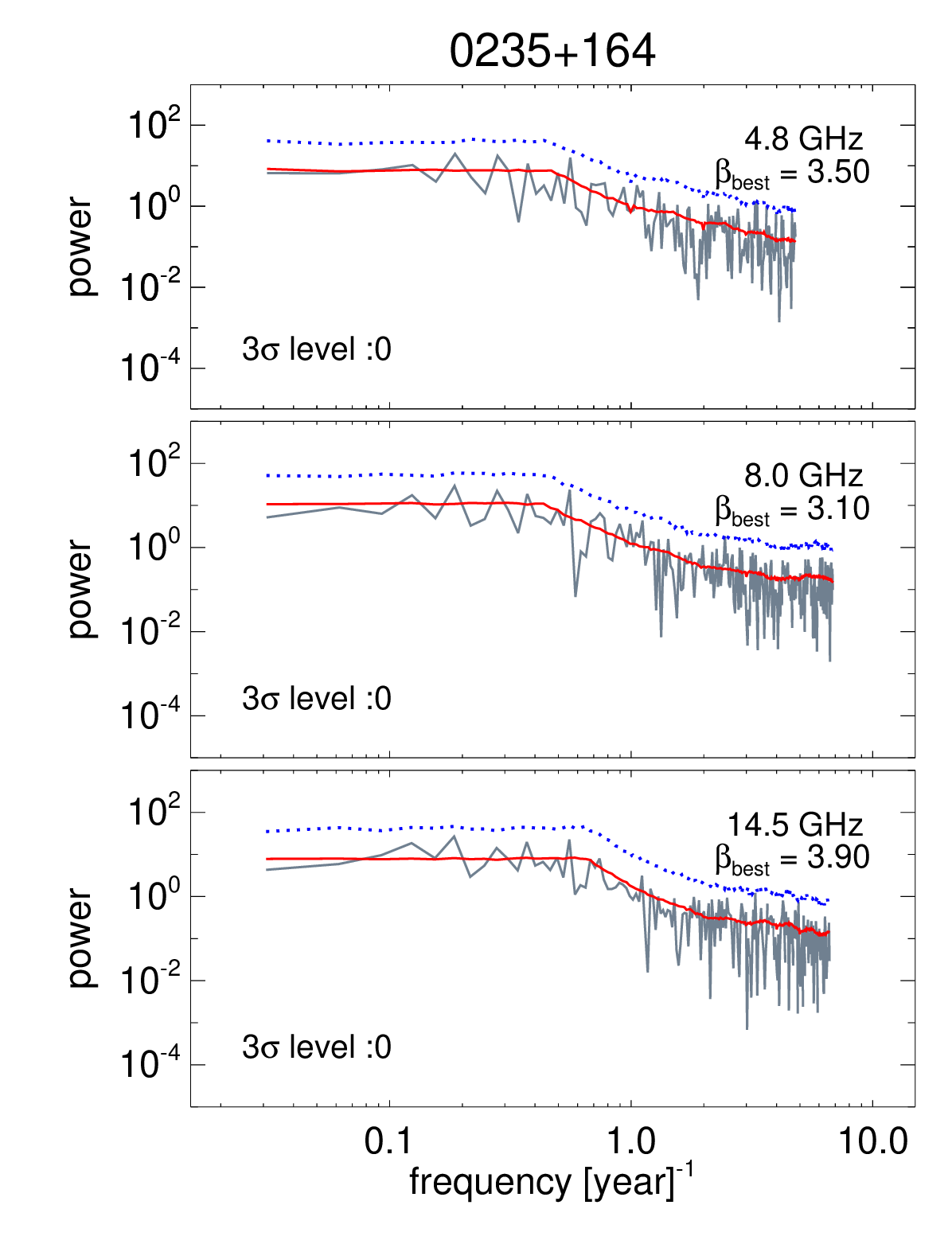}
\includegraphics[trim=8mm 0mm 4mm 0mm, clip, width = 44mm]{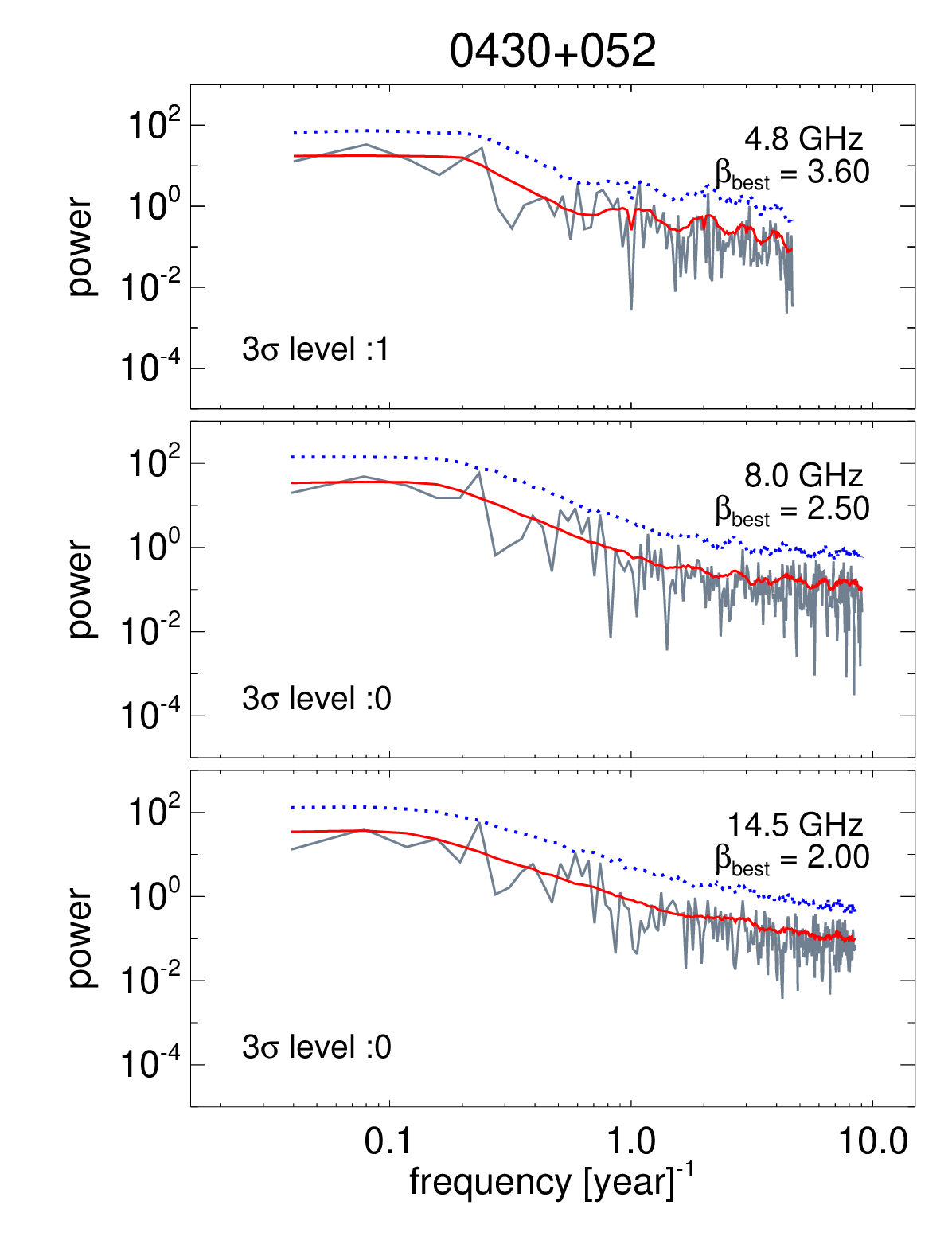}
\includegraphics[trim=8mm 0mm 4mm 0mm, clip, width = 44mm]{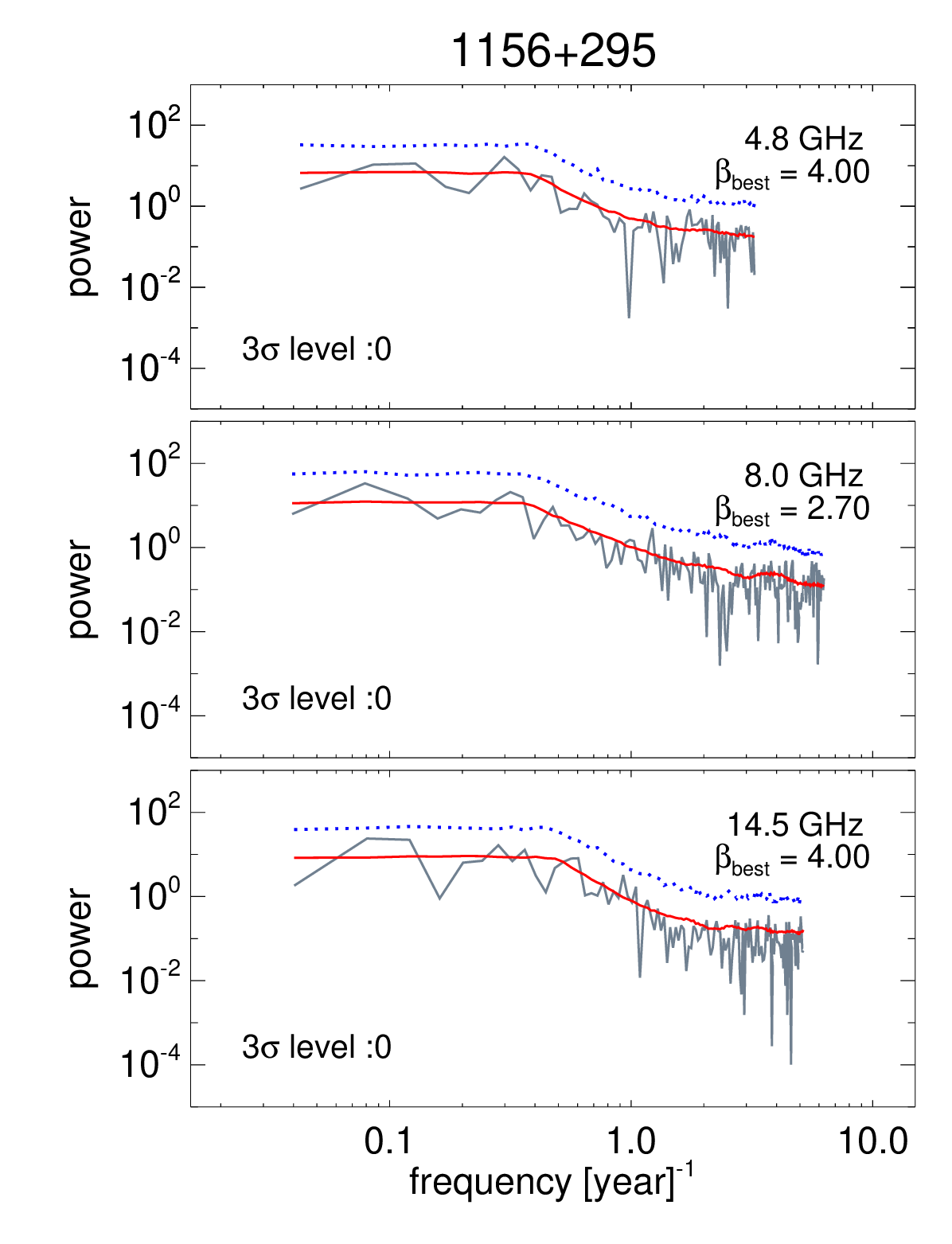}
\includegraphics[trim=8mm 0mm 4mm 0mm, clip, width = 44mm]{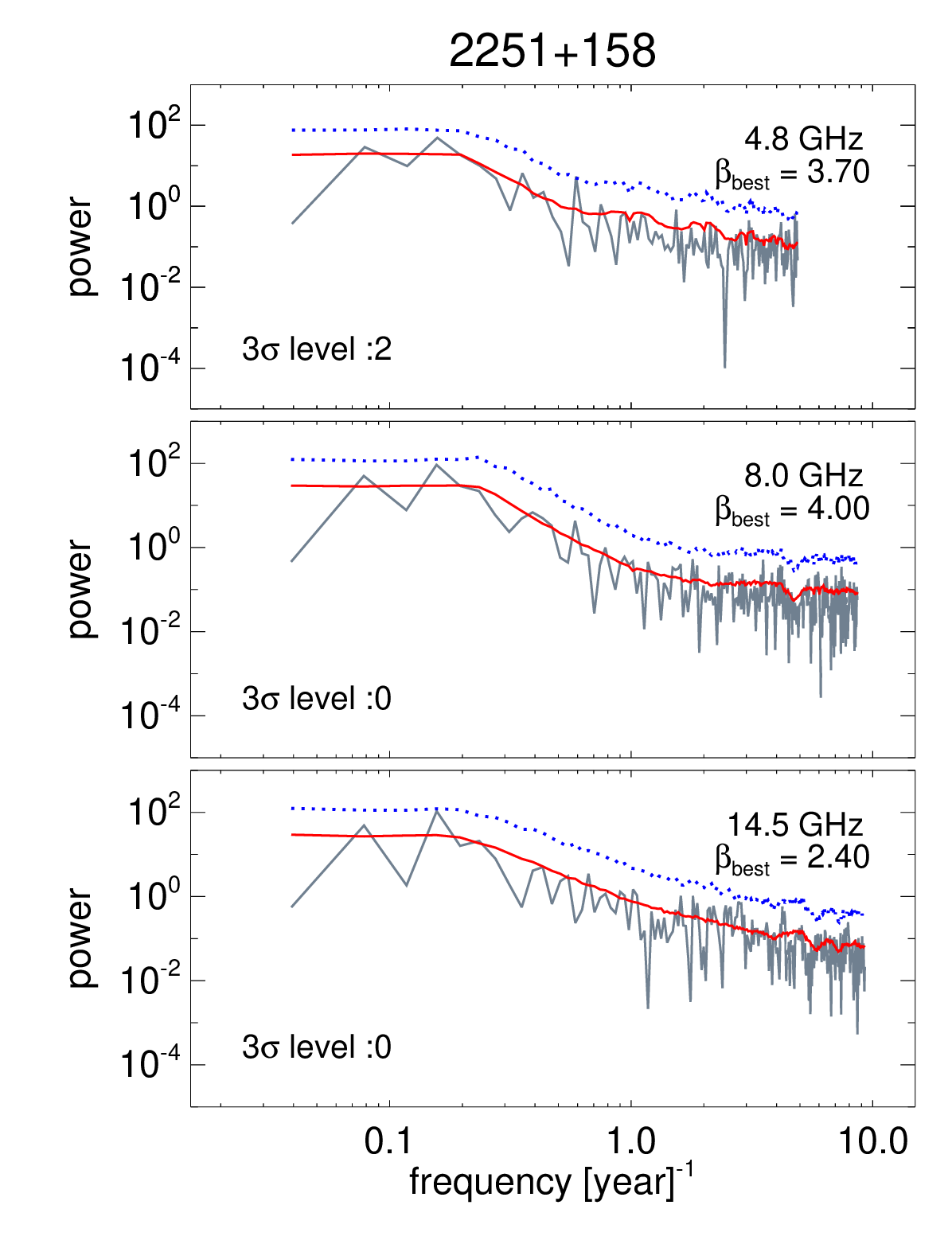}
\caption{Same as Figure~\ref{fig_QPO} but with the best-fit broken power-law models (red solid lines) instead of simple powerlaws. In each diagram the value of the best-fit slope $\beta$ above the break frequencies is noted. (The model curves are flat below the break frequencies.) The apparent excess spectral power noted in Figure~\ref{fig_QPO} is now modeled properly for each source. \label{fig_bp}}
\end{center}
\end{figure*}

\begin{figure}[!t]
\begin{center}
\includegraphics[trim=12mm 13mm 5mm 0mm, clip, width = 89mm]{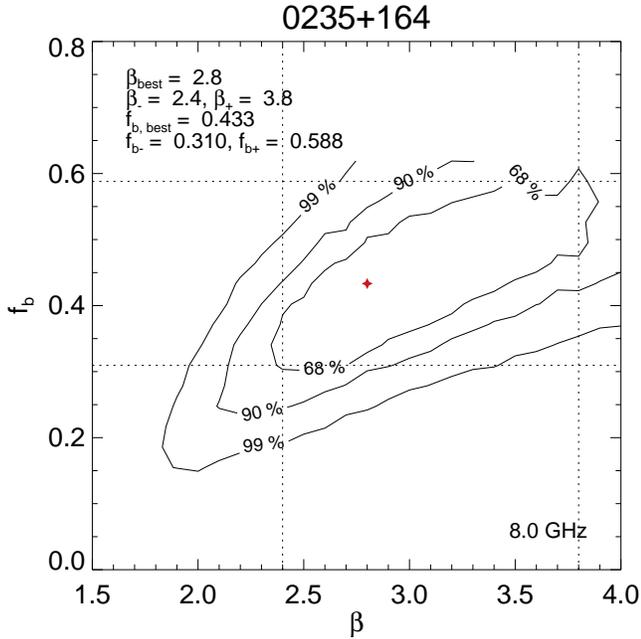}
\caption{$\chi^2$ contours of the broken power-law model for the periodogram of 0235+164 at 8.0 GHz. Parameters are break frequency, $f_b$, and power spectral index, $\beta$ (see Section~\ref{ssect_br} for details). \label{fig_2dchi}}
\end{center}
\end{figure}

\begin{figure*}[!t]
\centering
\includegraphics[trim=12mm 5mm 8mm 5mm, clip, width = 59mm]{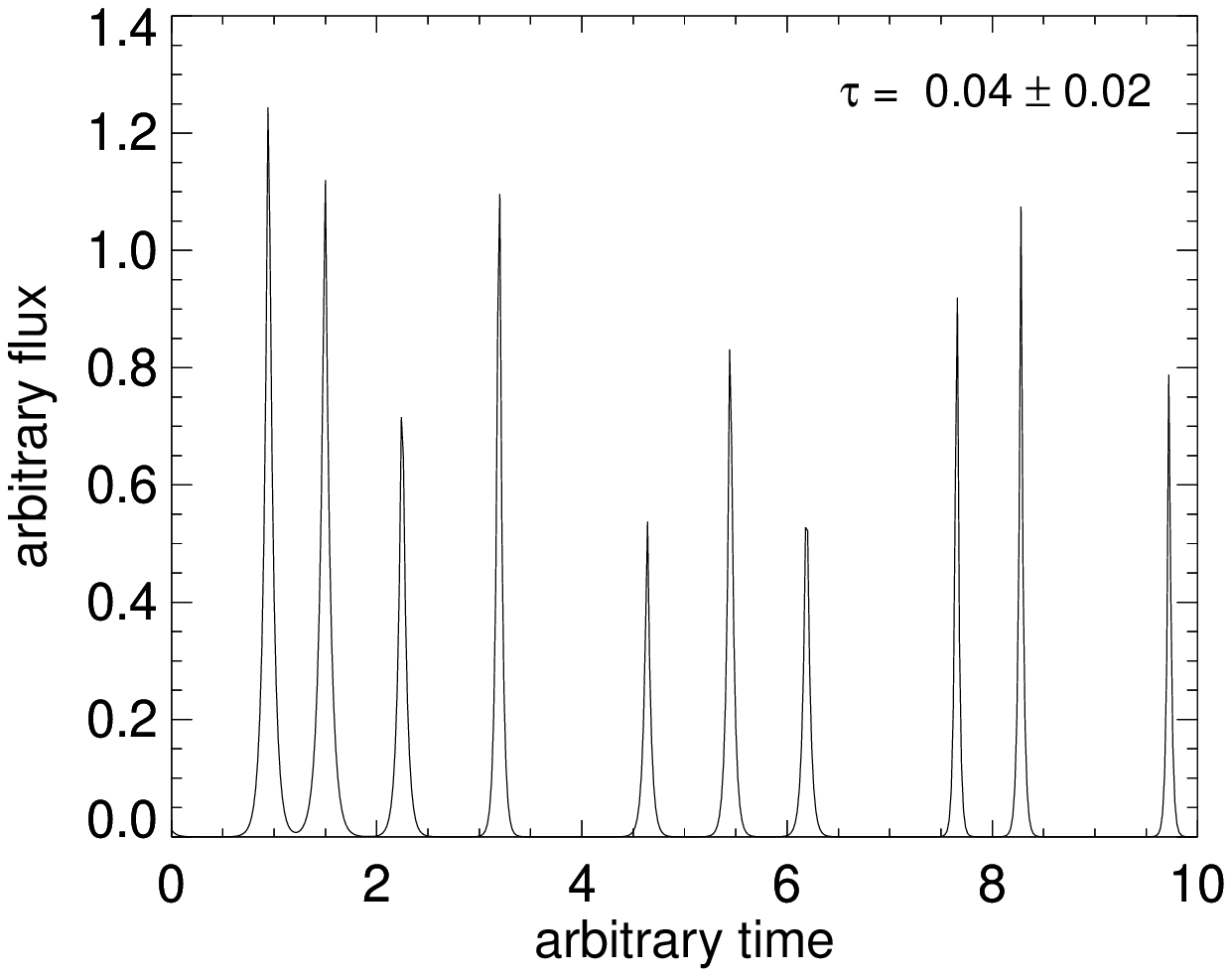}
\includegraphics[trim=12mm 5mm 8mm 5mm, clip, width = 59mm]{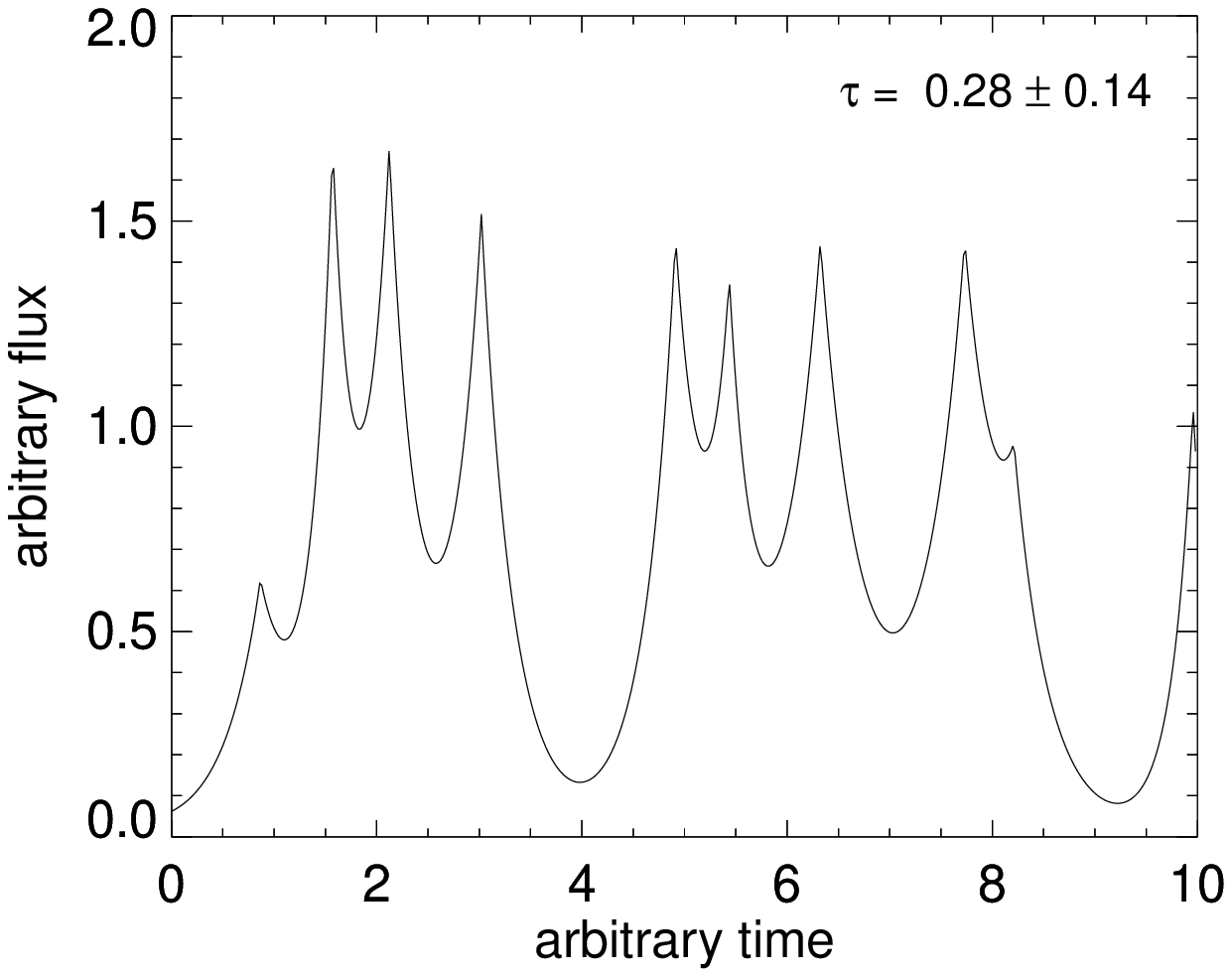}
\includegraphics[trim=12mm 5mm 8mm 5mm, clip, width = 59mm]{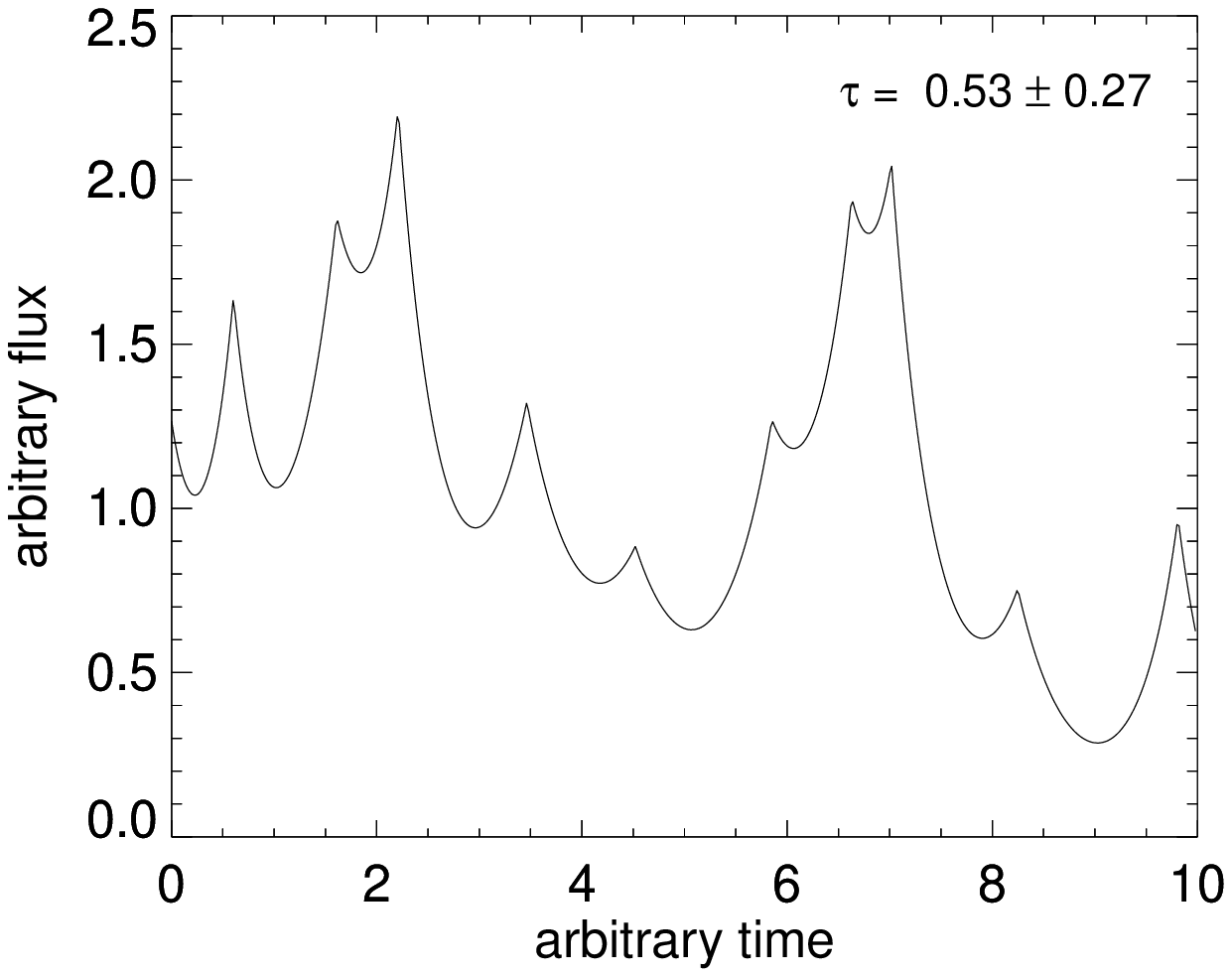}
\includegraphics[trim=12mm 5mm 8mm 5mm, clip, width = 59mm]{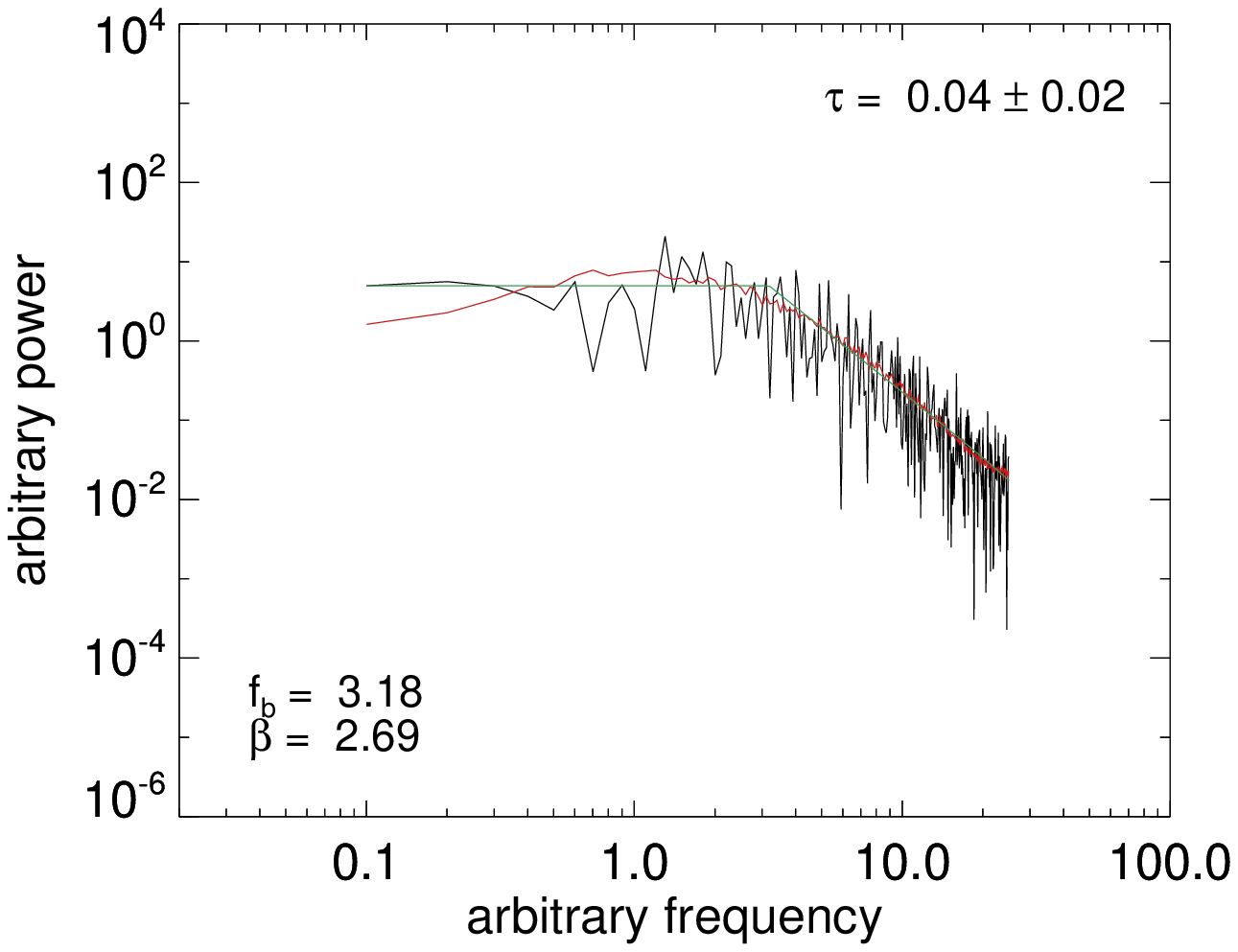}
\includegraphics[trim=12mm 5mm 8mm 5mm, clip, width = 59mm]{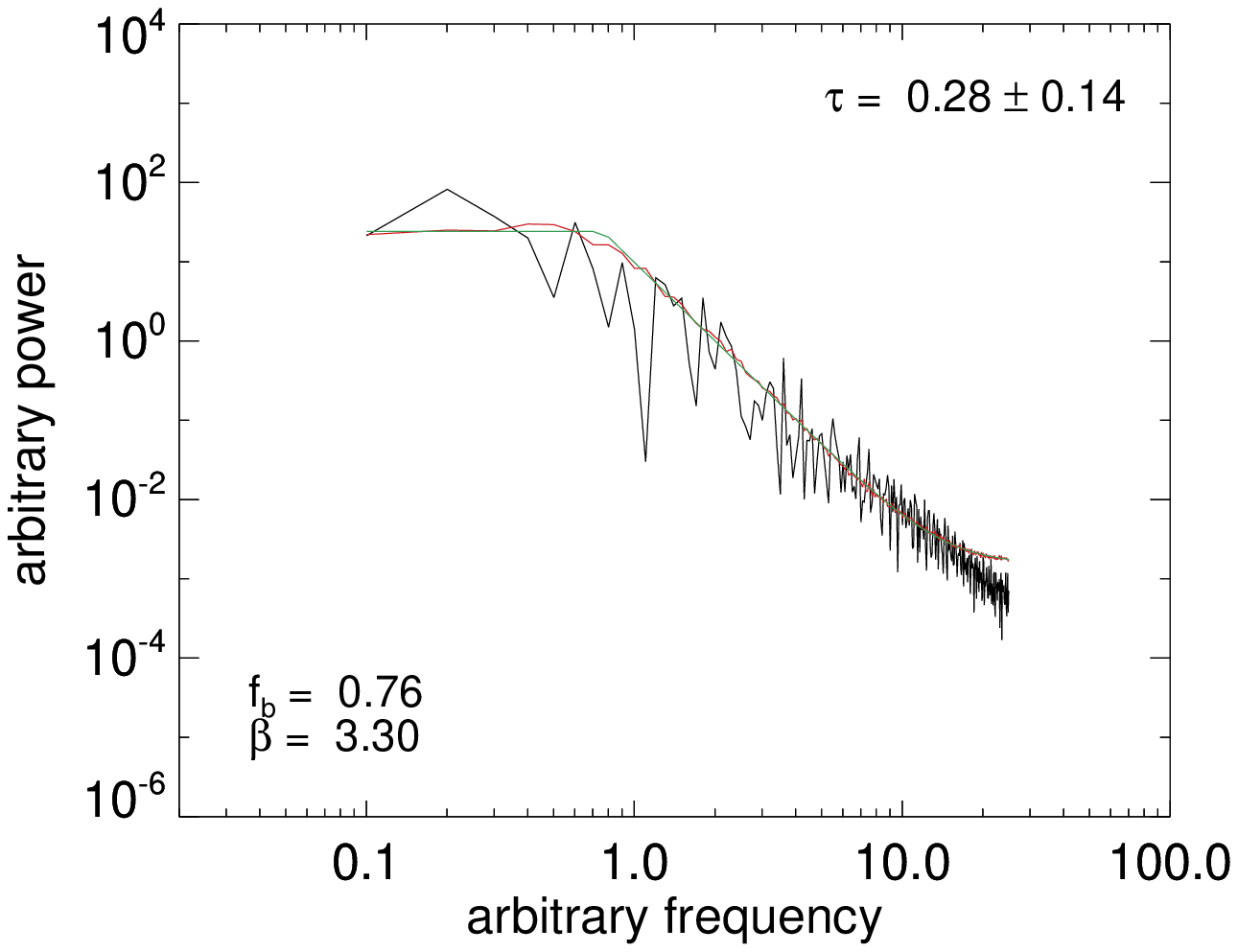}
\includegraphics[trim=12mm 5mm 8mm 5mm, clip, width = 59mm]{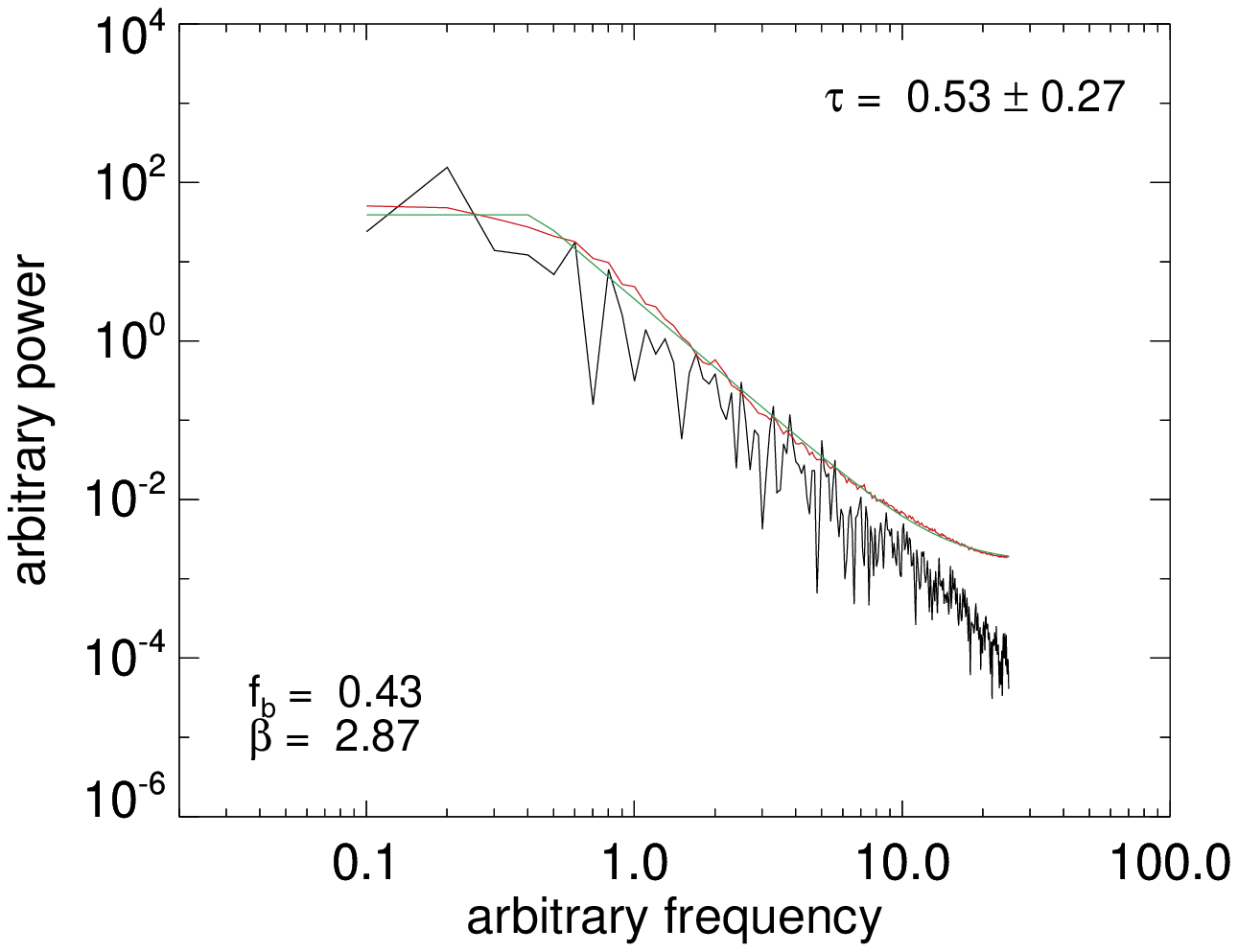}
\caption{\emph{Top panels}: Simulated lightcurves with a combination of exponential flares with randomized amplitude, duration, and separation between the flares for three different mean duration values (see Section~\ref{ssect_br} for details). \emph{Bottom panels}: One realization of the power spectra (black solid lines), mean power spectra of 100 realizations (red solid lines), and fitted lines with broken power-law models (green solid lines). \label{fig_shot}}
\end{figure*}

\subsection{Broken Power-law Periodograms\label{ssect_br}}

A simple power-law model explains the observed power spectra of most of our sources successfully without any indication for statistically significant QPO signals. However, there are four sources out of which each shows significant excess spectral power simultaneously at two or three observing frequencies and at similar sampling frequencies: 0235+164, 0430+052, 1156+295, and 2251+158. This excess power might indicate the presence of QPOs. We show the power spectra, the best-fit power-law models, and the corresponding $3\sigma$ significance levels in Figure~\ref{fig_QPO}.

Actually, the candidate QPO signals are located at rather low sampling frequencies and the power spectra appear to flatten below those frequencies. Thus, we tested if the periodograms can be (or have to be) modeled as broken power-laws with break frequencies $f_b$. We performed Monte Carlo simulations as we did in Section~\ref{ssect_lc} but this time with broken power-law models. We assumed that $\beta$ becomes zero below the break frequency.\footnote{In principle, both power spectral indices below and above the break frequency are free parameters (e.g., \citealt{Uttley2002}), but we aimed at models with the smallest number of free parameters that actually describe the data.} We computed sets of models with break frequencies ranging from $\approx$0.05 to $\approx$0.7~yr$^{-1}$ and power-law indices $\beta$ (above the break frequencies) ranging from 1.5 to 4.0. We obtained 1\,000 artificial lightcurves for each combination of $f_b$ and $\beta$. After mapping the observed sampling pattern into the artificial lightcurves, we obtained the binned logarithmic power spectra and took the average of them for each bin. Then, we calculated $\chi^2$ using Equation~\ref{eq1}. We show the $\chi^2$ contours for 0235+164 at 8.0 GHz -- $\chi^2_{\rm min} + 2.30, 4.61, 9.21$, corresponding to 68\%, 90\%, and 99\% significance levels, respectively -- (e.g., \citealt{Wall2012}), as function of $\beta$ and $f_b$ in Figure~\ref{fig_2dchi}. We obtained the values and the unmarginalized $1\sigma$ errors of the best-fit $\beta$ and $f_b$. The best-fit broken power-law periodograms for the four sources in question are shown in Figure~\ref{fig_bp}. Within errors, the periodograms are completely described by the models. We note that the $\chi^2 / \rm d.o.f$ values are reduced significantly when changing from simple to broken power-law models: from 3.03, 1.44, 2.21, and 2.51 to 0.77, 0.48, 0.69, and 1.22 for 0235+164, 0430+052, 1156+295, and 2251+158, respectively.

We note that the best-fit values for $\beta$ in the broken power-law models tend to be very high, up to the simulation limit of 4.0. The formal errors of these values are large, ranging from 0.3 to about 1.2 -- meaning that $\beta$ is not well constrained. Thus, using the broken power-law results for other analyses, e.g., the scaling relations of variability timescale with black hole mass, would lead to highly uncertain results. We suspect that our assumption of flat power spectra below the break frequencies results in the high power spectral indices we observe.
Our primary interest was to investigate whether there is indication for broken power-law periodograms at least for a few sources. For most of our targets, the two models -- simple vs. broken power-law -- are not distinguishable within errors. Therefore, we stick to the best-fit values of $\beta$ obtained with the simple power-law models throughout this paper.

As noted already in Section~\ref{ssect_bt}, the origin of the featureless red-noise power spectra of AGN (mostly blazars) radio lightcurves is not understood. This is in contrast to the case of broken power-law periodograms typically seen in the X-ray and optical lightcurves of non-blazars (i.e., Seyferts and quasars). One possibility is that the emission is correlated over very long timescales -- comparable to the observation time covered by UMRAO database, $\approx 30$ years. If the radio emission of AGNs is directly linked to the accretion flows, the spatial correlation of accretion flows (see e.g., \citealt{Kelly2009, Kelly2011}) for radio-bright AGNs is much stronger than for nonblazars. However, many of our sources, especially the FSRQs, are active at optical bands as well; they should have accretion disks which are similar to X-ray/optical bright nonblazars. An alternative scenario involves the special feature of blazars lightcurves: the flares. Even if the power spectra of accretion flows or matter injection flows into jets have break frequencies at relatively high sampling frequencies, the break frequencies can move toward low sampling frequencies if the duration of the flares is long enough to cause substantial overlap of individual emission events. Such overlap effectively increases the timescales for flux variations, resulting in higher spectral power at lower sampling frequencies. The location of the break frequencies would depend on the degree of superposition.

To test this scenario, we employed a simple simulation of lightcurves. We generated 100 artificial lightcurves $f(t)$, each composed of multiple exponential flares, according to
\begin{equation}
\begin{split}
f(t) &= f_{\rm max}\exp[(t - t_0)/T_r], \quad \rm{for\ t<t_0\ and} \\
&= f_{\rm max}\exp[-(t-t_0)/T_d], \quad \rm{for\ t>t_0},
\end{split}
\end{equation}
where $f_{\rm max}$ is the peak amplitude of the flare, $t$ is the time, $t_0$ is the time of the peak, and $T_r$ and $T_d$ are the rise and decay timescales, respectively \citep{Chatterjee2012}. This model is based on the assumption that AGN radio lightcurves can indeed be decomposed into exponential flux peaks whenever the overlap between flares is not too strong (e.g., \citealt{Valtaoja1999, Hovatta2009}). For each lightcurve, we initially generated 20 flares that span 20 units (that can be identified with years) in time. From this, we took the half of the data points located in the middle of each lightcurves to avoid having lightcurves that converge to zero at the beginning and at the end. As a result, we have 10 artificial flares that span 10 units on average for each lightcurve. For each flare, we used $f_{\rm max}$ uniformly randomly distributed from 0.5 to 1.5 units to randomize the amplitudes of flares, $t_0$ from $n - 0.5$ to $n + 0.5$ units for the $n$th flare to make aperiodic variability, and $T_r = T_d$ from 0.5 to 1.5, multiplied by a characteristic timescale $\tau$. This timescale controls the degree of overlap of flares. We varied $\tau$ from $\approx 0.03$ to $\approx 0.7$ units and obtained the average of the periodograms for each $\tau$ value. This averaged periodogram we fitted with a broken power-law model with three parameters: the power-law index $\beta$ above the break frequency $f_b$ (the slope below $f_b$ being zero), and a constant offset for taking into account aliasing. One realization of a lightcurve, the corresponding power spectrum, the averaged power spectrum, and the best-fit broken power-law model for each of three different values of $\tau$ are shown in Figure~\ref{fig_shot}. When $\tau$ is small compared to the average separation between two flares (i.e., one unit) there is almost no overlap between flares; the break frequency appears at sampling frequencies well above one frequency unit. As $\tau$ increases, it becomes more difficult to disentangle individual flares and the break frequency moves toward lower sampling frequencies -- as expected.

\begin{figure}[!t]
\begin{center}
\includegraphics[trim=12mm 7mm 5mm 0mm, clip, width = 89mm]{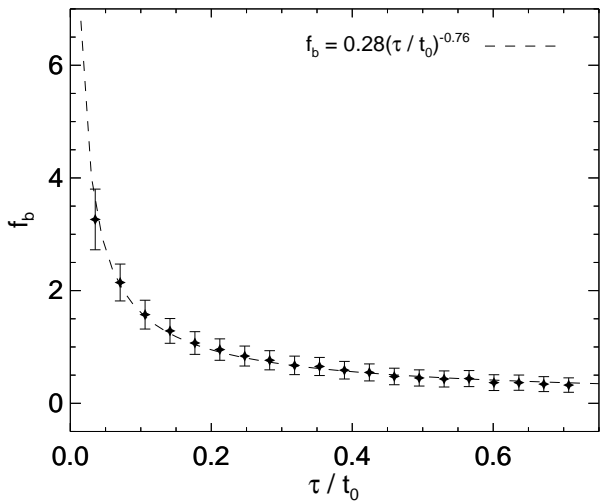}
\caption{Break frequency as function of the ratio of mean duration to mean separation of flares. The black dashed line indicates the best-fit power-law function; the corresponding formula is shown on the top right. Error bars are scaled to $\chi^2 / \rm d.o.f. \equiv 1$ for the best-fit model. \label{fig_break}}
\end{center}
\end{figure}

We obtained the break frequencies of the simulated power spectra as function of $\tau / t_0$. We found a power-law relation, $f_b = 0.28\times(\tau/t_0)^{-0.76}$. We scaled the errors on $f_b$ such that $\chi^2 \rm / d.o.f. = 1$ for the best-fit model.\footnote{Obtaining absolute errors on the break frequencies requires reliable error estimates for the simulated power spectra at each sampling frequency. This cannot be achieved in a straightforward manner because spectral powers do not follow Gaussian distributions .} As shown in Figure~\ref{fig_break}, the break frequency decreases with increasing $\tau / t_0$ and, at $\tau/t_0 \gtrsim 0.7$, converges to a value located close to the lowest sampling frequency. Accordingly, periodograms from lightcurves that show extensive overlap of flares appear as simple power-laws -- as is indeed the case for most of our target AGNs.

\begin{figure}[!t]
\begin{center}
\includegraphics[trim=12mm 5mm 5mm 0mm, clip, width = 89mm]{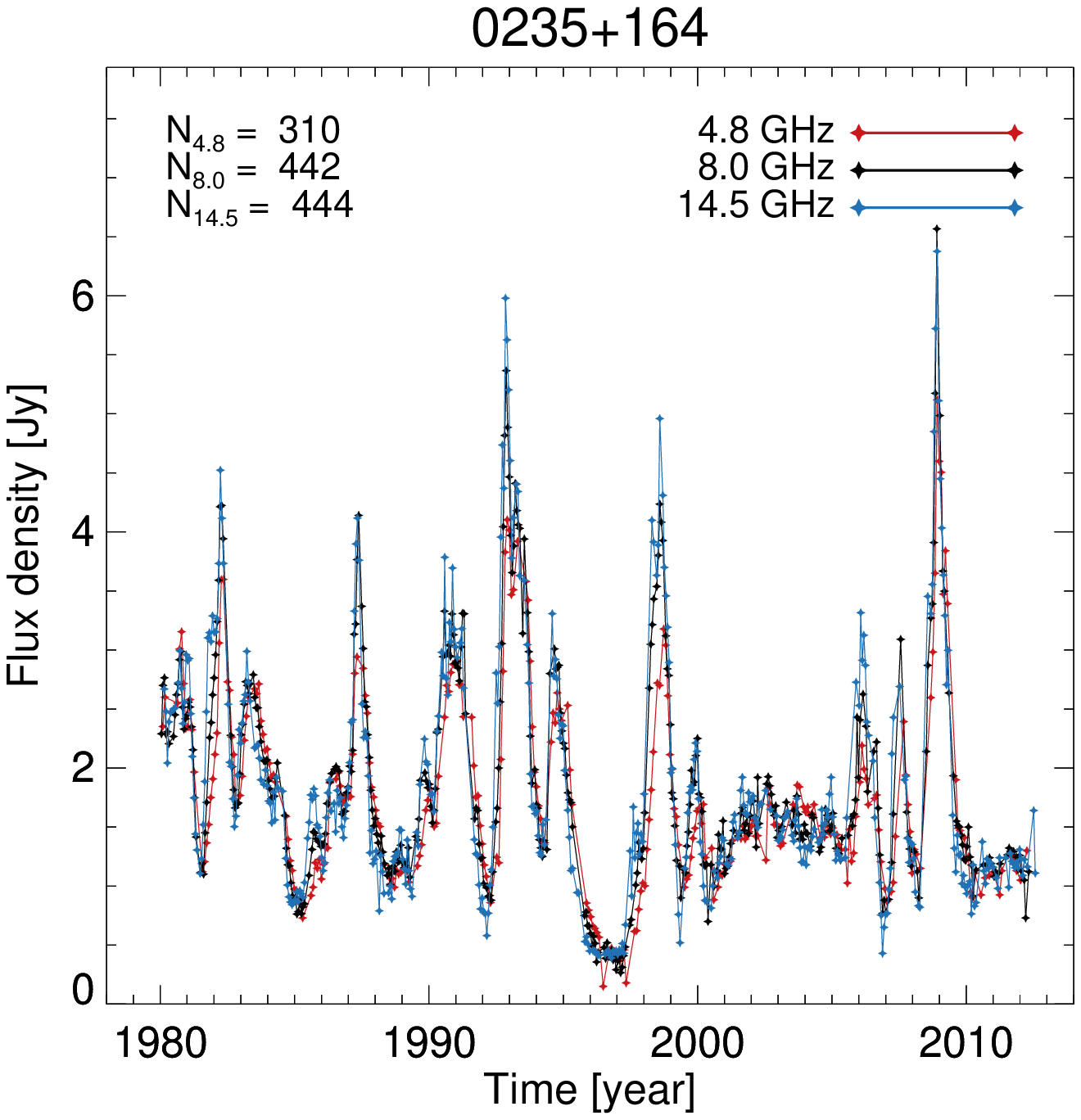}
\caption{Lightcurves of 0235+164 (after binning and flagging). Red, black, and blue solid lines indicate 4.8, 8, and 14.5 GHz data, respectively. The number of data points, $N_{\nu}$, is noted explicitly for each frequency $\nu$. \label{fig_0235}}
\end{center}
\end{figure}

It is now possible to check the degree of agreement between the observed break frequencies as function $\tau / t_0$ and the simulation results. On the one hand, the break frequencies found in 0430+052 and 2251+158 are very small, $\approx$0.2\,yr$^{-1}$, meaning that a wide range of $\tau / t_0$ is consistent with the observed value for $f_b$ (see also Figures \ref{fig_QPO} and \ref{fig_bp}). 0235+164 and 1156+295, on the other hand, show relatively large break frequencies, $\approx$0.5\,yr$^{-1}$ and $\approx$0.4\,yr$^{-1}$, respectively. We took the median duration and the median separation of the flares we obtained in Section~\ref{ssect_fit} (for each source and each observing frequency) and calculated the observational values for $\tau / t_0$. Since we used Gaussian flares in our lightcurve fitting but exponential flares in the simulations, we multiplied the observed $\tau / t_0$ by $\sqrt{2}$ to compare the \emph{e}-folding timescales of observed and simulated flares; these values are denoted $\tau / t_0 \rm(obs)$ in Table~\ref{table_shot}. We inserted the observed break frequencies into the theoretical $f_b$--$\tau/t_0$ power-law relation and obtained the theoretical ratio $\tau/t_0 \rm(sim)$. The difference between $\tau/t_0 \rm(obs)$ and $\tau/t_0 \rm(sim)$ is on the order of 10\% typically and reaches 31\% at most (see Table~\ref{table_shot}). Therefore, we conclude that the observation of red-noise periodograms for most of our target sources is consistent with being due to strong temporal overlap of flares. The main reason for this might be the relatively long variability timescales of AGNs at centimeter wavelengths. We show the lightcurves of 0235+164, for which the break in the periodogram is quite prominent, in Figure~\ref{fig_0235}. The duration of the individual flares is short compared to their typical separation; indeed, it seems that there is no substantial overlap between the flares -- which is consistent with our scenario.

\begin{deluxetable*}{cccccc}

\tablecaption{Variability timescales of two sources \label{table_shot}}
\tablehead{
\colhead{Source} & \colhead{Obs freq.} & \colhead{$f_b\rm (obs)$ [yr$^{-1}$]} & \colhead{$\tau / t_0 \rm(sim)$} &
\colhead{$\tau / t_0 \rm(obs)$} & \colhead{Difference [\%]}\\
 & & (1) & (2) & (3) & (4)
}
\startdata
\multirow{3}{*}{\centering 0235+164} & 4.8 & 0.47 & 0.51 & 0.59 & 13 \\
 & 8.0 & 0.43 & 0.57 & 0.57 & 0.41\\
 & 14.5 & 0.65 & 0.33 & 0.40 & 17\\
 \hline
 \multirow{3}{*}{\centering 1156+295} & 4.8 & 0.38 & 0.67 & 0.68 & 2 \\
 & 8.0 & 0.36 & 0.74 & 0.66 & 12\\
 & 14.5 & 0.48 & 0.49 & 0.71 & 31
\enddata

\tablecomments{(1) Break frequencies found from Monte Carlo simulations using broken power-law models. (2) Values of $\tau/t_0$ expected theoretically from the relation shown in Figure~\ref{fig_break} and the observed break frequencies. (3) Observed values of $\tau/t_0$ from fitting Gaussian flares piecewise to the lightcurves. (4) Relative difference between (2) and (3), in units of percent.}

\end{deluxetable*}

\subsection{Comparison with Other Studies\label{ssect_comp}}

So far, we discussed long-term UMRAO lightcurves of AGNs with jets approximately aligned with the line of sight (even for the radio galaxies). Naturally, we have to ask if the variability patterns we observe agree with those for nonblazars and for blazars at other observing frequencies. \cite{Kelly2011} concluded that the slope of X-ray periodograms (below the high-frequency break) of 10 Seyfert galaxies does not correlate with black hole mass. This indicates that, for Seyferts, factors such as the amplitude of the driving noise field are more important than black hole mass in determining the structure of flux variability. The noise field is arguably related to the viscous, thermal, and radiative response of accretion disks to perturbations \citep{Kelly2011}. In contrast, radio variability of AGNs is governed by the crossing time of radiation and/or disturbances through the emission region. This also explains the quite low break frequencies observed in AGN radio periodograms. In optical and X-ray power spectra of nonblazars, break frequencies are found at timescales of less than a few years \citep{Kelly2009, Kelly2011}.

Blazars usually show symmetric flares across multiple wavelengths -- see, e.g., \cite{Valtaoja1999} and \cite{Hovatta2009} for 22 and 37 GHz data respectively, \cite{Chatterjee2012} for optical and $\gamma$-rays, and \cite{Abdo2010} for $\gamma$-rays. As the observing frequency increases, the rise and decay times of flares become shorter (e.g., \citealt{Chatterjee2012, Rani2013}) due to the shorter cooling times of higher energy particles (cf., e.g., \citealt{Marscher1996}). Therefore, we may expect (1) smaller power spectral indices $\beta$ and (2) observations of broken power-law periodograms at higher frequencies. \cite{Trippe2011} showed that the power spectra of six radio bright AGNs at millimeter wavelengths have $\beta\approx0.5$, which is much smaller than the values we find in this work. However, windowing effects (especially red-noise leak and aliasing) were not taken into consideration then, making it hard to conclude on the general behavior of mm-radio periodograms. 

\begin{figure}[!t]
\begin{center}
\includegraphics[trim=5mm 3mm 10mm 5mm, clip, width = 89mm]{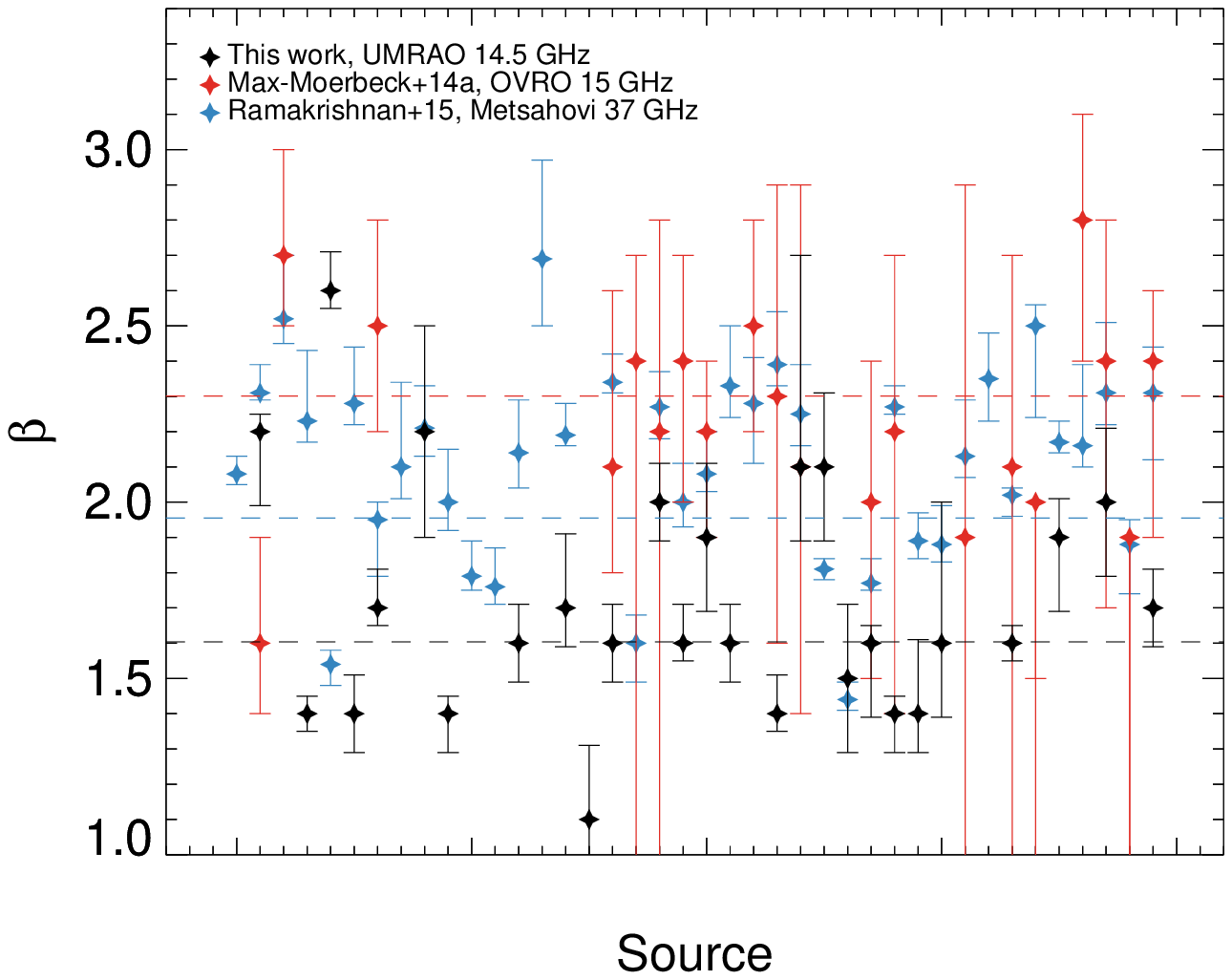}
\caption{ Observed values of $\beta$ for all sources covered by three different studies: this work, using UMRAO 14.5 GHz data (black); \cite{MaxM2014a} using OVRO 15 GHz data (red); and \cite{Ramakrishnan2015} using Mets\"ahovi 37 GHz data (blue). A given abscissa value indicates a given source. The horizontal dashed lines indicate weighted averages of the $\beta$ values found in each study. \label{comparison}}
\end{center}
\end{figure}

Several recent studies concluded that the power spectra of many radio bright AGNs are quite steep ($\beta>2$) at 15 GHz (from the Owens Valley Radio Observatory (OVRO) 40 m monitoring program; \citealt{MaxM2014a}) as well as at 37 GHz (from the Mets\"{a}hovi AGN monitoring program; \citealt{Ramakrishnan2015}). This behaviour differs from our expectation in that (i) the OVRO result seems to disagree with our result even though the observing frequency is very similar and (ii) the 37 GHz power spectra appear to be steeper than ours even though the observing frequency is higher. We present a comparison in Figure~\ref{comparison}. Although the errors of \cite{MaxM2014a} are too large for a quantitative one-to-one comparison, the average $\beta$ from the OVRO observations (weighted mean $\beta_{\rm wmean} = 2.30$) seem to be larger than that from the Mets\"{a}hovi observations ($\beta_{\rm wmean} = 1.95$), which is in agreement with our expectation. However, the systematic difference between these studies and our work ($\beta_{\rm wmean} = 1.60$) needs to be investigated.

There are important differences in the methodology for estimating $\beta$ between the various works. The OVRO and Mets\"ahovi studies employed linear interpolation of lightcurves (whenever there is a gap in the data) after binning and convolved the resulting lightcurves with a Hanning sampling window function, which reduces red-noise leak \citep{MaxM2014b}. Interpolation leads to a suppression of power at high sampling frequencies because it correlates adjacent data points, thus resulting in an artificial steepening of power spectra. Our work did not use interpolation of data. We employed the Scargle periodogram which can be applied to unevenly sampled lightcurves \citep{Scargle1982}. In case of uneven sampling, the periodogram suffers from red noise leak and aliasing, leading to characteristic distortions as shown in \cite{MaxM2014b}. However, our simulated power spectra are distorted in the same way because we mapped the sampling patterns of the observed lightcurves into the simulated ones (cf., Figure~\ref{ex} and the bottom right panel of Figure~\ref{pd}). We also note that uncertainties in the \emph{absolute} values of $\beta$ do not affect our results, as long as the \emph{relative} (i.e., source-to-source) differences in $\beta$ are preserved -- as demonstrated by the correlation between $\beta$ and fractal dimension of a lightcurve (Figure~\ref{fig_Tfrac}). A more detailed dedicated investigation will be necessary in the future for making an informed choice from among the various approaches.

In the $\gamma$ ray regime, \cite{Abdo2010} found periodogram slopes of $1.4\pm0.1$ and $1.7\pm0.3$ averaged over nine FSRQs and six BLOs, respectively. These values are compatible to the values which we find in this work, whereas we would actually expect smaller values of $\beta$ in the high-energy regime. We note however that averaging power spectra of different sources is likely to lead to spurious results -- as noted in Section~\ref{ssect_beta}, even for sources of the same AGN type $\beta$ covers a wide range of values (from $\approx$1 to $\approx$3). \cite{Sobolewska2014} used longer \emph{Fermi} lightcurves than those used in \cite{Abdo2010} (4 years vs. 11 months) and modeled them in the time domain based on the assumption of a mixed OU process. For their sample of 13 blazars, they obtained $\beta \lesssim 1$. They also found low-frequency breaks for two sources, 3C~66A and PKS~2155. \cite{Shimizu2013} found slopes around 0.85 for the hard X-ray lightcurves of three blazars and a broken power-law periodogram for 3C 273 (see also \citealt{McHardy2008}). The above results are in agreement with our scenario of $\beta$ becoming smaller, and broken power-law periodograms becoming more prominent, at higher observing frequencies. Contrary to this trend, \cite{Chatterjee2012} found slopes up to 2.3 for a few blazars at optical wavelengths. We also note that, when comparing radio, optical, and high-energy emission, we are looking at radiation from different physical emission mechanisms and/or different emission regions. For example, \cite{Ramakrishnan2016} showed that the variability of blazars at optical and $\gamma$ ray wavelengths is correlated, while the presence of a correlation between radio and optical bands is unclear. Accordingly, a comparison of radio and optical lightcurves may be of questionable value, while a comparison between optical and X-rays/$\gamma$-ray lightcurves is more straightforward.

\section{Conclusions}
\label{sect5}

We studied long-term (25--32 years), high-quality radio lightcurves of 43 radio bright AGNs -- 27 FSRQs, 13 BLOs, and 3 radio galaxies -- at 4.8, 8, and 14.5\,GHz. We investigated the physical origin of different variability patterns found in the radio lightcurves of different sources by means of peridogram analyses. Our work leads us to the following principal conclusions: 

\begin{enumerate}

\item The power spectra of 39 out of 43 sources are in agreement with simple power-law periodograms without any indication for (quasi-)periodic signals. Power spectral indices range from $\approx$1 to $\approx$3. We find a strong anti-correlation between the power spectral index and the fractal dimension of the lightcurves, thus quantifying the one-to-one relation between the geometry of lightcurves and the slopes of periodograms as $\beta \propto -4.43d_f$, where $\beta$ is the power spectral index and $d_f$ is the fractal dimension.

\item We find that $\beta$ is a proxy for the variability timescale $\tau_{\rm var}$. We discover a strong correlation between $\beta$ and the median duration of flares. We apply an improved measure for variability timescales, the width of the distribution of the derivatives of lightcurves, $\sigma_{\mathrm{der}}$. We find the relation $\beta\propto-1.39\log\sigma_{\mathrm{der}}$.

\item When taking into account relativistic Doppler boosting and cosmological redshift, $\beta$ shows a correlation with the accretion rate. We find the relation $\beta\propto1.39\alpha\log \dot{M}_{\rm acc}$, corresponding to $\tau_{\rm var}\propto \dot{M}_{\rm acc}^{\alpha}$, with $\alpha = 0.25 \pm 0.03$. At this point, we cannot explain the specific value $\alpha \approx 1/4$.

\item For four sources in our sample -- 0235+164, 0430+052, 1156+295, and 2251+158 -- we find that broken power-law models provide significantly better fits to the observed periodograms than simple power-law models. From random realizations of lightcurves composed of sequences of exponential flares, we obtain a theoretical power-law relation between break frequency $f_b$ and scaled duration of flares $\tau/t_0$. We find that, within errors, our observed values for $f_b$ and $\tau/t_0$ agree with the theoretical relation.

\item We conclude that the periodograms of AGN lightcurves follow broken power-laws intrinsically. The strong overlap of subsequent flares in cm-radio lightcurves leads to correlation of the observed flux over long timescales and thus to red-noise power spectra (simple power-law periodograms). Accordingly, we expect observations of smaller $\beta$ values and broken power-law periodograms at higher observing frequencies that probe shorter cooling timescales. This is indeed observed for the X/$\gamma$ ray lightcurves of blazars; for optical lightcurves, the case remains ambiguous.

\end{enumerate}

\acknowledgments
This work is based on observations obtained by the University of Michigan Radio Astronomy Observatory (UMRAO) supported by the National Science Foundation (NSF) of the U.S.A.. We are most grateful to \name{Margo F. Aller} and \name{Talvikki Hovatta} for making the data available to us. We thank \name{Markus B\"{o}ttcher} and \name{Talvikki Hovatta} for valuable discussions, which improved the paper. This study made use of the NASA/IPAC Extragalactic Database (NED). We acknowledge financial support from the Korean National Research Foundation (NRF) via Global Ph.D. Fellowship Grant 2014H1A2A1018695 (Jongho Park) and Basic Research Grant NRF-2015R1D1A1A01056807 (Sascha Trippe). Correspondence should be addressed to S.T.

~



\LongTables
\begin{deluxetable*}{ccccccc}

\tablecaption{Black hole masses and disk luminosities (in Eddington units) \label{table_BH}}
\small
\tablehead{
\colhead{Object} & \colhead{$\log M_{\rm BH}$} &
\colhead{$M_{\rm BH}$ Estimator} & \colhead{Ref.} &
\colhead{Adopted} & $L_{\rm disk} / L_{\rm Edd}$\\
& (1) & (2) & (3) & (4) & (5)
}
\startdata
\multirow{2}{*}{\centering 0133+476} & 8.75 & $L(\rm H\beta) = 20.97\times10^{42}\ erg\ s^{-1}$, $\rm FWHM(H\beta) = 4223\ km\ s^{-1}$ & \multirow{2}{*}{\centering Tor12} & \multirow{2}{*}{\centering 8.88} & \multirow{2}{*}{\centering 4.3e-2} \\
 & 9.01 & $L(\rm Mg\ II) = 28.10\times10^{42}\ erg\ s^{-1}$, $\rm FWHM(Mg\ II) = 5367\ km\ s^{-1}$ & &  &  \\
\hline
0235+164 & 9.00 & Fitting the spectral energy distribution (SED) & Ghi09 & 9.00 & 1.0e-2$^a$ \\
\hline
\multirow{3}{*}{\centering 0316+413} & 8.49 & $M_{\rm BH}$-$\sigma$ relation, $\sigma = 246 \rm \ km\ s^{-1}$ & Heck85 & \multirow{3}{*}{\centering 8.51} & \multirow{3}{*}{\centering 6.8e-4} \\
 & 8.51 & $M_{\rm BH}$-$\sigma$ relation, $\sigma = 248 \rm \ km\ s^{-1}$ & NW95 &  &  \\
 & 8.53 & H$_2$ rotation curve & Wil05 & & \\
 \hline
\multirow{2}{*}{\centering 0333+321} & 9.25 & Single epoch Mg II line & Liu06 & \multirow{2}{*}{\centering 9.56} & \multirow{2}{*}{\centering 1.0e-0} \\
 & 9.86 & $L(\rm Mg\ II) = 6091\times10^{42}\ erg\ s^{-1}$, $\rm FWHM(Mg\ II) = 3735\ km\ s^{-1}$ & Tor12 &  &  \\
\hline
\multirow{2}{*}{\centering 0336$-$019} & 8.89 & Single epoch H$\beta$ line & Liu06 & \multirow{2}{*}{\centering 8.97} & \multirow{2}{*}{\centering 6.6e-2} \\
 & 9.05 & $L(\rm Mg\ II) = 52.68\times10^{42}\ erg\ s^{-1}$, $\rm FWHM(Mg\ II) = 4781\ km\ s^{-1}$ & Tor12 &  &  \\
\hline
0415+379 & 8.46 & $L(\rm H\beta) = 7.85\times10^{42}\ erg\ s^{-1}$, $\rm FWHM(H\beta) = 4100\ km\ s^{-1}$ & Tor12 & 8.46 & 4.6e-2 \\
\hline
\multirow{2}{*}{\centering 0420$-$014} & 8.41 & Single epoch H$\beta$ line & Liu06 & \multirow{2}{*}{\centering 8.72} & \multirow{2}{*}{\centering 9.9e-2} \\
 & 9.02 & $L(\rm Mg\ II) = 43.37\times10^{42}\ erg\ s^{-1}$, $\rm FWHM(Mg\ II) = 4846\ km\ s^{-1}$ & Tor12 &  &  \\
\hline
\multirow{5}{*}{\centering 0430+052} & 7.36 & H$\beta$ reverberation mapping & Kas00 & \multirow{5}{*}{\centering 7.70} & \multirow{5}{*}{\centering 4.0e-2} \\
 & 8.13 &  $M_{\rm BH}$-$\sigma$ relation, $\sigma = 200 \rm \ km\ s^{-1}$  & Woo02 &  &  \\
 & 8.03 & H$\beta$ reverberation mapping & Kol14 &  &  \\
 & 7.62 & $L(\rm H\beta) = 1.29\times10^{42}\ erg\ s^{-1}$, $\rm FWHM(H\beta) = 2750\ km\ s^{-1}$ & Tor12 & & \\
 & 7.57 & $L(\rm H\beta) = 1.82\times10^{42}\ erg\ s^{-1}$, $\rm FWHM(H\beta) = 2328\ km\ s^{-1}$ & Ves06 & &\\
 \hline
 0607$-$157 & 7.78 & $L(\rm H\beta) = 1.44\times10^{42}\ erg\ s^{-1}$, $\rm FWHM(H\beta) = 3200\ km\ s^{-1}$ & Tor12 & 7.78 & 4.0e-2 \\
\hline
\multirow{4}{*}{\centering 0923+392} & 9.09 & Single epoch H$\beta$ line & Liu06 & \multirow{4}{*}{\centering 9.15} & \multirow{4}{*}{\centering 1.2e-1} \\
 & 9.57 & $L(\rm H\beta) = 77.18\times10^{42}\ erg\ s^{-1}$, $\rm FWHM(H\beta) = 7200\ km\ s^{-1}$ & Wang04 &  &  \\
 & 8.81 & $L(\rm H\beta) = 25.11\times10^{42}\ erg\ s^{-1}$, $\rm FWHM(H\beta) = 4250\ km\ s^{-1}$ & Tor12 &  &  \\
 & 9.14 & $L(\rm Mg\ II) = 73.05\times10^{42}\ erg\ s^{-1}$, $\rm FWHM(Mg\ II) = 4927\ km\ s^{-1}$ & Tor12 & & \\
 \hline
  1055+018 & 9.16 & $L(\rm Mg\ II) = 34.07\times10^{42}\ erg\ s^{-1}$, $\rm FWHM(Mg\ II) = 6039\ km\ s^{-1}$ & Tor12 & 9.16 & 2.6e-2 \\
\hline
\multirow{2}{*}{\centering 1101+384} & 8.29 & $M_{\rm BH}$-$\sigma$ relation, $\sigma = 219 \rm \ km\ s^{-1}$ & Bar03 & \multirow{2}{*}{\centering 8.35} & \multirow{2}{*}{\centering 1.5e-4$^a$} \\
 & 8.42 & $M_{\rm BH}$-$\sigma$ relation, $\sigma = 236 \rm \ km\ s^{-1}$ & Fal02 &  &  \\
 \hline
 \multirow{2}{*}{\centering 1156+295} & 8.54 & Single epoch H$\beta$ line & Liu06 & \multirow{2}{*}{\centering 8.68} & \multirow{2}{*}{\centering 8.9e-2} \\
 & 8.81 & $L(\rm Mg\ II) = 28.77\times10^{42}\ erg\ s^{-1}$, $\rm FWHM(Mg\ II) = 4245\ km\ s^{-1}$ & Tor12 &  &  \\
\hline
 \multirow{4}{*}{\centering 1226+023} & 8.74 & H$\alpha$, H$\beta$, H$\gamma$ reverberation mapping & Kas00 & \multirow{4}{*}{\centering 8.97} & \multirow{4}{*}{\centering 2.2e-1} \\
 & 8.92 & Single epoch H$\beta$ line & Liu06 &  &  \\
 & 9.00 & $L(\rm H\beta) = 85.17\times10^{42}\ erg\ s^{-1}$, $\rm FWHM(H\beta) = 3800\ km\ s^{-1}$ & Tor12 &  &  \\
 & 9.22 & $L(\rm H\beta) = 186.21\times10^{42}\ erg\ s^{-1}$, $\rm FWHM(H\beta) = 3627\ km\ s^{-1}$ & Ves06 & & \\
\hline
 1253$-$055 & 8.70 & $L(\rm C\ IV) = 26.61\times10^{42}\ erg\ s^{-1}$, $\rm FWHM(C\ IV) = 8613\ km\ s^{-1}$ & Tor12 & 8.70 & 3.1e-2 \\
\hline
\multirow{2}{*}{\centering 1308+326} & 8.77 & $L(\rm Mg\ II) = 29.95\times10^{42}\ erg\ s^{-1}$, $\rm FWHM(Mg\ II) = 4016\ km\ s^{-1}$ & Wang04 & \multirow{2}{*}{\centering 8.85} & \multirow{2}{*}{\centering 3.9e-2} \\
 & 8.93 & $L(\rm Mg\ II) = 21.33\times10^{42}\ erg\ s^{-1}$, $\rm FWHM(Mg\ II) = 5267\ km\ s^{-1}$ & Tor12 &  &  \\
\hline
1335$-$127 & 8.64 & $L(\rm Mg\ II) = 9.33\times10^{42}\ erg\ s^{-1}$, $\rm FWHM(Mg\ II) = 4602\ km\ s^{-1}$ & Wang04 & 8.64 & 2.4e-2 \\
\hline
\multirow{3}{*}{\centering 1510$-$089} & 8.20 & Single epoch H$\beta$ line & Liu06 & \multirow{3}{*}{\centering 8.40} & \multirow{3}{*}{\centering 1.3e-1} \\
 & 8.46 & $L(\rm H\beta) = 17.68\times10^{42}\ erg\ s^{-1}$, $\rm FWHM(H\beta) = 3180\ km\ s^{-1}$ & Wang04 &  &  \\
 & 8.54 & $L(\rm H\beta) = 21.94\times10^{42}\ erg\ s^{-1}$, $\rm FWHM(H\beta) = 3250\ km\ s^{-1}$ & Tor12 & & \\
 \hline
 \multirow{3}{*}{\centering 1633+382} & 8.67 & Single epoch Mg II line & Liu06 & \multirow{3}{*}{\centering 9.05} & \multirow{3}{*}{\centering 2.2e-1} \\
 & 9.27 & $L(\rm Mg\ II) = 78.33\times10^{42}\ erg\ s^{-1}$, $\rm FWHM(Mg\ II) = 5583\ km\ s^{-1}$ & Tor12 &  &  \\
 & 9.20 & $L(\rm C\ IV) = 318.55\times10^{42}\ erg\ s^{-1}$, $\rm FWHM(C\ IV) = 6499\ km\ s^{-1}$ & Tor12 & & \\
 \hline
  \multirow{5}{*}{\centering 1641+399} & 9.27 & Single epoch H$\beta$ line & Liu06 & \multirow{5}{*}{\centering 9.05} & \multirow{5}{*}{\centering 9.2e-2} \\
 & 9.14 & $L(\rm H\beta) = 46.51\times10^{42}\ erg\ s^{-1}$, $\rm FWHM(H\beta) = 5140\ km\ s^{-1}$ & Wang04 &  &  \\
 & 8.46 & $L(\rm H\beta) = 10.80\times10^{42}\ erg\ s^{-1}$, $\rm FWHM(H\beta) = 3700\ km\ s^{-1}$ & Tor12 & & \\
 & 9.32 & $L(\rm Mg\ II) = 102.18\times10^{42}\ erg\ s^{-1}$, $\rm FWHM(Mg\ II) = 5520\ km\ s^{-1}$ & Tor12 & & \\
 & 9.07 & $L(\rm C\ IV) = 192.96\times10^{42}\ erg\ s^{-1}$, $\rm FWHM(C\ IV) = 6710\ km\ s^{-1}$ & Tor12 & & \\
 \hline
 \multirow{2}{*}{\centering 1652+398} & 8.78 & $M_{\rm BH}$-$\sigma$ relation, $\sigma = 291 \rm \ km\ s^{-1}$ & Fal02 & \multirow{2}{*}{\centering 9.00} & \multirow{2}{*}{\centering 1.3e-4$^a$} \\
 & 9.21 & $M_{\rm BH}$-$\sigma$ relation, $\sigma = 372 \rm \ km\ s^{-1}$ & Bar03 &  &  \\
 \hline
1749+096 & 8.66 &  $M_{\rm bh}$-$L_{\rm bulge}$ relation & Fal03a & 8.66 & 8.8e-3$^a$ \\
\hline
 \multirow{3}{*}{\centering 1803+784} & 7.92 & Single epoch H$\beta$ line & Liu06 & \multirow{3}{*}{\centering 8.26} & \multirow{3}{*}{\centering 1.2e-1} \\
 & 8.17 & $L(\rm Mg\ II) = 13.38\times10^{42}\ erg\ s^{-1}$, $\rm FWHM(Mg\ II) = 2451\ km\ s^{-1}$ & Wang04 &  &  \\
 & 8.69 & $L(\rm H\beta) = 15.59\times10^{42}\ erg\ s^{-1}$, $\rm FWHM(H\beta) = 4320\ km\ s^{-1}$ & Tor12 & & \\
 \hline
 \multirow{2}{*}{\centering 1807+698} & 8.74 & $M_{\rm BH}$-$\sigma$ relation, $\sigma = 284 \rm \ km\ s^{-1}$ & Fal03b & \multirow{2}{*}{\centering 8.62} & \multirow{2}{*}{\centering 8.0e-5} \\
 & 8.51 & $M_{\rm BH}$-$\sigma$ relation, $\sigma = 249 \rm \ km\ s^{-1}$ & Bar03 &  &  \\
 \hline
 1921$-$293 & 9.14 & $L(\rm H\beta) = 7.35\times10^{42}\ erg\ s^{-1}$, $\rm FWHM(H\beta) = 9134\ km\ s^{-1}$ & Wang04 & 9.14 & 9.0e-3 \\
\hline
 \multirow{3}{*}{\centering 1928+738} & 8.35 & Single epoch H$\beta$ line & Liu06 & \multirow{3}{*}{\centering 8.57} & \multirow{3}{*}{\centering 1.9e-1} \\
 & 8.64 & $L(\rm H\beta) = 28.60\times10^{42}\ erg\ s^{-1}$, $\rm FWHM(H\beta) = 3360\ km\ s^{-1}$ & Wang04 &  &  \\
 & 8.73 & $L(\rm H\beta) = 39.20\times10^{42}\ erg\ s^{-1}$, $\rm FWHM(H\beta) = 3360\ km\ s^{-1}$ & Tor12 & & \\
 \hline
 2007+777 & 8.80 & $M_{\rm bh}$-$L_{\rm bulge}$ relation & Fal03a & 8.80 & 1.4e-3$^a$ \\
\hline
 \multirow{3}{*}{\centering 2134+004} & 8.52 & $L(\rm Mg II) = 39.48\times10^{42}\ erg\ s^{-1}$, $\rm FWHM(Mg II) = 2800\ km\ s^{-1}$ & Wang04 & \multirow{3}{*}{\centering 9.18} & \multirow{3}{*}{\centering 1.7e-1} \\
 & 9.44 & $L(\rm Mg\ II) = 235.34\times10^{42}\ erg\ s^{-1}$, $\rm FWHM(Mg\ II) = 5194\ km\ s^{-1}$ & Tor12 &  &  \\
 & 9.58 & $L(\rm C\ IV) = 794.82\times10^{42}\ erg\ s^{-1}$, $\rm FWHM(H\beta) = 7418\ km\ s^{-1}$ & Tor12 & & \\
 \hline
  \multirow{3}{*}{\centering 2145+067} & 8.87 & Single epoch Mg II line & Liu06 & \multirow{3}{*}{\centering 9.31} & \multirow{3}{*}{\centering 2.1e-1} \\
 & 9.64 & $L(\rm Mg\ II) = 457.33\times10^{42}\ erg\ s^{-1}$, $\rm FWHM(Mg\ II) = 5517\ km\ s^{-1}$ & Tor12 &  &  \\
 & 9.41 & $L(\rm C\ IV) = 663.85\times10^{42}\ erg\ s^{-1}$, $\rm FWHM(C\ IV) = 6423\ km\ s^{-1}$ & Tor12 & & \\
 \hline
  2200+420 & 8.77 & $M_{\rm bh}$-$L_{\rm bulge}$ relation & Fal03a & 8.77 & 3.8e-4$^a$ \\
\hline
 2223$-$052 & 8.57 & $L(\rm C\ IV) = 247.89\times10^{42}\ erg\ s^{-1}$, $\rm FWHM(C\ IV) = 3449\ km\ s^{-1}$ & Wang04 & 8.57 & 3.9e-1 \\
\hline
\multirow{2}{*}{\centering 2230+114} & 9.08 & $L(\rm Mg\ II) = 71.21\times10^{42}\ erg\ s^{-1}$, $\rm FWHM(Mg\ II) = 4583\ km\ s^{-1}$ & \multirow{2}{*}{\centering Tor12} & \multirow{2}{*}{\centering 9.07} & \multirow{2}{*}{\centering 9.7e2} \\
 & 9.06 & $L(\rm C\ IV) = 255.71\times10^{42}\ erg\ s^{-1}$, $\rm FWHM(C\ IV) = 5990\ km\ s^{-1}$ & &  &  \\
\hline
 \multirow{3}{*}{\centering 2251+158} & 8.86 & Single epoch H$\beta$ line & Liu06 & \multirow{3}{*}{\centering 9.15} & \multirow{3}{*}{\centering 1.8e-1} \\
 & 9.30 & $L(\rm Mg\ II) = 125.98\times10^{42}\ erg\ s^{-1}$, $\rm FWHM(Mg\ II) = 5162\ km\ s^{-1}$ & Tor12 &  &  \\
 & 9.29 & $L(\rm C\ IV) = 250.01\times10^{42}\ erg\ s^{-1}$, $\rm FWHM(C\ IV) = 7875\ km\ s^{-1}$ & Tor12 & &
\enddata
\tablecomments{(1) Logarithm of the black hole mass calculated by us or taken from the literature. (2) Methods used to determine black hole masses. We do not name the method if we adopt literature values without modification. We converted the luminosities provided in \cite{Wang2004} into the values in the table by converting their $L_{\rm BLR}$ into line luminosities and updated cosmological parameters. (3) References for the black hole masses (listed below). (4) Adopted values for the (logarithmic) black hole masses, which are the averages of (1) for each source. (5) Disk luminosities in Eddington units.\\ $^a$ $L_{\rm disk} / L_{\rm Edd}$ calculated from the $L_{\rm disk}$ values in \cite{Ghisellini2011} or from using their $L_{\rm BLR}$ values, assuming $L_{\rm disk} \approx 10L_{\rm BLR}$ \citep{Ghisellini2011, Calderone2013} if $L_{\rm disk}$ is not available.\\\\ REFERENCES.--- Bar03: \cite{Barth2003}; Fal02: \cite{Falomo2002}; Fal03a: \cite{Falomo2003a}; Fal03b: \cite{Falomo2003b}; Ghi09: \cite{Ghisellini2009b}; Heck85: \cite{Heckman1985}; Kas00: \cite{Kaspi2000}; Kol14: \cite{Kollatschny2014}; Liu06: \cite{Liu2006}; NW95: \cite{Nelson1995}; Tor12: \cite{Torrealba2012}; Ves06: \citealt{Vestergaard2006}; Wang04: \cite{Wang2004}; Wil05: \cite{Wilman2005}}

\end{deluxetable*}


\begin{thebibliography}{}


\bibitem[Abdo et al.(2010)]{Abdo2010} Abdo, A. A., Ackermann, M., Ajello, M., et al. 2010, \apj, 722, 520
\bibitem[Abramowicz et al.(1991)]{Abramowicz1991} Abramowicz, M.A., Bao, G., Lanza, A., \& Zhang, X. -H. 1991, \aap, 245, 454
\bibitem[Aller et al.(1985)]{Aller1985} Aller, H. D., Aller, M. F., Latimer, G. E., \& Hodge, P. E. 1985, \apjs, 59, 513
\bibitem[Aller et al.(2003)]{Aller2003} Aller, M. F., Aller, H. D., \& Hughes, P. A. 2003, \apj, 586, 33
\bibitem[Arshakian et al.(2010)]{Arshakian2010} Arshakian, T. G., Le\'{o}n-Tavares, J., Lobanov, A. P., et al. 2010, \mnras, 401, 1231
\bibitem[Barth et al.(2003)]{Barth2003} Barth, A. J., Ho, L., C., \& Sargent, W. L. W. 2003, \apj, 583, 134
\bibitem[Baum et al.(1995)]{Baum1995} Baum, S. A., Zirbel, E. L., \& O'Dea, C. P. 1995, \apj, 451, 88
\bibitem[Benlloch et al.(2001)]{Benlloch2001} Benlloch, S., Wilms, J., Edelson, R., et al. 2001, \apj, 562, L121
\bibitem[Bettoni et al.(2003)]{Bettoni2003} Bettoni, D., Falomo, R., Fasano, G., \& Govoni, F. 2003, \aap, 399, 869
\bibitem[Blandford \& K\"{o}nigl(1979)]{BK1979} Blandford, R. D., \& K\"{o}nigl, A. 1979, \apj, 232, 34
\bibitem[Bloom \& Marscher(1996)]{BM1996} Bloom, S. D., \& Marscher, A. P. 1996, \apj, 461, 657
\bibitem[B\"{o}ttcher \& Chiang(2002)]{BC2002} B\"{o}ttcher, M. \& Chiang, J. 2002, \apj, 581, 127
\bibitem[Calderone et al.(2013)]{Calderone2013} Calderone, G., Ghisellini, G., Colpi, M., \& Dotti, M. 2013, \mnras, 431, 210
\bibitem[Cavaliere \& D'Elia(2002)]{Cavaliere2002} Cavaliere, A., \& D'Elia, V. 2002, \apj, 571, 226
\bibitem[Cawthorne(2006)]{Cawthorne2006} Cawthorne, T. V. 2006, \mnras, 367, 851
\bibitem[Cawthorne et al.(2013)]{Cawthorne2013} Cawthorne, T. V., Jorstad, S. G., \& Marscher, A. P. 2013, \apj, 772, 14
\bibitem[Celotti et al.(1997)]{Celotti1997} Celotti, A., Padovani, P., \& Ghisellinni, G. 1997, \mnras, 286, 415
\bibitem[Chatterjee et al.(2008)]{Chatterjee2008} Chatterjee, R., Jorstad, S. G., Marscher, A. P., et al. 2008, \apj, 689, 79
\bibitem[Chatterjee et al.(2009)]{Chatterjee2009} Chatterjee, R., Marscher, A. P., Jorstad, S. G., et al. 2009, \apj, 704, 1689
\bibitem[Chatterjee et al.(2011)]{Chatterjee2011} Chatterjee, R., Marscher, A. P., Jorstad, S. G., et al. 2011, \apj, 734, 43
\bibitem[Chatterjee et al.(2012)]{Chatterjee2012} Chatterjee, R., Bailyn, C. D., Bonning, E. W., et al. 2012, \apj, 749, 191
\bibitem[Ciaramella et al.(2004)]{Ciaramella2004} Ciaramella, A., Bongardo, C., Aller, H.D., et al. 2004, \aap, 419, 485
\bibitem[Cohen et al.(2014)]{Cohen2014} Cohen, M. H., Meier, D. L., Arshakian, T. G., et al. 2014, \apj, 787, 151
\bibitem[Daly \& Marscher(1988)]{DM1988} Daly, R. A., \& Marscher, A. P. 1988, \apj, 334, 539
\bibitem[Do et al.(2009)]{Do2009} Do, T., Ghez, A.M., Morris, M.R., et al. 2009, \apj, 691, 1021
\bibitem[Emmanoulopoulos et al.(2010)]{Emmanoulopoulos2010} Emmanoulopoulos, D., M$^{\rm c}$Hardy, I. M., \& Uttley, P. 2010, \mnras, 404, 931
\bibitem[Falconer(1990)]{Falconer1990} Falconer, K. 1990, Fractal Geometry: Mathematical Foundations and Applications (Chicester: Wiley)
\bibitem[Falomo et al.(2002)]{Falomo2002} Falomo, R., Kotilainen, J. K., \& Treves, A. 2003, \apj, 569, L35
\bibitem[Falomo et al.(2003a)]{Falomo2003a} Falomo, R., Carangelo, N., \& Treves, A. 2003, \mnras, 343, 505
\bibitem[Falomo et al.(2003b)]{Falomo2003b} Falomo, R., Kotilainen, J. K., Carangelo, N., \& Treves, A. 2003, \apj, 595, 624
\bibitem[Fan(1999)]{Fan1999} Fan, J. H. 1999, \mnras, 308, 1032
\bibitem[Ferrarese \& Ford(2005)]{Ferrarese2005} Ferrarese, L., \& Ford, H. 2005, Space Sci. Rev. 116, 523
\bibitem[Finke \& Becker(2014)]{Finke2014} Finke, J. D., \& Becker, P. A. 2014, \apj, 791, 21
\bibitem[Finke \& Becker(2015)]{Finke2015} Finke, J. D., \& Becker, P. A. 2015, \apj, 809, 85
\bibitem[Francis et al.(1991)]{Francis1991} Francis, P. J., Hewett, P. C., Foltz, C. B., et al. 1991, \apj, 373, 465
\bibitem[Fromm et al.(2011)]{Fromm2011} Fromm, C. M., Perucho, M., Ros, E., et al. 2011, \aap, 531, 95
\bibitem[Gu et al.(2001)]{Gu2001} Gu, M., Cao, X., \& Jiang, D. R. 2001, \mnras, 327, 1111
\bibitem[Gupta et al.(2012)]{Gupta2012} Gupta, A. C., Krichbaum, T. P., Wiita, P. J., et al. 2012, \mnras, 425, 1357
\bibitem[Gebhardt et al.(2000)]{Gebhardt2000} Gebhardt, K., Bender, R., Bower, G., et al. 2000, \apj, 539, L13
\bibitem[Ghisellini et al.(2009a)]{Ghisellini2009a} Ghisellini, G., Maraschi, L., \& Tavecchio, F. 2009, \mnras, 396, L105
\bibitem[Ghisellini et al.(2009b)]{Ghisellini2009b} Ghisellini, G., Tavecchio, F., \& Ghirlanda, G. 2009, \mnras, 399, 2041
\bibitem[Ghisellini et al.(2011)]{Ghisellini2011} Ghisellini, G., Tavecchio, F., Foschini, L., \& Ghirlanda, G. 2011, \mnras, 414, 2674
\bibitem[Ghisellini et al.(2014)]{Ghisellini2014} Ghisellini, G., Tavecchio, F., Maraschi, L., et al. 2014, \nat, 515, 376
\bibitem[Heckman et al.(1985)]{Heckman1985} Heckman, T. M., Illingworth, G. D., Miley, G. K., et al. 1985, \apj, 299, 41
\bibitem[Heckman \& Best(2014)]{Heckman2014} Heckman, T. M., \& Best, P. N. 2014, \araa, 52, 589
\bibitem[Ho et al.(2012)]{Ho2012} Ho, L. C., Goldoni, P., Dong, X.-B., et al. 2012, \apj, 754, 11
\bibitem[Hovatta et al.(2007)]{Hovatta2007} Hovatta, T., Tornikoski, M., Lainela, M., et al. 2007, \aap, 469, 899
\bibitem[Hovatta et al.(2008)]{Hovatta2008} Hovatta, T., Nieppola, E., Tornikoski, M., et al. 2008, \aap, 485, 51
\bibitem[Hovatta et al.(2009)]{Hovatta2009} Hovatta, T., Valtaoja, E., Tornikoski, M., \& L\"ahteenm\"aki, A., 2009, \aap, 494, 527
\bibitem[Hughes et al.(1985)]{Hughes1985} Hughes, P. A., Aller, H. D., \& Aller, M. F. 1985, \apj, 298, 301
\bibitem[Hughes et al.(1992)]{Hughes1992} Hughes, P. A., Aller, H. D., \& Aller, M. F. 1992, \apj, 396, 469
\bibitem[Isobe et al.(2015)]{Isobe2015} Isobe, N., Sato, R., Ueda, Y., et al. 2015, \apj, 798, 27
\bibitem[Jorstad et al.(2005)]{Jorstad2005} Jorstad, S. G., Marscher, A. P., Lister, M. L., et al. 2005, \apj, 130, 1418
\bibitem[Jorstad et al.(2010)]{Jorstad2010} Jorstad, S. G., Marscher, A. P., Larionov, V. M., et al. 2010, \apj, 715, 362
\bibitem[Kaspi et al.(2000)]{Kaspi2000} Kaspi, S., Smith, P. S., Netzer, H., et al. 2000, \apj, 533, 631
\bibitem[Kaspi et al.(2005)]{Kaspi2005} Kaspi, S., Maoz, D., Netzer, H., et al. 2005, \apj, 629, 61
\bibitem[Katarzy\'{n}ski et al.(2001)]{Katarzynski2001} Katarzy\'{n}ski, K., Sol, H., \& Kus, A. 2001, \aap, 367, 809
\bibitem[Kelly(2007)]{Kelly2007} Kelly, B. C. 2007, \apj, 665, 1489
\bibitem[Kelly et al.(2009)]{Kelly2009} Kelly, B. C., Bechtold, J., \& Siemiginowska, A. 2009, \apj, 698, 895
\bibitem[Kelly et al.(2011)]{Kelly2011} Kelly, B. C., Sobolewska, M., \& Siemiginowska, A. 2011, \apj, 730, 52
\bibitem[Kollatschny et al.(2014)]{Kollatschny2014} Kollatschny, W., Ulbrich, K., Zetzl, M., et al. 2014, \aap, 566, 106
\bibitem[Kormendy \& Ho(2013)]{Kormendy2013} Kormendy, J., \& Ho, L. C. 2013, \araa, 51, 511
\bibitem[Kormendy \& Richstone(1995)]{Kormendy1995} Kormendy, J., \& Richstone, D. 1995, \araa, 33, 581
\bibitem[L\"{a}hteenmaki et al.(1999)]{Lahteenmaki1999} L\"{a}hteenm\"{a}ki, A., Valtaoja, E., \& Wiik, K. 1999, \apj, 511, 112
\bibitem[L\"{a}hteenmaki \& Valtaoja(1999)]{LV1999} L\"{a}hteenm\"{a}ki, A., \& Valtaoja, E. 1999, \apj, 521, 493
\bibitem[Lawrence et al.(1987)]{Lawrence1987} Lawrence, A., Watson, M. G., Pounds, K. A., \& Elvis, M. 1987, Nature, 325, 694
\bibitem[Le\'{o}n-Tavares et al.(2010)]{Leon-Tavares2010} Le\'{o}n-Tavares, J., Lobanov, A. P., Chavushyan, V. H., et al. 2010, \apj, 715, 335
\bibitem[Lico et al.(2012)]{Lico2012} Lico, R., Giroletti, M., Orienti, M., et al. 2012, \aap, 545, 117
\bibitem[Liodakis \& Pavlidou(2015)]{Liodakis2015} Liodakis, I., \& Pavlidou, V. 2015, \mnras, 454, 1767
\bibitem[Lister \& Homan(2005)]{LH2005} Lister, M. L., \& Homan, D. C. 2005, \apj, 130, 1389
\bibitem[Lister et al.(2013)]{Lister2013} Lister, M. L., Aller, M. F., Aller, H. D., et al. 2013, \apj, 146, 120
\bibitem[Liu et al.(2006)]{Liu2006} Liu, Y., Jiang, D. R., \& Gu, M. F. 2006, \apj, 637, 669
\bibitem[Lyubarskii(1997)]{Lyubarskii1997} Lyubarskii, Y. E. 1997, \mnras, 292, 679
\bibitem[Lyutikov et al.(2005)]{Lyutikov2005} Lyutikov, M., Pariev, V. I., \& Gabuzda, D. C. 2005, \mnras, 360, 869
\bibitem[Macchetto et al.(1997)]{Macchetto1997} Macchetto, F., Marconi, A., Axon, D. J., et al. 1997, \apj, 489, 579
\bibitem[Maraschi \& Rovetti(1994)]{Maraschi1994} Maraschi, L., \& Rovetti, F. 1994, \apj, 436, 79
\bibitem[Marscher \& Gear(1985)]{Marscher1985} Marscher, A. P., \& Gear, W. K. 1985, \apj, 298, 114
\bibitem[Marscher(1996)]{Marscher1996} Marscher, A. P. 1996, in ASP Conf. Ser. 110, Blazar Continuum Variability, ed.
H. R. Miller, J. R. Webb, \& J. C. Noble (San Francisco, CA: ASP), 248
\bibitem[Marscher \& Travis(1996)]{MT1996} Marscher, A. P., \& Travis, J. P. 1996, \aaps, 120, 537
\bibitem[Marscher et al.(2002)]{Marscher2002} Marscher, A. P., Jorstad, S. G., Mattox, J. R., \& Wehrle, A. E. 2002, \apj 577, 85
\bibitem[Marscher(2006)]{Marscher2006} Marscher, A. P. 2006, in AIP Conf. Proc. 856, Relativistic Jets: The Common
Physics of AGN, Microquasars, and Gamma Ray Bursts, ed. P. A. Hughes \& J. N. Bregman (Melville, NY: AIP), 1
\bibitem[Marscher et al.(2008)]{Marscher2008} Marscher, A. P., Jorstad, S. G., D'Arcangelo, F. D., et al. 2008, \nat, 452, 966
\bibitem[Marscher et al.(2010)]{Marscher2010} Marscher, A. P., Jorstad, S. G., Larionov, V. M., et al. 2010, \apj, 710, L126
\bibitem[Marscher et al.(2011)]{Marscher2011} Marscher, A. P., Jorstad, S. G., Agudo, I., MacDonald, N. R., \& Scott, T. L.
2012, arXiv:1204.6707
\bibitem[Marscher(2013)]{Marscher2013} Marscher, A. P. 2013, EPJWC, 61, 04001
\bibitem[Marscher(2014)]{Marscher2014} Marscher, A. P. 2014, \apj, 780, 87
\bibitem[Mastichiadis \& Kirk(1997)]{MK1997} Mastichiadis, A., \& Kirk, J. G. 1997, \aap, 320, 19
\bibitem[Max-Moerbeck et al.(2014a)]{MaxM2014a} Max-Moerbeck, W., Hovatta, T., Richards, J. L., et al. 2014, \mnras, 445, 428
\bibitem[Max-Moerbeck et al.(2014b)]{MaxM2014b} Max-Moerbeck, W., Richards, J. L., Hovatta, T., et al. 2014, \mnras, 445, 437
\bibitem[M$^{\rm c}$Hardy et al.(2004)]{McHardy2004} M$^{\rm c}$Hardy, I. M., Papadakis, I. E., Uttley, P., et al. 2004, \mnras, 348, 783
\bibitem[M$^{\rm c}$Hardy et al.(2006)]{McHardy2006} M$^{\rm c}$Hardy, I. M., Koerding, E., Knigge, C., et al. 2006, \nat, 444, 730
\bibitem[M$^{\rm c}$Hardy(2008)]{McHardy2008} M$^{\rm c}$Hardy, I. 2008, Proc. Blazar Variability across the Electromagnetic Spectrum, PoS (BLAZARS2008) (Trieste: SISSA Proc. Sci), 14
\bibitem[McLure \& Dunlop(2001)]{McLure2001} McLure, R. J., \& Dunlop, J. S. 2001, \mnras, 327, 199
\bibitem[McLure \& Jarvis(2002)]{McLure2002} McLure, R. J., \& Jarvis, M. J. 2002, \mnras, 337, 109
\bibitem[Merritt \& Ferrarese(2001)]{Merritt2001} Merritt, D., \& Ferrarese, L. 2001, \apj, 547, 140
\bibitem[Mohan et al.(2015)]{Mohan2015} Mohan, P., Agarwal, A., Mangalam, A., et al. 2015, \mnras, 452, 2004
\bibitem[Nelson \& Whittle(1995)]{Nelson1995} Nelson, C. H., \& Whittle, M. 1995, \apjs, 99, 67
\bibitem[Netzer(2013)]{Netzer2013} Netzer, H. 2013, The Physics and Evolution of Active Galactic Nuclei (Cambridge: Cambridge University Press)
\bibitem[Nieppola et al.(2009)]{Nieppola2009} Nieppola, E., Hovatta, T., Tornikoski, M., et al. 2009, \aj, 137, 5022
\bibitem[O'Sullivan \& Gabuzda(2009)]{OG2009} O'Sullivan, S. P. \& Gabuzda, D. C. 2009, MNRAS, 393, 429
\bibitem[Padovani(1992)]{Padovani1992} Padovani, P. 1992, \mnras, 257, 404
\bibitem[Papadakis \& Lawrence(1993)]{Papadakis1993} Papadakis, I. E., \& Lawrence, A. 1993, \mnras, 261, 612
\bibitem[Park et al.(2012)]{Park2012} Park, D., Woo, J.-H., Treu, T., et al. 2012, \apj, 747, 30
\bibitem[Park \& Trippe(2014)]{Park2014} Park, J.-H., \& Trippe, S. 2014, \apj, 785, 76
\bibitem[Press(1978)]{Press1978} Press, W. H. 1978, Comment. Astrophys., 7, 103
\bibitem[Press \& Rybicki(1989)]{Press1989} Press, W. H., \& Rybicki, G. B. 1989, \apj, 338, 277
\bibitem[Press et al.(1992)]{Press1992} Press, W. H., Teukolsky, S. A., Vetterling, W. T., \& Flannery, B. P. 1992, Numerical Recipes (2nd ed.; Cambridge: Cambridge Univ. Press)
\bibitem[Priestley(1981)]{Priestley1981} Priestley, M. B. 1981, Spectral Analysis and Time Series (London: Elsevier)
\bibitem[Pyatunina et al.(2006)]{Pyatunina2006} Pyatunina, T. B., Kudryavtseva, N. A., Gabuzda, D. C., et al. 2006, \mnras, 373, 1470
\bibitem[Pyatunina et al.(2007)]{Pyatunina2007} Pyatunina, T. B., Kudryavtseva, N. A., Gabuzda, D. C., et al. 2007, \mnras, 381, 797
\bibitem[Ramakrishnan et al.(2015)]{Ramakrishnan2015} Ramakrishnan, V., Hovatta, T., Nieppola, E., et al. 2015, \mnras, 452, 1280
\bibitem[Ramakrishnan et al.(2016)]{Ramakrishnan2016} Ramakrishnan, V., Hovatta, T., Tornikoski, M., et al. 2016, \mnras, 456, 171
\bibitem[Rani et al.(2009)]{Rani2009} Rani, B., Wiita, P. J., \& Gupta, A. C. 2009, \apj, 696, 2170
\bibitem[Rani et al.(2010)]{Rani2010} Rani, B., Gupta, A. C., Joshi, U. C., et al. 2010, \apj, 719, L153
\bibitem[Rani et al.(2013)]{Rani2013} Rani, B., Krichbaum, T. P., Fuhrmann, L., et al. 2013, \aap, 552, 11
\bibitem[Readhead(1994)]{Readhead1994} Readhead, A. C. S. 1994, \apj, 426, 51
\bibitem[S\'{a}nchez et al.(2005)]{Sanchez2005} S\'{a}nchez, N., Alfaro, E. J., \& P\'{e}rez, E. 2005, \apj, 625, 849
\bibitem[S\'{a}nchez et al.(2010)]{Sanchez2010} S\'{a}nchez, N., A\~{n}ez, N., Alfaro, E. J., \& Odekon, M. C. 2010, \apj, 720, 541
\bibitem[Savolainen et al.(2002)]{Savolainen2002} Savolainen, T., Wiik, K., Valtaoja, E., et al. 2002, \aap, 394, 851
\bibitem[Savolainen et al.(2006)]{Savolainen2006} Savolainen, T., Wiik, K., Valtaoja, E., \& Tornikoski, M. 2006, \aap, 446, 71
\bibitem[Savolainen et al.(2010)]{Savolainen2010} Savolainen, T., Homan, D. C., Hovatta, T., et al. 2010, \aap, 512, 24
\bibitem[Scargle(1982)]{Scargle1982} Scargle, J. D. 1982, \apj, 263, 835
\bibitem[Shakura \& Sunyaev(1973)]{Shakura1973} Shakura, N. I., \& Sunyaev, R. A. 1973, \aap, 24, 337
\bibitem[Shimizu \& Mushotzky(2013)]{Shimizu2013} Shimizu, T. T., \& Mushotzky, R. F. 2013, \apj, 770, 60
\bibitem[Simonetti et al.(1985)]{Simonetti1985} Simonetti, J. H., Cordes, J. M., \& Heeschen, D. S. 1985, \apj, 296, 46
\bibitem[Sobolewska et al.(2014)]{Sobolewska2014} Sobolewska, M. A., Siemiginowska, A., Kelly, B. C., \& Nalewajko, K. 2014, \apj, 786, 143
\bibitem[Tavecchio et al.(1998)]{Tavecchio1998} Tavecchio, F., Maraschi, L., \& Ghisellini, G. 1998, \apj, 509, 608
\bibitem[Tchekhovskoy et al.(2011)]{Tchekhovskoy2011} Tchekhovskoy, A., Narayan, R., \& McKinney, J. C. 2011, \mnras, 418, L79
\bibitem[Timmer \& K\"{o}nig(1995)]{Timmer1995} Timmer, J., \& K\"{o}nig, M. 1995, \apj, 300, 707
\bibitem[Torrealba et al.(2012)]{Torrealba2012} Torrealba, J., Chavushyan, V., Cruz-Gonz\'{a}lez, I., et al. 2012, Rev. Mex. Astron. Astrofis., 48, 9
\bibitem[Tremaine et al.(2002)]{Tremaine2002} Tremaine, S., Gebhardt, K., Bender, R., et al. 2002, \apj, 574, 740
\bibitem[Trippe et al.(2011)]{Trippe2011} Trippe, S., Krips, M., Pi\'{e}tu, V., et al. 2011, \aap, 533, 97
\bibitem[Trippe(2015)]{Trippe2015} Trippe, S. 2015, JKAS, 48, 203
\bibitem[Ulrich et al.(1997)]{Ulrich1997} Ulrich, M.-H., Maraschi, L., \& Urry, C. M. 1997, ARAA, 35, 445
\bibitem[Uttley et al.(2002)]{Uttley2002} Uttley, P., M$^{\rm c}$Hardy, I. M., \& Papadakis, I. E. 2002, \mnras, 332, 231
\bibitem[Uttley \& M$^{\rm c}$Hardy(2005)]{UM2005} Uttley, P., \& M$^{\rm c}$Hardy, I. M. 2005, \mnras, 363, 586
\bibitem[Valtaoja et al.(1992)]{Valtaoja1992} Valtaoja, E., Ter\"asranta, H., Urpo, S., et al. 1992, \aap, 254, 71
\bibitem[Valtaoja et al.(1999)]{Valtaoja1999} Valtaoja, E., L\"ahteenm\"aki, A., Ter\"asranta, H., \& Lainela, M. 1999, \apjs, 120, 95
\bibitem[Vaughan(2005)]{Vaughan2005} Vaughan, S. 2005, \aap, 431, 391
\bibitem[Vaughan(2010)]{Vaughan2010} Vaughan, S. 2010, \mnras, 402, 307
\bibitem[Vestergaard \& Peterson(2006)]{Vestergaard2006} Vestergaard, M., \& Peterson, B. M. 2006, \apj, 641, 689
\bibitem[Vestergaard \& Osmer(2009)]{Vestergaard2009} Vestergaard, M., \& Osmer, P. S. 2009, \apj, 699, 800
\bibitem[Wall \& Jenkins(2012)]{Wall2012} Wall, J. V., \& Jenkins, C. R. 2012, Practical Statistics for Astronomers (Cambridge: Cambridge Univ. Press)
\bibitem[Wang et al.(2004)]{Wang2004} Wang, J.-M., Luo, B., \& Ho, L. C. 2004, \apj, 615, L9
\bibitem[Wilman et al.(2005)]{Wilman2005} Wilman, R. J., Edge, A. C., \& Johnstone, R. M. 2005, \mnras, 359, 755
\bibitem[Woo \& Urry(2002)]{Woo2002} Woo, J.-H., \& Urry, C. M. 2002, \apj, 579, 530
\bibitem[Zamaninasab et al.(2014)]{Zamaninasab2014} Zamaninasab, M., Clausen-Brown, E., Savolainen, T., \& Tchekhovskoy, A. 2014, \nat, 510, 126

\end{thebibliography}
\end{document}